\documentclass[runningheads]{llncs}
\usepackage{graphicx}
\usepackage{booktabs}
\usepackage[round]{natbib}
\usepackage{float}
\usepackage[caption=false]{subfig} 

\begin{document}
\title{Automatic Monitoring Social Dynamics During Big Incidences: A Case Study of COVID-19 in Bangladesh}
%
\titlerunning{Automatic Monitoring Social Dynamics During Big Incidences}

\author{Fahim Shahriar
\inst{1} \and
Md Abul Bashar \inst{2}}
%
%
\institute{Comilla University, Cumilla, Bangladesh \\ 
\email{imnirobs15@gmail.com} \and
Queensland University of Technology, Brisbane, Australia \\
\email{m1.bashar@qut.edu.au} }
\maketitle              
%
\begin{abstract}
Newspapers are trustworthy media where people get the most reliable and credible information compared with other sources. On the other hand, social media often spread rumors and misleading news to get more traffic and attention. Careful characterization, evaluation, and interpretation of newspaper data can provide insight into intrigue and passionate social issues to monitor any big social incidence. This study analyzed a large set of spatio-temporal Bangladeshi newspaper data related to the COVID-19 pandemic. The methodology included volume analysis, topic analysis, automated classification, and sentiment analysis of news articles to get insight into the COVID-19 pandemic in different sectors and regions in Bangladesh over a period of time. This analysis will help the government and other organizations to figure out the challenges that have arisen in society due to this pandemic, what steps should be taken immediately and in the post-pandemic period, how the government and its allies can come together to address the crisis in the future, keeping these problems in mind.
\end{abstract}


%
%
\section{Introduction}
The outbreak of COVID-19 has brought serious health and economic consequences to society. It triggered one of the largest recessions in the world. Travel and currency companies lost billions of dollars, global stock markets plummeted, schools were closed, and the health care system was exhausted. Mental and social problems arose as people started to worry about infection, losing friends and family, losing their jobs, or isolation. 

Bangladesh has not been rid of this terrible virus. The virus has had major impacts on people's lives and significantly degraded quality of life. There were significant cases of infections and deaths. The hospital did not have adequate treatment facilities, including doctors, beds, and emergency supplies. Besides the health crisis, people have suffered enormous economic losses. Many people have lost their jobs; companies lost revenues, many of them go bankrupt. The most affected were the day-laborer and low-income workers. The lockdown in the pandemic suppressed their income. Many workers starved since their livelihood was cut-off. Working people took to the streets in search of their livelihood. They started protesting in the streets for relief. Seeing their plight, many people, including the government, came forward to help them.  Because of the lockdown, the international transport system was shut down, and stopped imports and exports. As a result, the country's industry suffered miserably. Objective monitoring and analysis of social dynamics during such a big incident can help the government and other authorities decide and take initiatives where required. This research proposes utilizing articles published in newspapers to objectively monitor and analyze social dynamics during a big incidence, such as the COVID-19 pandemic.

Newspapers are one of the most popular mass media in our daily life. Newspapers provide information on all the country's financial, political, social, environmental, etc. Whether it is a public campaign, an emergency, or a provocation, newspapers are a great resource for keeping track of internal and external events and stories. This mass media generally provide authentic information, whereas social media such as Facebook and Twitter often spread rumors and cannot be relied upon for authentic news. Effective classification, analysis, and interpretation of newspaper data can provide a deep understanding of any big incident in a society.

In this research, we analyzed a large spatio-temporal dataset of Bangladeshi Daily Newspapers related to COVID-19. The approach incorporated volume analysis, topic analysis, automatic classification of news articles, and sentiment analysis to better understand the COVID-19 pandemic in Bangladesh's divisions and districts over time. The experimental results and analysis will give an objective insight into the COVID-19 pandemic in Bangladesh that will benefit the government and other authorities for disseminating resources. This paper especially shows how to utilize automatic techniques for monitoring social dynamics in big incidents such as a pandemic, natural disaster, and social unrest. 

This research makes the following main contributions. (1) It collects, manually classifies, and publishes a large collection of COVID-19 related Bangladeshi news articles in Bengali and English. (2) It investigates the topics discussed during the COVID-19 pandemic in Bangladesh and how they have changed over time using manual and automatic techniques. (3) It designs a CNN-BiLSTM architecture for analyzing sentiment in Bengali text. (4) It analyzes COVID-19 related sentiments in the community over time and space. (5) It automatically categorizes documents into classes of observation interest for monitoring social interests.

The rest of the paper is organized as follows: Section \ref{sec:related_work} discusses related work, Sections \ref{sec:method_data} discussed methodology and data collection, Section \ref{sec:exp_res} presents experimental results, and Section \ref{sec:conclusion} concludes the paper. 

\section{Related Work}
\label{sec:related_work}
In this segment, we will discuss some past related works done by different analysts. We will divide it into four sections: Static Topic Modeling, Dynamic Topic Modeling, Sentiment Analysis, and Text Classification.

\subsection{Static Topic Modeling}
Topic modeling is a process of discovering hidden topics in a collection of texts \cite{bashar2020topic,balasubramaniam2020understanding}. It can be considered as a factual show of topics through text mining. One of the most popular topics modeling technique Latent Dirichlet Allocation (LDA) \citep{blei2003latent,bashar2020topic} discovers topics based on word recurrence in a set of documents. LDA is incredibly valuable for finding a sensibly precise blend of topics inside a given record. 

Topic modeling has been well studied for English text mining. For instance, \cite{zhao2011comparing} used unsupervised topic modeling in their research and compared the content of Twitter with the traditional news media \emph{“New York Times”}. They used the Twitter-LDA model to find topics from a representative sample of the entire Twitter and then used text mining techniques to compare these Twitter topics with \emph{New York Times}’ topics, taking into account the topic category and type. \cite{wang2011collaborative} developed an algorithm to recommend scientific articles to users in online communities. Their method combines the advantages of traditional collaborative filtering and probabilistic topic modeling. They applied collaborative topic modeling for recommending scientific articles. \cite{wayasti2018mining} applied the Latent Dirichlet Allocation function in the research and extracted topics based on ride-hailing customers’ posts on Twitter. In their research, they used 40 parameter combinations of LDA to obtain the best combination of topics. According to the perplexity value, the customers discussed 9 topics in the post, including keywords for each topic. \cite{tong2016text} recommended two experiments to build topic models on Wikipedia articles and Twitter users’ tweets.

However, topic modeling has not been well studied for Bengali text mining, unlike English text mining. \cite{das2010topic} used topic wise opinion summarization from Bengali text. They applied K-Means clustering and \emph{document-level theme relational graph} representation. However, they did not use any topic modeling technique, such as LDA.  \cite{rakshit2015automated} applied a Multi-class SVM classifier for analyzing Bengali poetry and poet relations. They performed a subject-wise classification of poems into foreordained categories. \cite{hasan2019topic} compared the performance of the LDA and LDA2vec topic model in Bengali Newspaper. \cite{al2018topic} used LDA for detecting the primary topics from a Bengali news corpus. However, they did not directly apply LDA in the Bengali text. Instead, they translated the Bengali text into English and then applied LDA to detect the topics. \cite{rahman2019sentence} used lexical analysis for sentence wise topic modeling. Their topic modeling was based on sentiment analysis. None of the existing works used Bengali text topic modeling for monitoring a pandemic or a major event.

In addition to English and Bengali, topic modeling in various languages is also studied. \cite{de2020infoveillance} analyzed a system that uses NLP pipelines, a theoretical framework for content aging to determine the qualitative parameters of tweets, and co-occurrence analysis to build topic maps chart splits to identify topics related to posts from Italian Twitter users. \cite{han2020using} extracted topics related to COVID-19 from Sina Weibo(Chinese microblogging website) text dataset through the LDA topic model.

\subsection{Dynamic Topic Modeling}
The dynamic topic model is a cumulative model that can be used to analyze changes in document collection over time \cite{bashar2020topic}. There are many studies on dynamic topic modeling for the English language. For example, \cite{alsumait2008line} showed that the LDA model could be extended to the online version by gradually updating the current model with new data, and the model has the ability to capture the dynamic changes of the topics. \cite{dieng2019dynamic} researched D-ETM on three data sets and discovered the word probabilities of eight different topics that D-ETM learned over time. \cite{nguyentopics} discovered latent topics from the financial reports of listed companies in the United States and studied the evolution of the themes discovered through dynamic topic modeling methods. \cite{marjanen2020topic} discussed humanistic interpretation's role in analyzing discourse dynamics through historical newspapers' topic models. \cite{bashar2020topic}  extracted five COVID-19 related topics from the Twitter dataset through LDA topic modeling, and they showed the changes in the extracted topics over time. Nevertheless, for the Bengali language, so far, there is no research on dynamic topic modeling. In this study, we study the evolution of the extracted COVID-19 related topics over time using dynamic topic modeling.

\subsection{Text Classification}
Text classification, moreover known as text labeling or text categorization, is categorizing content into organized bunches \cite{bashar2020regularising,bashar2020qutnocturnal,bashar2018misogynistic,bashar2020qutnocturnal}. By utilizing NLP, classifiers can naturally label text and, after that, relegate a set of predefined labels or categories based on its substance. 

Many researchers worked on text classification in English. For example, \cite{patil2012automated} used the Naive Bayes algorithm to classify website content. They divided the website content into ten categories, and the average accuracy of the ten categories was almost 80\%. \cite{bijalwan2014knn} used K-Nearest Neighbors, Naive Bayes, and Term-gram to classify text. They showed that in their research, K-Nearest Neighbors’ accuracy was better than Naive Bayes and Term-gram. \cite{tam2002comparative} showed that K-Nearest Neighbors was superior to NNet and Naive Bayes for English documents. \cite{pawar2012comparative} showed that Support Vector Machines’ performance is far superior to Decision Trees, Naive Bayes, K-Nearest Neighbors, Rocchio’s algorithms, and Backpropagation networks. \cite{liu2010study} showed that Support Vector Machines is better than K-Nearest Neighbors and Naive Bayes.

In addition to English text classification, some researchers have also classified Bengali text. For example, \cite{mandal2014supervised} applied four supervised learning methods: (Naive Bayes, k nearest neighbor, Decision Tree classifier, and Support Vector Machine) for labeled web documents. They classified the documents into five categories: (Business, Sports, Health, Technology, Education). \cite{chy2014bangla} applied a Naive Bayes classifier to categorized Bengali news. \cite{pal2015automatic} described Naive Bayes classifier for Bengali sentence classification. They used over 1747 sentences in their experiment and got an accuracy of 84\%. \cite{kabir2015bangla} used Stochastic Gradient Descent (SGD) classifier to categorize Bengali documents. \cite{eshan2017application} created an application that identifies abusive texts in Bengali. They applied Naive Bayes, Random Forest, Support Vector Machine (SVM) with Radial Basis Function (RBF), Linear, Polynomial, and Sigmoid kernel to classify the texts and compare the results among them. \cite{islam2017comparative} applied SVM, Naive Bayes, and Stochastic Gradient Descent(SGD) to classify Bengali documents and compare results of those classifiers. However, non of the existing works used Bengali text classification for monitoring a pandemic or a major event. 

\subsection{Sentiment Analysis}
Sentiment Analysis refers to computationally recognizing and categorizing opinions communicated in a chunk of text. It is successfully used in commerce where they use it to track online discussions to identify social estimation of their brand, item, or benefit. 

A lot of research work has been done in sentiment analysis for the English language. For example, \cite{cui2006comparative} have reviewed about 100,000 product reviews from various websites. They divided reviews into two main categories: positive and negative. \cite{jagtap2014svm} applied the Support Vector Machine and Hidden Markov Model, and the Hybrid classification model is well suited for extracting teacher feedback and evaluating sentiments. \cite{alm2005emotions} divided the seven emotional words into three polarity categories: positive emotion, negative emotion, and neutral, and the Winnow parameter adjustment method used can reach 63\% accuracy. For extracting the Twitter sentiment, \cite{agarwal2011sentiment} applied unigram, tree model, and feature-based model. \cite{bashar2020topic} used Convolutional Neural Network to extract sentiments related to COVID-19 from the Twitter dataset.

Some research used sentiment analysis in Bengali texts. For instance, \cite{das2010sentiwordnet} classified emotions into six categories: Happy, Sad, Anger, Disgust, Fear, and Surprise. \cite{chowdhury2014performing} used sentiment analysis in Bangla Microblog Posts. They applied a semi-supervised bootstrapping method utilizing SVM and Maximum Entropy. \cite{hasan2014sentiment} proposed a strategy to identify sentiments in Bengali texts by Contextual Valency Analysis. They employed the methodology of POS Tagger in their approach. \cite{hassan2016sentiment} used recurrent neural networks to Romanize Bengali texts and analyze sentiments. In their experiments, they used Bangla and Romanized Bangla Text (BRBT) dataset. For Sentiment Analysis of Bangla Microblogs, \cite{asimuzzaman2017sentiment} used Adaptive Neuro-Fuzzy Deduction Framework to anticipate extremity and utilized fluffy rules of speech in semantic rules. \cite{mahtab2018sentiment} designed a model for sentiment analysis on Bangladesh Cricket news. They applied TF-IDF and SVM (Support Vector Machine) in their model and found 64.596\% accuracy. \cite{tripto2018detecting} used sentiment analysis on Youtube comments. Their research built a model based on deep learning that classifies a Bengali text into three classes and five sentiment classes. \cite{tabassum2019design} used the Random Forest Algorithm to classify the sentiments in Bengali texts. \cite{tuhin2019automated} applied Naive Bayes and a topic modeling approach to design an Automated System of Sentiment Analysis in Bengali Text. Their system classifies emotions into six categories: happy, sad, tender, excited, angry, and scared. However, non of the existing works used Bengali text sentiment analysis for monitoring a pandemic or a major event.

\section{Experimental Methodology}
\label{sec:method_data}

\subsection{Methodology}
This pandemic situation has changed society and the country by a significant margin. The whole face of the country has changed completely. Some significant sectors of the nation, such as economic, social, political, have been affected massively. The education systems have been hit particularly hard. 
This research aims to automatically analyze the daily newspapers in Bangladesh to reveal what is going on in society and gain knowledge to comprehend the fundamental topics (or subjects) and sentiment arising and advance in the discussion. 

\begin{figure}
    \centering
    \includegraphics[width=1\textwidth]{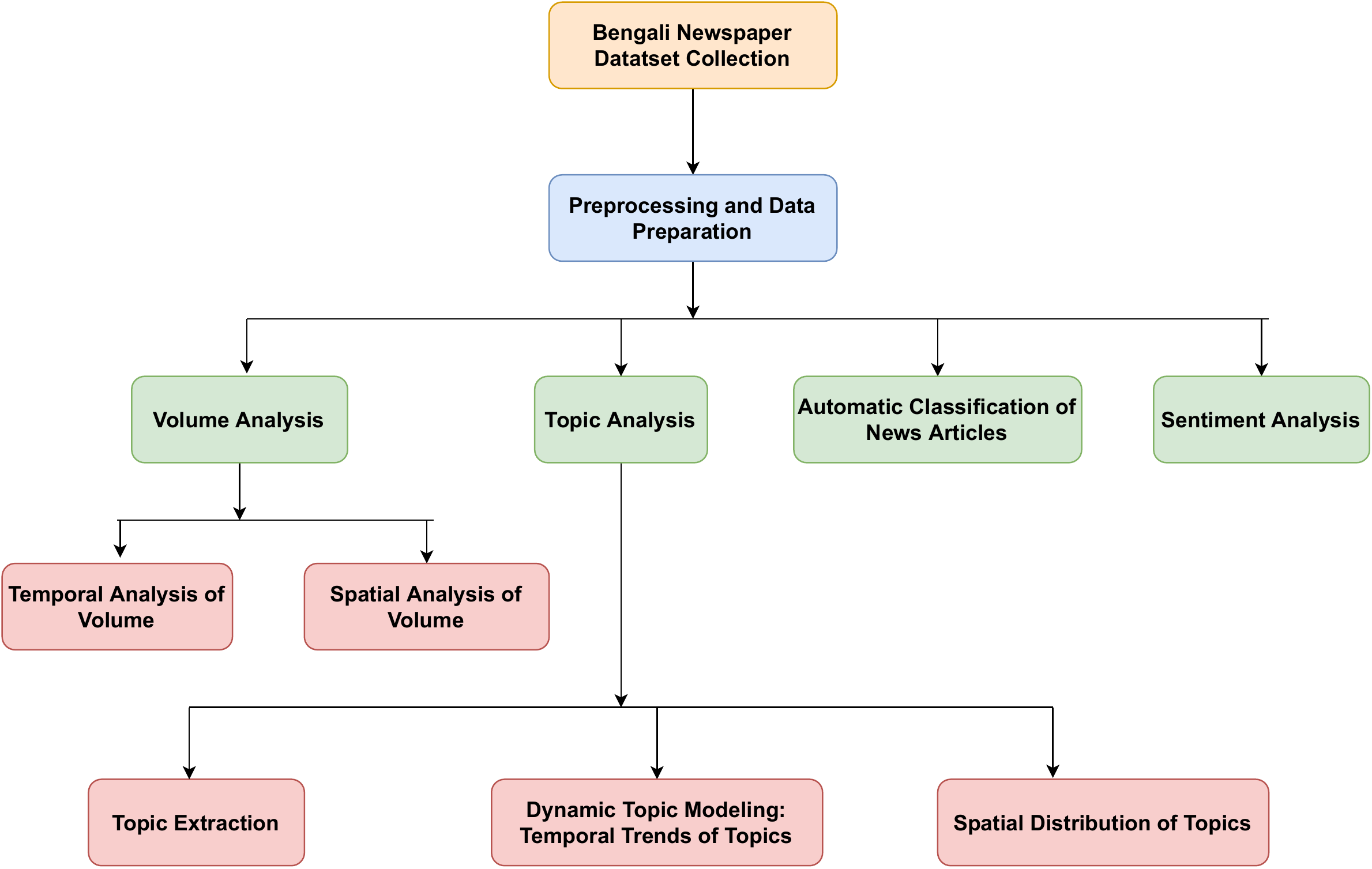}
    \caption{Experimental Work
    flow}
    \label{methodology}
\end{figure}

This study will conduct a topic and sentiment analysis on a large collection of COVID-19 related news articles published in Bangladesh both in Bengali and English texts. The study will focus the analysis on both spatial and temporal dimensions. In the topic analysis, we used LDA-based topic modeling and dynamic topic modeling to find the topics, their evolution over time, and their time and space (location). We also analyzed what impact each topic had on particular areas. Then we analyzed the sentiment distribution over time and space to identify social sentiment in space and time. The experimental workflow of this study is shown in Figure \ref{methodology}. 

First, we manually gathered a large collection of COVID-19 related news articles from Bangladeshi six most circulated daily newspapers. Along with the news, the collection contains geospatial and temporal information on the news. The dataset was then preprocessed by removing HTML, markers, and other non-relevant information such as adverts. 

Then, we manually organized the news articles in a set of classes and sub-classes. Then we extracted the topics and the subtopics from the dataset. We used these classes and sub-classes to perform basic analysis such as comparing similarity and diversity in the news. These classes and sub-classes have also been used to qualitatively evaluate the accuracy of the topics discovered by LDA and labels predicted by classifiers before LDA and classifiers are employed for detailed analysis.

\subsection{Data Collection and Preparation}
These publicly available News articles related to COVID-19 have been collected from the six most popular newspapers in Bangladesh from 21 January 2020 to 19 May 2020. The six newspapers are \emph{The Daily Prothom Alo}, \emph{Bangladesh Pratidin}, \emph{Kaler Kantho}, \emph{The Daily Star}, \emph{Newage}, and \emph{The Daily Observer}. A total of 15,565 news articles are collected from these six newspapers. From every news article, we extracted the news title, the main body of the news, a summary of the news (i.e., first few lines of the news body), the published date, and the news incident's location. We used Python's \emph{BeautifulSoup} and \emph{Newspaper3k} tool for extracting the news content. \emph{BeautifulSoup} is a popular Python package for parsing HTML and XML archives and one of the most popular web scraping tools. \emph{Newspaper3k} is a user-friendly library for scraping the news articles and other related data from newspaper portals. It is built upon request and used to parse LXML. This module is an improved version of the \emph{Newspaper} module and is also used for the same purpose. Table~\ref{my_table01} summarizes the statistics of the collection. We call this collection \emph{Comilla University COVID-19 News Collection} (CoU-CNC). We made it available online\footnote{CoU-CNC Dataset: \url{https://cutt.ly/djGILi2}} for anyone for further analysis.


\begin{table}
    \begin{center}
        \begin{tabular}{l|c|c}
    \toprule
    Article Source & Language & Article Count\\
    \midrule
    The Daily Prothom Alo &	Bengali &	4169\\
    \hline
    Bangladesh Pratidin &	Bengali &	5584\\
    \hline
    Kaler Kantho &	Bengali &	1160\\
    \hline
    The Daily Star &	English &	1278\\
    \hline
    The Daily Observer &	English &	1191\\
    \hline
    New Age &	English &	2183\\
    \bottomrule
\end{tabular}
    \end{center}
    \caption{CoU-CNC Dataset Statistics}
    \label{my_table01}
\end{table}

Out of these six newspapers, news articles in three newspapers (\emph{The Daily Prothom Alo}, \emph{Bangladesh Pratidin}, \emph{Kaler Kantho}) are composed in the Bengali language, and in the other three newspapers (\emph{The Daily Star}, \emph{The Daily Observer}, \emph{New Age}) articles are composed in the English language. There are 10,913 news articles in the Bengali language, and the remaining 4652 news articles are in the English language.

As the dataset has 4,652 articles in English and we wanted all articles in the same language to be better parsed, so we translated the English articles into Bengali via Python’s \emph{googletrans} module. As a result, after translating these articles, all the articles are in the Bengali language. Then, we applied tokenization to split a string of text into smaller tokens. The news articles are split into sentences, and sentences are tokenized into words. Then, we applied noise removal (e.g., removing HTML tags, extra white spaces, special characters, numbers) to clean up the text. Then, we removed the stopwords from the document. As there is no build-in stopwords module for Bengali nltk, we manually created a stopword list and made it available online\footnote{Bengali-Stopwords: \url{https://cutt.ly/2jXbDRB}}. Then, we expanded contraction. We set the minimum letter length to 6. We also removed all the words that were below the minimum letter length.
There are no good resources for stemming and lemmatization in the Bengali language. So, we applied stemming and lemmatization to the tokens in our own process. After removing all the stopwords and other noises, there were a total of 80,693 tokens. There are some specific suffixes for the Bengali language. The suffix removal from the word has also been done with the help of Python.
We used \emph{Bangla\_Steamer.Steamer} library of Python to improve accuracy. However, it did not show the expected results as the library is effective for a small number of Bengali words. To increase the accuracy of this 80,693 sizes lemmatized dictionary, we manually verified about 30000 most frequent tokens from 80693 words. We lemmatized where we needed to lemmatized manually, and we also corrected the incorrect and misspelled words where it was needed. Many more words are manually lemmatized and corrected through this manually 30,000 words check. We have published verified Bengali words on the Internet and titled \emph{“Modified Bengali Words”} \footnote{Modified Bengali Words: \url{https://cutt.ly/8jE6GIC}} for further analysis.

To compare the number of news published and the COVID-19 cases of Bangladesh, we collected an open-source dataset\footnote{\url{https://www.worldometers.info/coronavirus/country/bangladesh/}} of confirmed COVID-19 cases and death cases of Bangladesh from March 8 to May 19.
We also collected another open-source dataset\footnote{\url{https://data.humdata.org/dataset/district-wise-quarantine-for-covid-19}} of confirmed cases based on divisions and districts of Bangladesh from March 8 to May 19.

\subsubsection{Class Distribution in News Articles}
\begin{figure}
    \centering
    \includegraphics[width=.8\textwidth]{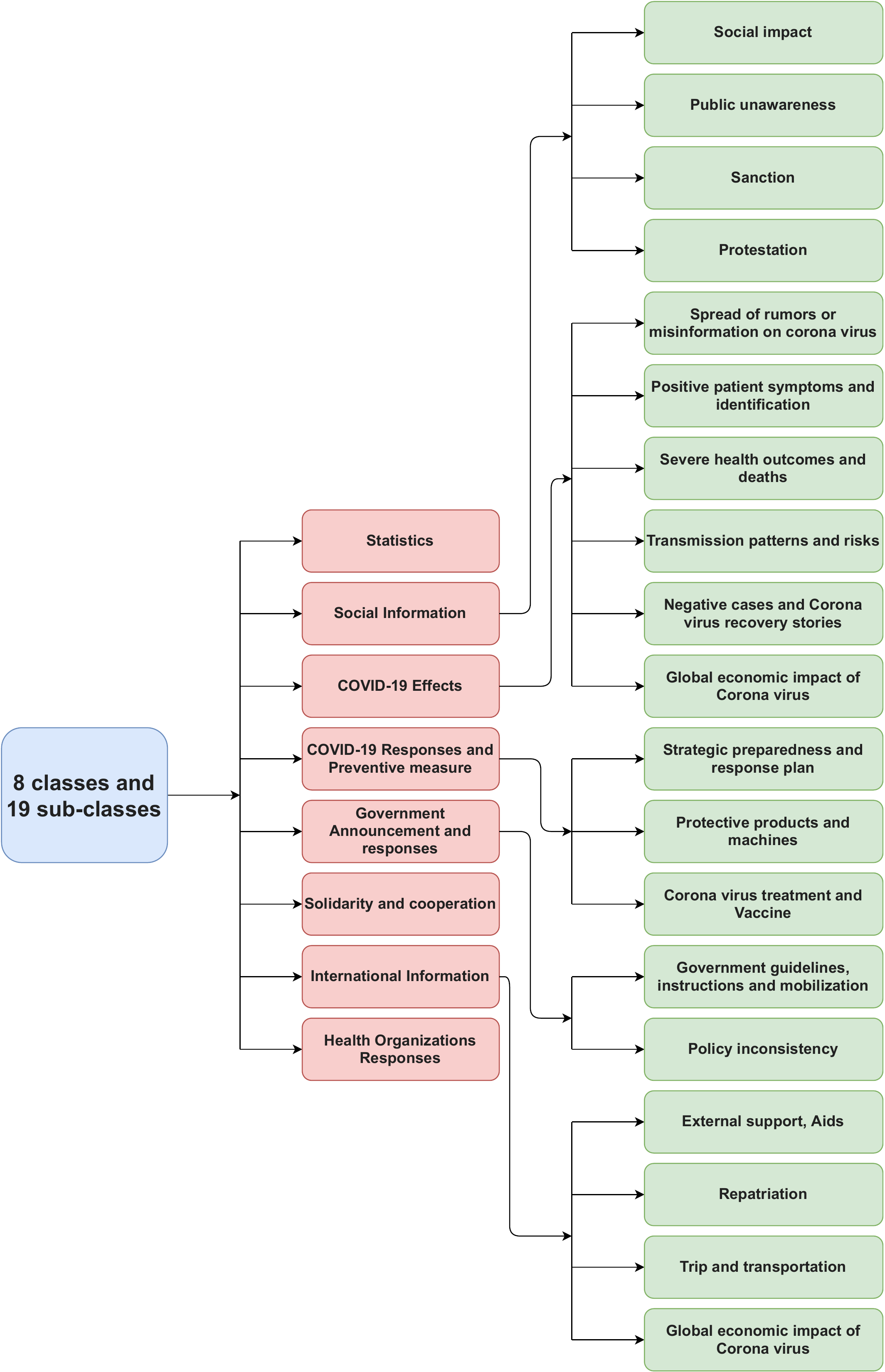}
    \caption{Manually Extracted Classes and Sub-classes}
    \label{topsub1}
\end{figure}

\begin{figure}
    \centering
    \includegraphics[width=1\textwidth]{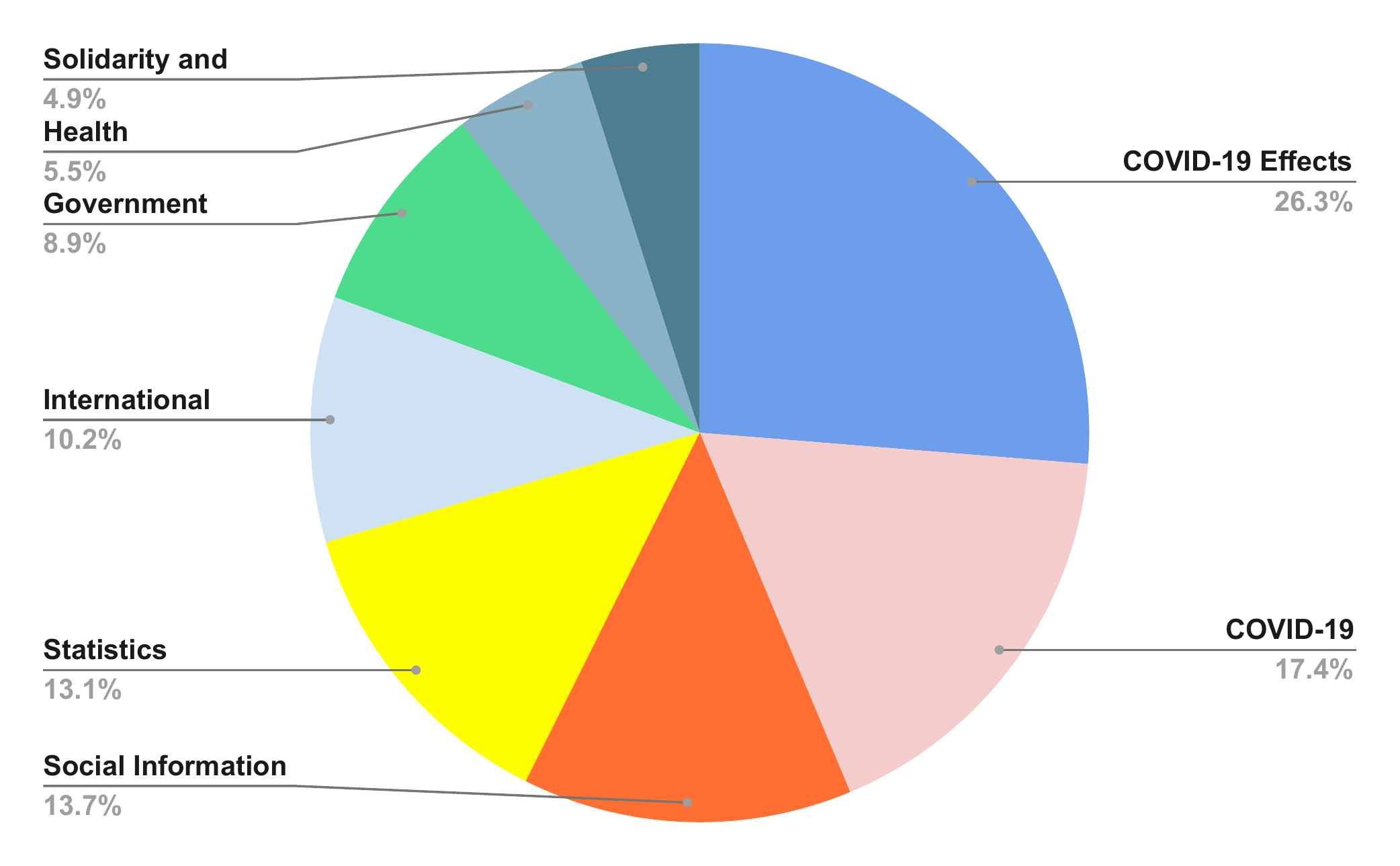}
    \caption{Distribution of Manually Extracted Classes over News Articles}
    \label{topsub2}
\end{figure}

\begin{figure}
    \centering
    \includegraphics[width=1\textwidth]{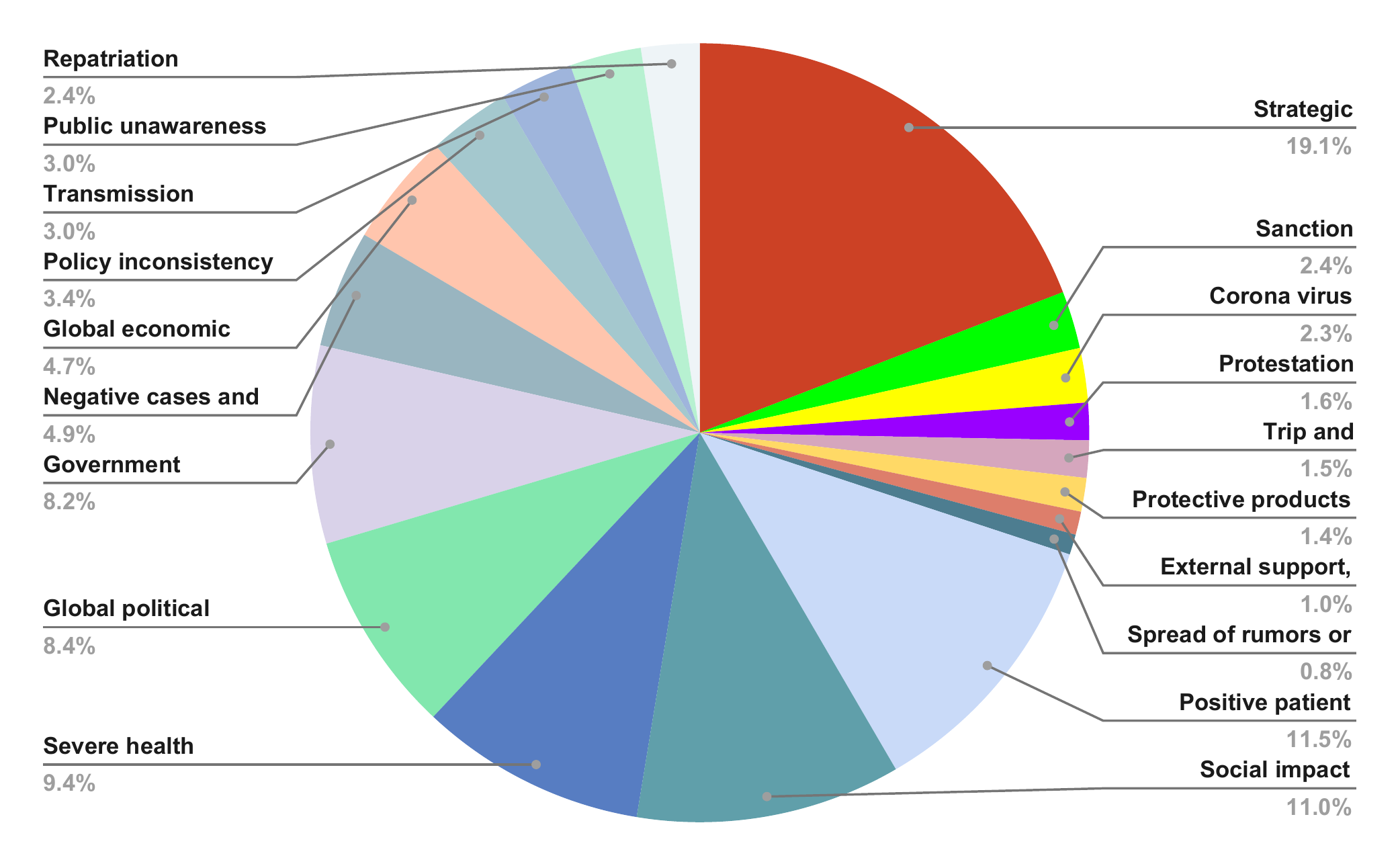}
    \caption{Distribution of Manually Extracted Sub-classes over News Articles}
    \label{topsub3}
\end{figure}

After collecting the new articles, first, we analyzed them manually. In this process, we extracted eight classes (shown in Table \ref{tab:manual_8_topics}) and 19 sub-classes from the news articles. 
The representation of the extracted eight classes and the hierarchical organization of sub-classes are shown in Figure \ref{topsub1}. The distribution of the extracted classes over news articles is shown in Figure \ref{topsub2} and the distribution of the extracted sub-classes over news articles is shown in Figure \ref{topsub3}. 

\begin{table}[htbp]
  \centering
  \caption{Eight Classes Extracted from the Collected News Articles}
    \begin{tabular}{p{\textwidth}}
    \toprule
    (1) Statistics, (2) Social Information, (3) COVID-19 Effects, (4) COVID-19 Responses and Preventive Measure, (5) Government Announcement and Responses, (6) Solidarity and Cooperation, (7) International Information, and (8) Health Organization Responses \\
    \bottomrule
    \end{tabular}%
  \label{tab:manual_8_topics}%
\end{table}%

\subsection{Spatio-Temporal Analysis}
Time Series or temporal analysis of newspaper articles is utilized to observe the transient expansion during the pandemic. Time series decomposition includes considering a series of components in the time dimension: Level, Trend, Seasonality, and Noise segments. Level refers to the average value in the series, Trend refers to the increasing or decreasing value in the series, Seasonality refers to the repeating short-term cycle in the series, and finally, Noise refers to the random variation in the series.
Decomposition gives a powerful supportive model for pondering time series and better arrangement issues during time series analysis and decision making. The additive model \citep{dagum2010time} suggests that the segments are added  as the following formula:
\begin{equation}
    y(x) = l(x) + t(x) + s(x) + n(x)
    \label{eq-1}
\end{equation}
where $y(x)$ represents the additive model, $l(x)$ represents the observed level, $t(x)$ represents the trend, $s(x)$ represents the seasonality and $n(x)$ represents the noise or residual in the signal $x$. This model is linear. The change over a period of time is reliably affected by the similar sum of the linear trend as a straight line. A linear seasonality has a similar recurrence and abundance. On the other hand, A multiplicative model \citep{dagum2010time} recommends that the components are multiplied together as the following formula:
\begin{equation}
    y(x) = l(x) \times t(x) \times s(x) \times n(x)
    \label{eq0}
\end{equation}
where $y(x)$ represents the multiplicative model, $l(x)$ represents the observed level, $t(x)$ represents the trend, $s(x)$ represents the seasonality and $n(x)$ represents the noise in the signal $x$. A multiplicative model is exponential or quadratic when expanding or diminishing over the long run. A nonlinear pattern is a bent line. In this examination, we disintegrated the time series utilizing the multiplicative model. For Spatial Analysis, we used Tableau Software to compare the number of news published, and the number of COVID-19 confirmed cases geographically.

\subsection{Static Topic Analysis}
Analyzing the topics of news articles published during a major incident or a pandemic like COVID-19 can help monitor the situation and understand the public concerns, which is critical for government authorities and charity organizations to disseminate required resources and aids. However, in such a situation, a large number of news articles are published in various newspapers. We observed that as the situation deteriorated during the pandemic, newspapers had to publish much news on various topics. Manually analyzing the topics by reading a large number of articles is time-consuming and expensive. 
We utilized two unsupervised machine learning techniques: (a) LDA \citep{blei2003latent}, a popular topic modeling technique, as static topic modeling to automatically find topics of articles published in newspapers, and (b) dynamic topic modeling in \citep{blei2006dynamic} to see how those topics evolve over the long haul.

LDA is a Bayesian probabilistic model that discovers topics and provides topic distribution over documents and word distribution over topics. It has two phases: (a) the first phase models each document as a composition of topics, and (b) the second phase models each topic as a composition of words. LDA utilizes word co-occurrences inside documents for discovering topics in a document assortment. Words occurring in an equivalent document are practically coming from the same topics, and documents containing comparative words will undoubtedly include comparable topics.
In this research, the \emph{Gensim} package in Python was utilized to execute the LDA model. We utilized every news article as a document in the topic modeling. Before applying the LDA topic model, we manually associated documents into general classes and sub-classes to know about the quality of LDA extracted fine-grained topics.

Then, we analyzed each LDA extracted topic's temporal trends to see when a topic has been discussed more or published more in the newspapers. Finally, we analyzed each topic's spatial distribution to see what effect each had in a particular place. We used Tableau software to analyze the spatial distribution of each topic.

\subsection{Dynamic Topic Analysis}
The static topic modeling treats words as interchangeable and indeed treats documents as interchangeable. However, the presumption of replaceable documents is impractical for some assortments when accumulating along the time. For example, tweets, news articles, and insightful articles as they are advancing substance along time. The subjects in a newspaper article assortment develop, and it is essential to display the elements of the fundamental topics unequivocally.

Dynamic topic modeling extends the static theme, which illustrates the progression of the theme in consolidate. Dynamic topic modeling can catch the development of topics in a successively coordinated assortment of news articles. In this research, the articles are synchronized by week. We used the dynamic topic model to analyze discussion topics and topic changes over time.

\subsection{Text Classification}
\label{subsec:Text Class}
Then we built a text classifier to verify their performance and predict the class, sub-class, and topics in the unknown (upcoming in the future) news articles. Such classification is important when we need to monitor a specific category (or class or group) of news. We made Long Short-Term Memory (LSTM) Recurrent Neural Network (RNN) models in Python utilizing Keras deep learning library for text classification. RNN is a special kind of neural network where the previous step’s output will be used as the current step’s input. In a traditional neural network, not all inputs and outputs are interdependent. However, interdependence is an important part of text data. In such cases, the model needs to predict the next word given the previous words, so the previous word must be stored. Thus RNN was born, which solved the problem with the help of hidden layers. The primary function of RNN is also essential, namely the \emph{hidden state}. It can remember some information about the sequence. 

\begin{figure}
    \centering
    \includegraphics[width=1\textwidth]{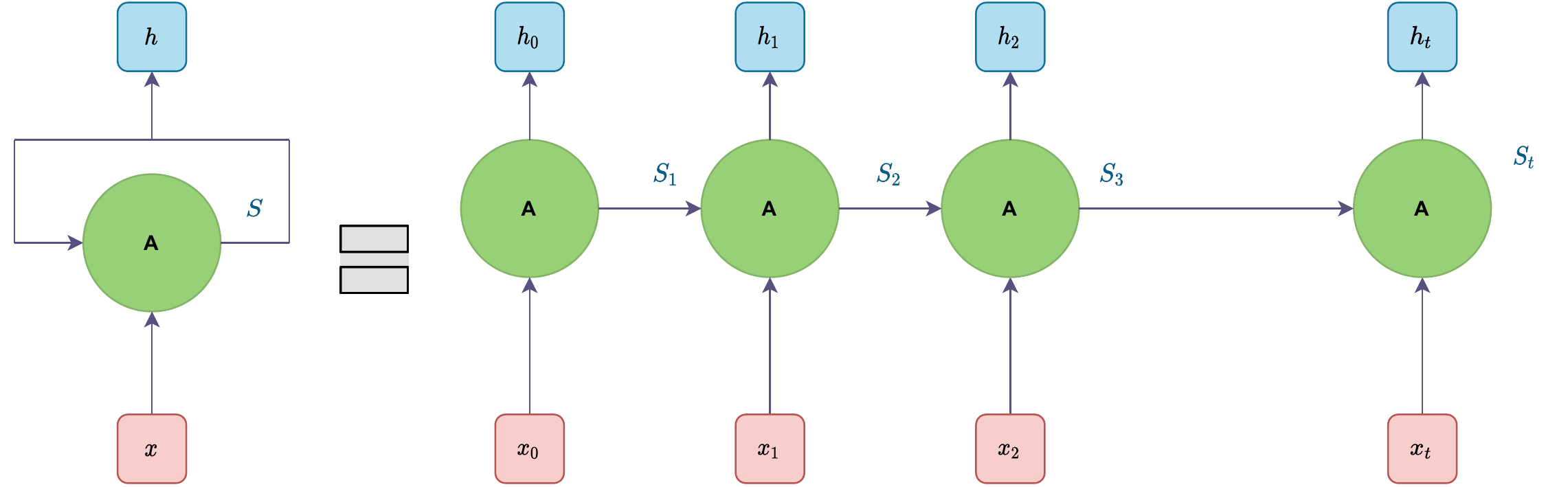}
    \caption{Unrolled Recurrent Neural Network}
    \label{fig_rnn}
\end{figure}

RNN is a neural feedback network that operates on the internal memory. Since the RNN has a similar function for each piece of information and the current range’s output is based on the last count, the RNN is essentially recursive. When there is an output, it is copied and sent back to the relay network. The current input and the output of the previous input are taken into account in determining the prediction of the next word. Unlike direct feedback neural networks, RNNs can use their internal state (memory) to manage the input elements' interdependence. That makes them useful for text data, handwriting recognition, or speech recognition. The architecture of an unrolled recurrent neural network is shown in Figure \ref{fig_rnn}.

In Figure \ref{fig_rnn}, first the model gets $x_0$ from the input sequence. Then it produces $h_0$, which is used in the next input to the model along with $x_1$. That is, both $h_0$ and $x_1$ become inputs to the next step. Then, $h_1$ and $x_2$ are input to the next step, and so on. Like this, RNN continues summarizing the unique circumstance in the hidden state while training. Then, it uses the summarized hidden state to classify the sequence \citep{bashar2020regularising}.

\subsection{Sentiment Detection}
We proposed a hybrid neural network model based on Convolutional Neural Network (CNN) and LSTM for sentiment analysis in Bengali texts.

Integrated models are used to solve various vision and NLP problems and improve a single model's performance. The following subsections provide an overview of the LSTM and CNN models offered. In subsection \ref{subsec:Text Class}, we described LSTM. In this research, we used two-layer bi-LSTM, word embedding include words in the news articles and provide sentiments. 

\begin{figure}
    \centering
    \includegraphics[width=1\textwidth]{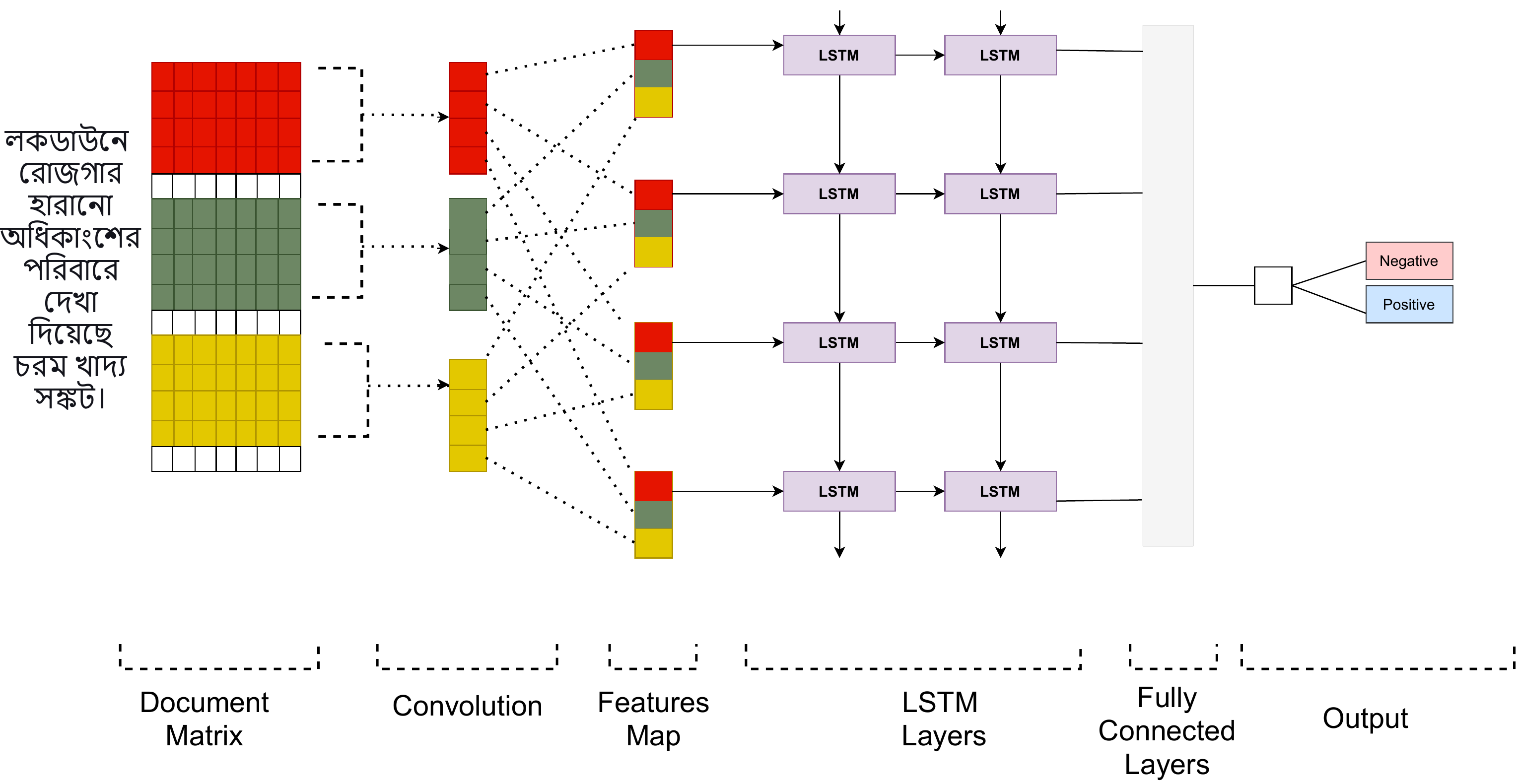}
    \caption{The architecture of CNN-BiLSTM Hybrid network for Sentiment Identification}
    \label{fig_cnn}
\end{figure}

Another part of our proposed structure is based on Convolutional Neural Network (CNN). CNN has very successful in various image processing and NLP tasks these last years. They are powerful in exploration, achieving local relevance, and data standards through learning. Generally, to rank text on CNN, different words in sentences (or paragraphs) need to be placed. Stacked to form a two-dimensional matrix, pleated filters (different lengths) are applied to the window. To use CNN for text classification, the different words stacked in a sentence are usually stacked in a two-dimensional matrix, and afterward, a convolution is applied to the word in the window in one word to be created applied a new function declaration. Then, a max-pooling is applied to the new function, and the combined functions of different filters are combined to shape a concealed portrayal. Completely associated layers trail these portrayals for the last estimate. The architecture of our CNN-BiLSTM Hybrid network model is shown in Figure~\ref{fig_cnn}.

We created a sequential model that includes an LSTM layer. Then we made our model sequential and adding layers. In the first layer, we applied a conv1D with 200 as a filter for CNN. After that, we applied two Bi-LSTMs on the second and third layer with an error of 0.5. Then we applied a dense network on the remaining levels. We also used \emph{Adam} as an optimizer with tight hyperparameters and applied \emph{L2} adjustments to reduce overfitting as much as possible. We only used five epochs, as using more epochs resulted in overfitting and kept the stack size of 256, as it worked very well.

\section{Experimental Results}
\label{sec:exp_res}

\subsection{Volume Analysis}
\subsubsection{Temporal Analysis of Volume}
\begin{figure}
\centering
\subfloat[Observed\label{timeseries1}]{%
  \fbox{\includegraphics[width=0.70\textwidth]{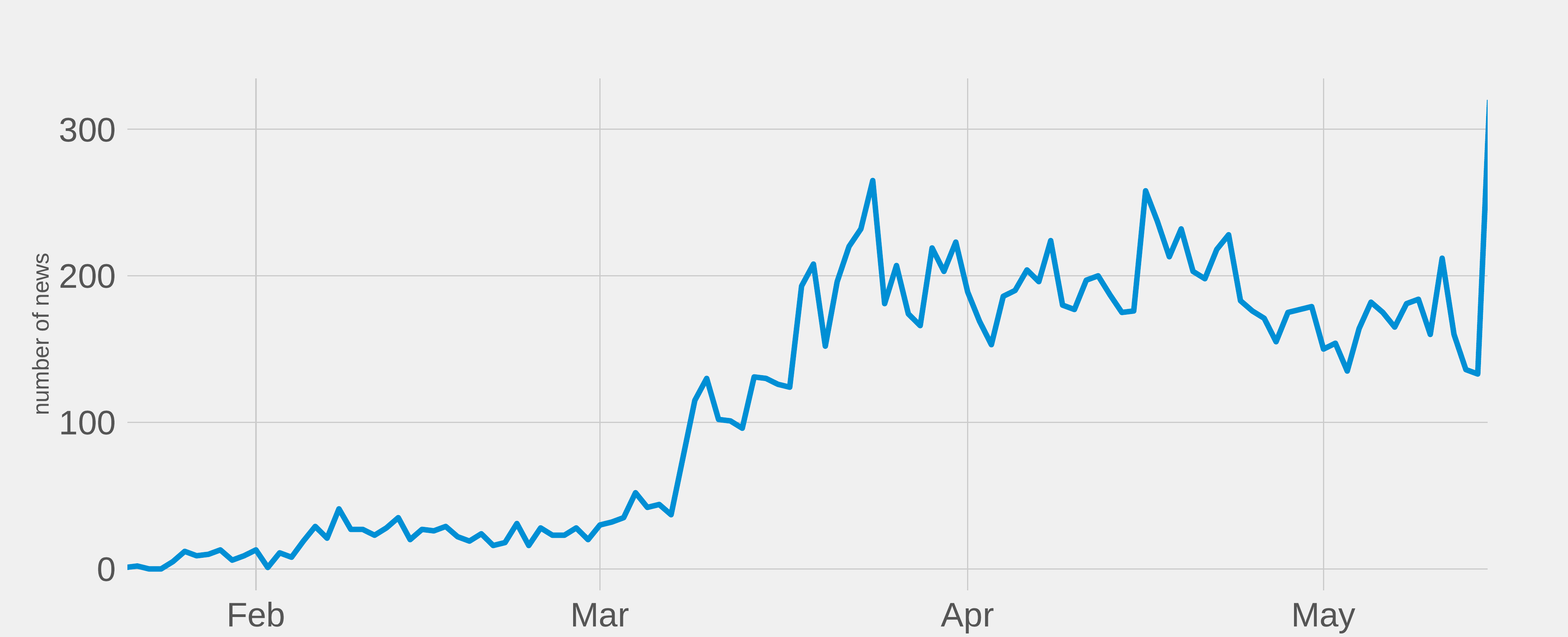}}%
}\hfill 
\subfloat[Trend\label{timeseries2}]{%
  \fbox{\includegraphics[width=0.70\textwidth]{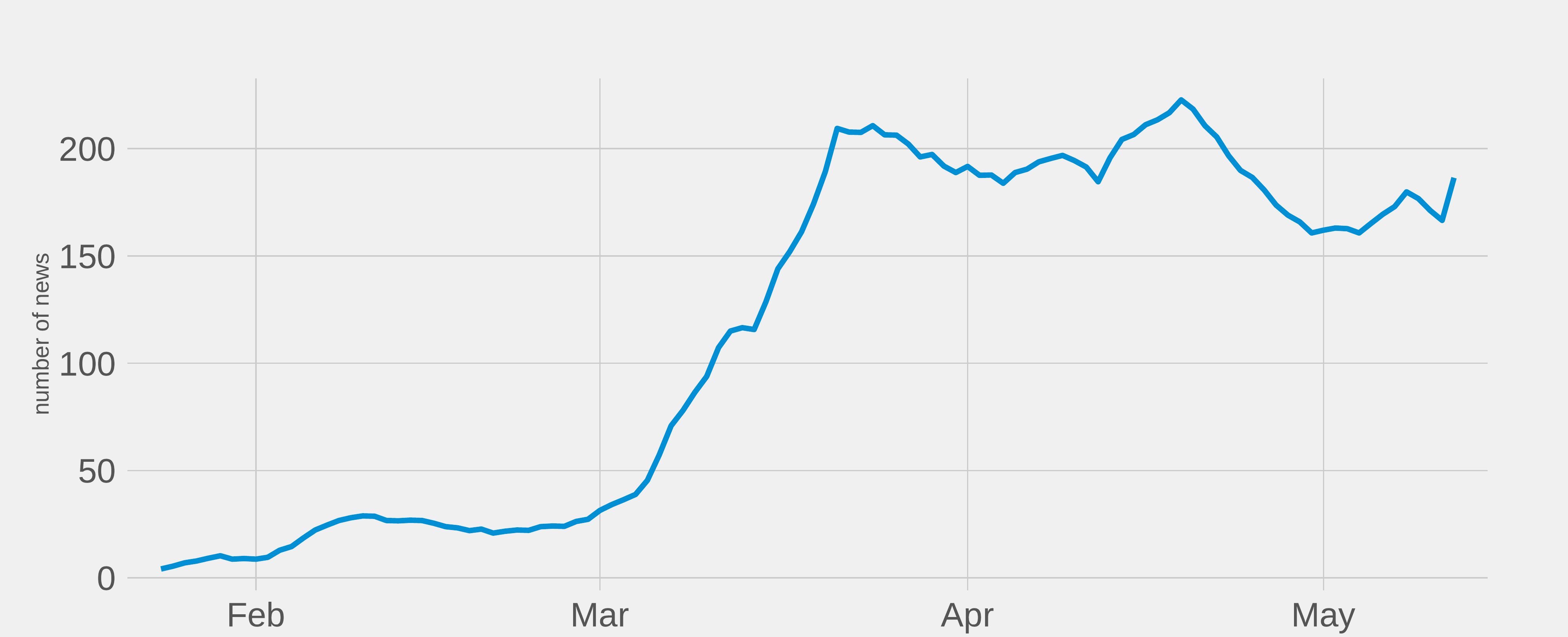}}%
}\hfill
\subfloat[Seasonal\label{timeseries3}]{%
  \fbox{\includegraphics[width=0.70\textwidth]{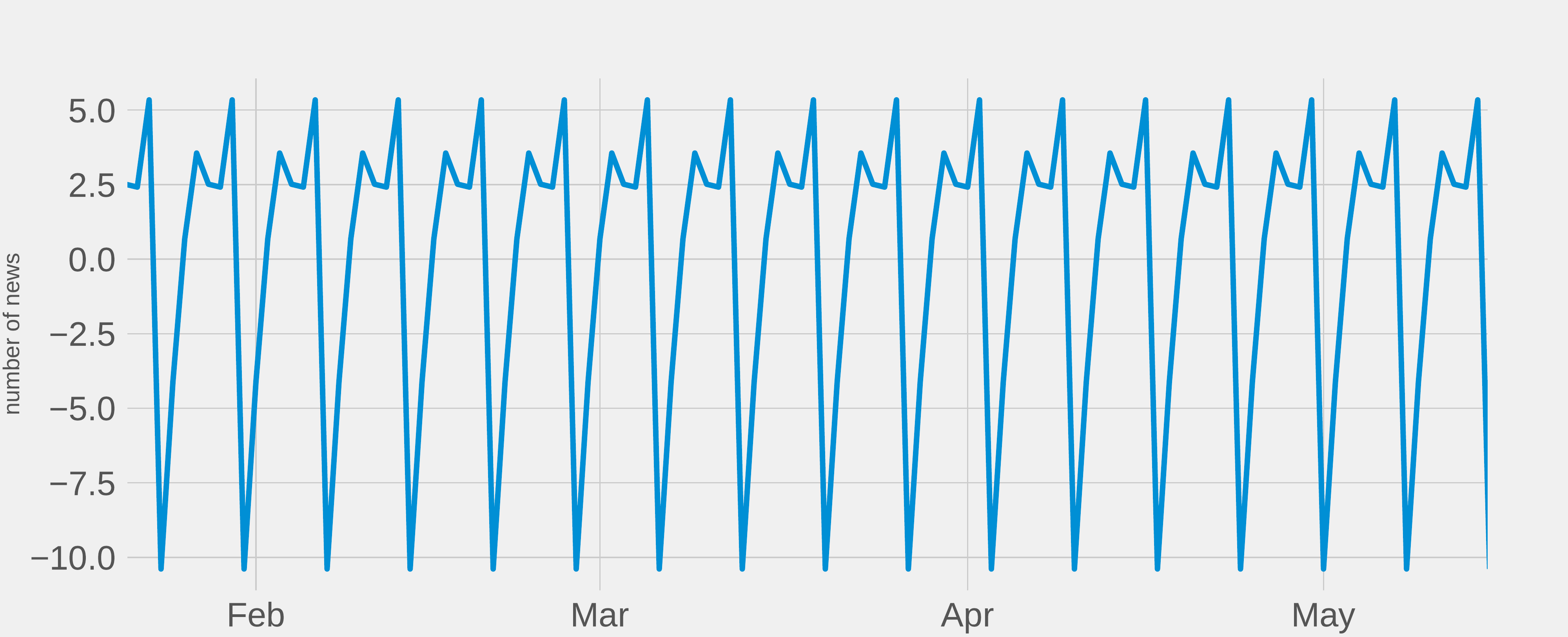}}%
}\hfill
\subfloat[Residual\label{timeseries4}]{%
  \fbox{\includegraphics[width=0.70\textwidth]{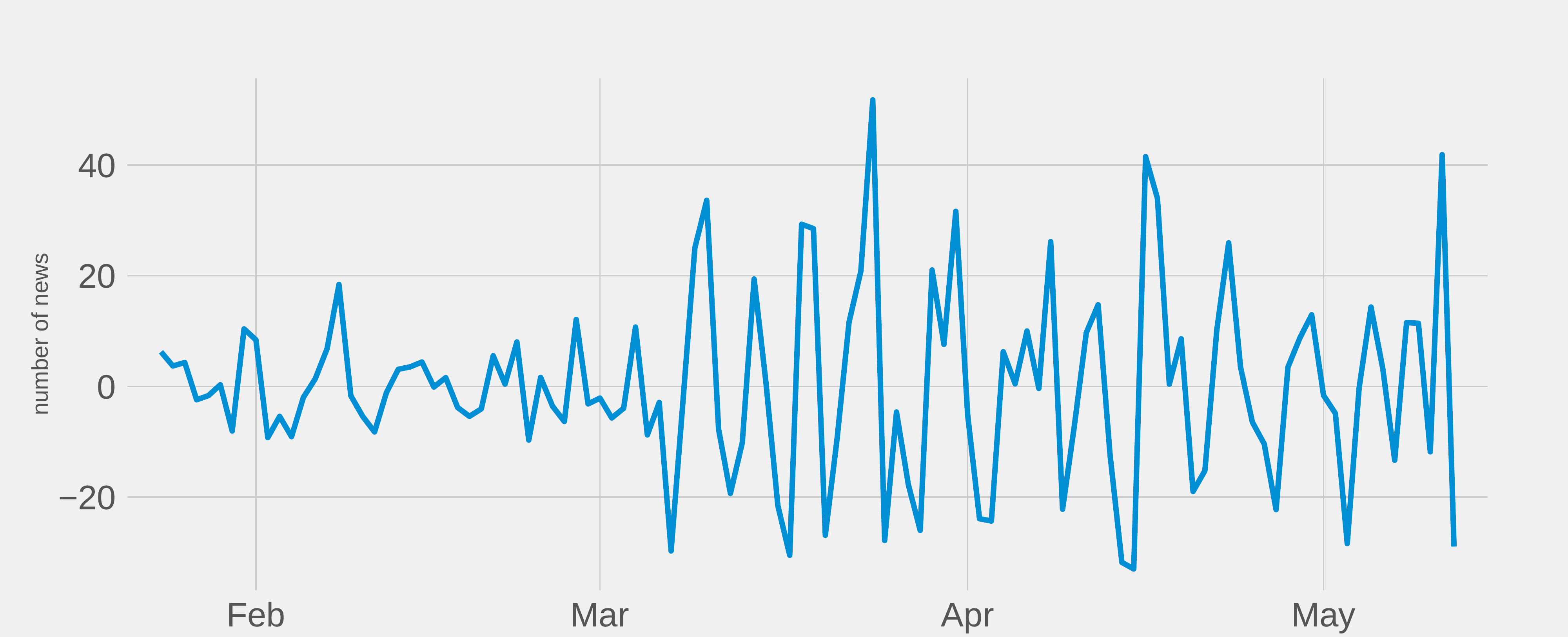}}%
}\hfill

\caption{Time Series Decomposition}
\label{timeseries-x}
\end{figure}

The time series volume analysis of newspapers is shown in Figure \ref{timeseries-x}. The figure has four plots, namely observed level, trend, seasonal, and noise or residual. The first plot Figure \ref{timeseries1} shows the original volume, i.e., the number of COVID-19 related news articles in a time point. It shows that the curve began to rise from January when some COVID-19 cases were found in China and other countries. The plot increased sharply in early March when a few instances of COVID-19 cases were identified in Bangladesh. The curve remained high onward with some fluctuations. 
The second plot Figure \ref{timeseries2} shows the trend of the COVID-19 related news publication volume. It shows that the COVID-19 related news started becoming trendy by the end of January, and the trend increased significantly in early March. The trend stayed high through the rest of the time with some fluctuations. The third plot Figure \ref{timeseries3} shows the seasonal, cyclical change in the volume. Moreover, the fourth plot Figure \ref{timeseries4} shows a residual or random variation in the volume.


\begin{figure}
    \centering
    \includegraphics[width=1\textwidth]{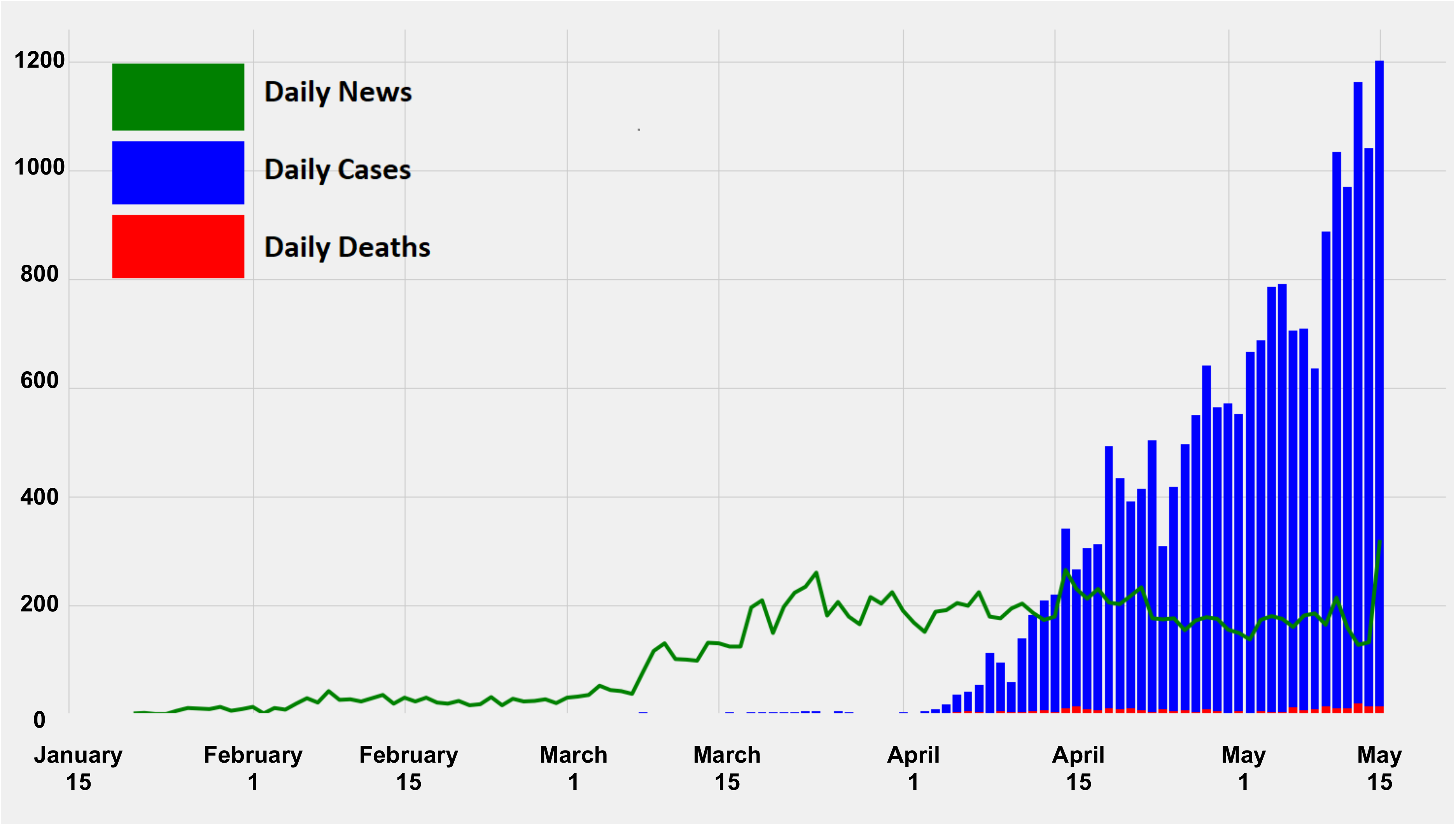}
    \caption{Comparison of Daily News Article Counts and Daily Cases (21 January 2020 - 19 May 2020)}
    \label{xyz2}
\end{figure}

To see how newspapers reacted during the COVID-19 pandemic, we tracked COVID-19 cases, death from COVID-19, and COVID-19 news volume in Figure \ref{xyz2}. The figure shows that the newspapers were vigilant from the beginning of the pandemic. The newspaper journalists increased COVID-19 related news coverage exponentially as soon as COVID-19 cases were found in Bangladesh in early March. The news volume continued increasing until the last quarter of March. This part of the news volume shows the newspapers reacted from about COVID-19 from the very early pandemic stage. They significantly covered the pandemic during the early period of the COVID-19 cases.

The number of identified cases increased significantly by the second quarter of April, and it continued to increase. However, the number of COVID-19 related news articles did not increase during this time. Even in some cases, the news article volume decreases marginally. The possible reasons might be: (a) Because Bangladesh is a developing country, to survive at this point, people had to think more about earnings than pandemic. As a result, pandemic news did not increase attention, and newspapers did not increase COVID-19 related articles. (b) Some other big incidences gained more attention than COVID-19. (c) The newspapers reached their allocated space for pandemic news already.
 
\subsubsection{Spatial Analysis of Volume}
\begin{figure}
\centering
\subfloat[Number of News Published Up to 19 May 2020\label{spatial1a}]{%
  \fbox{\includegraphics[width=.95\textwidth]{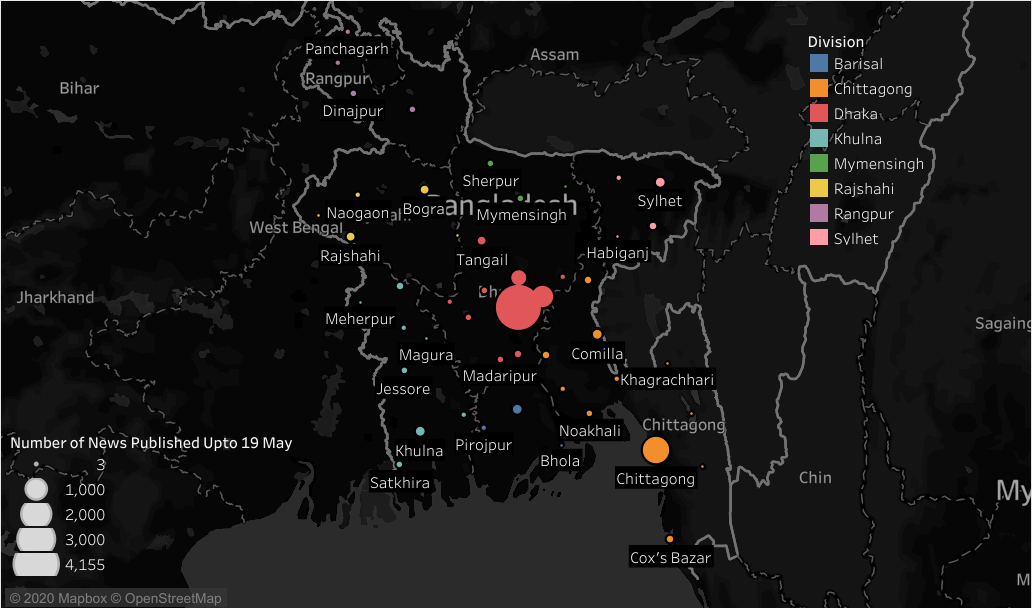}}%
}\hfill 
\subfloat[Confirmed Cases Up to 19 May 2020\label{spatial1b}]{%
  \fbox{\includegraphics[width=.95\textwidth]{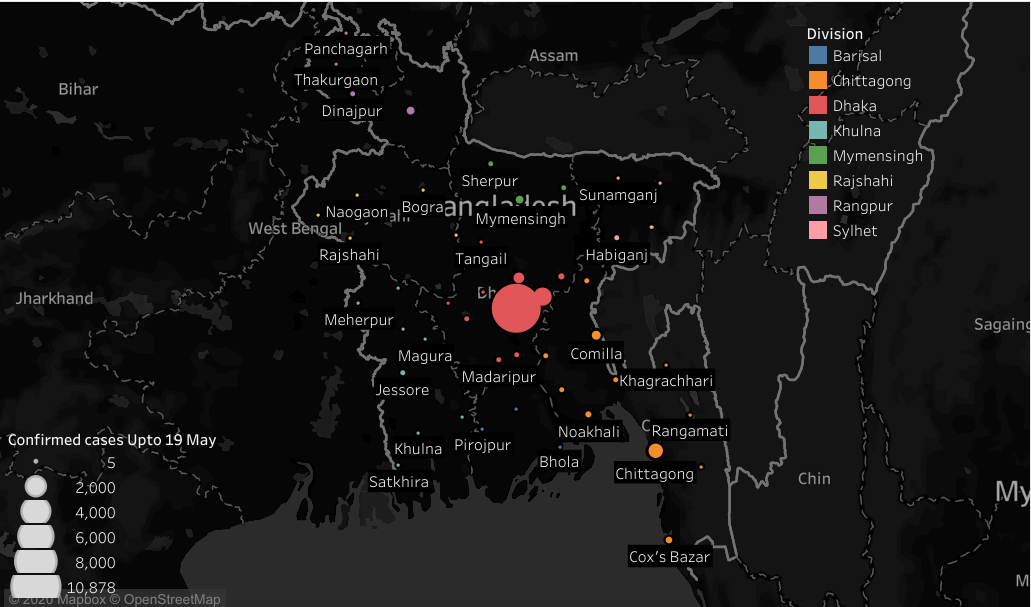}}%
}\hfill

\caption{Spatial Analysis of News Article Volume}
\label{spatial1}
\end{figure}

The spatial Distribution of Bengali newspapers is shown in Figure~\ref{spatial1a}. The number of news articles was concentrated on the central part of Bangladesh, mainly Dhaka, Narayanganj, and Gazipur. More than 6000 COVID-19 related news articles were published in Bangladeshi newspapers related to Bangladesh’s central part. More than 2000 news articles related to the southern part of Bangladesh, mainly Chittagong and Cox’s Bazar.


The spatial distribution of confirmed cases of COVID-19 is also shown in Figure~\ref{spatial1b}. The central part of Bangladesh is the most affected area. More than 10,000 COVID-19 patients were identified in Dhaka during this time. Outbreaks have been reported in the surrounding areas of Dhaka, mainly Narayanganj and Gazipur. After the Dhaka division, we can see the highest infection rate in the southern part of Bangladesh, mainly Chittagong. Figure \ref{spatial1} shows a correlation between the number of confirmed COVID-19 cases in an area and the published news volume related to that area. This means automatic monitoring of news article volume can give a clear view of the severity of a pandemic or big instances in a society.

\begin{figure}
    \centering
    \includegraphics[width=0.6\textwidth]{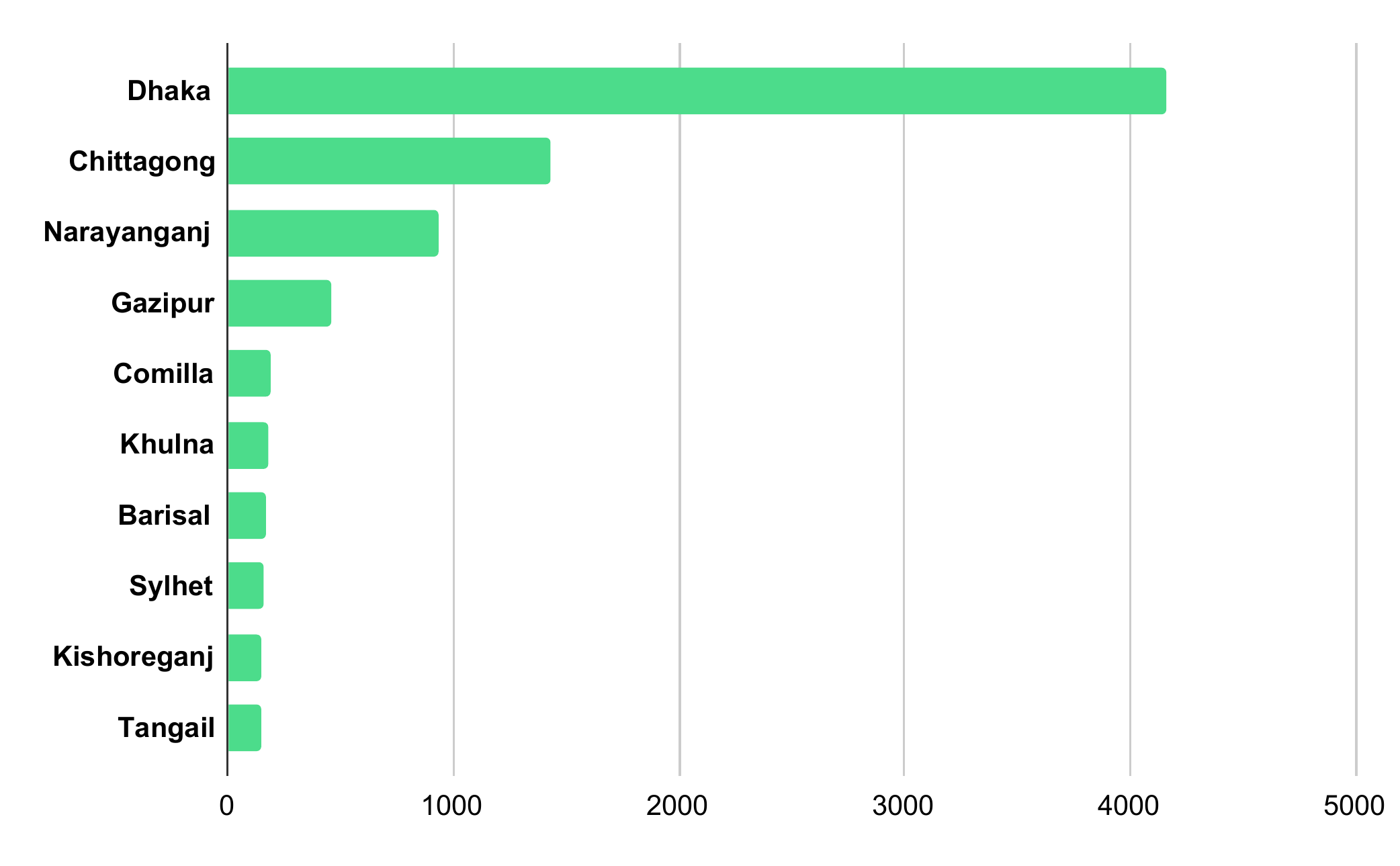}
    \caption{District-wise Distribution of News Articles}
    \label{spatial3}
\end{figure}

\begin{figure}
    \centering
    \includegraphics[width=0.6\textwidth]{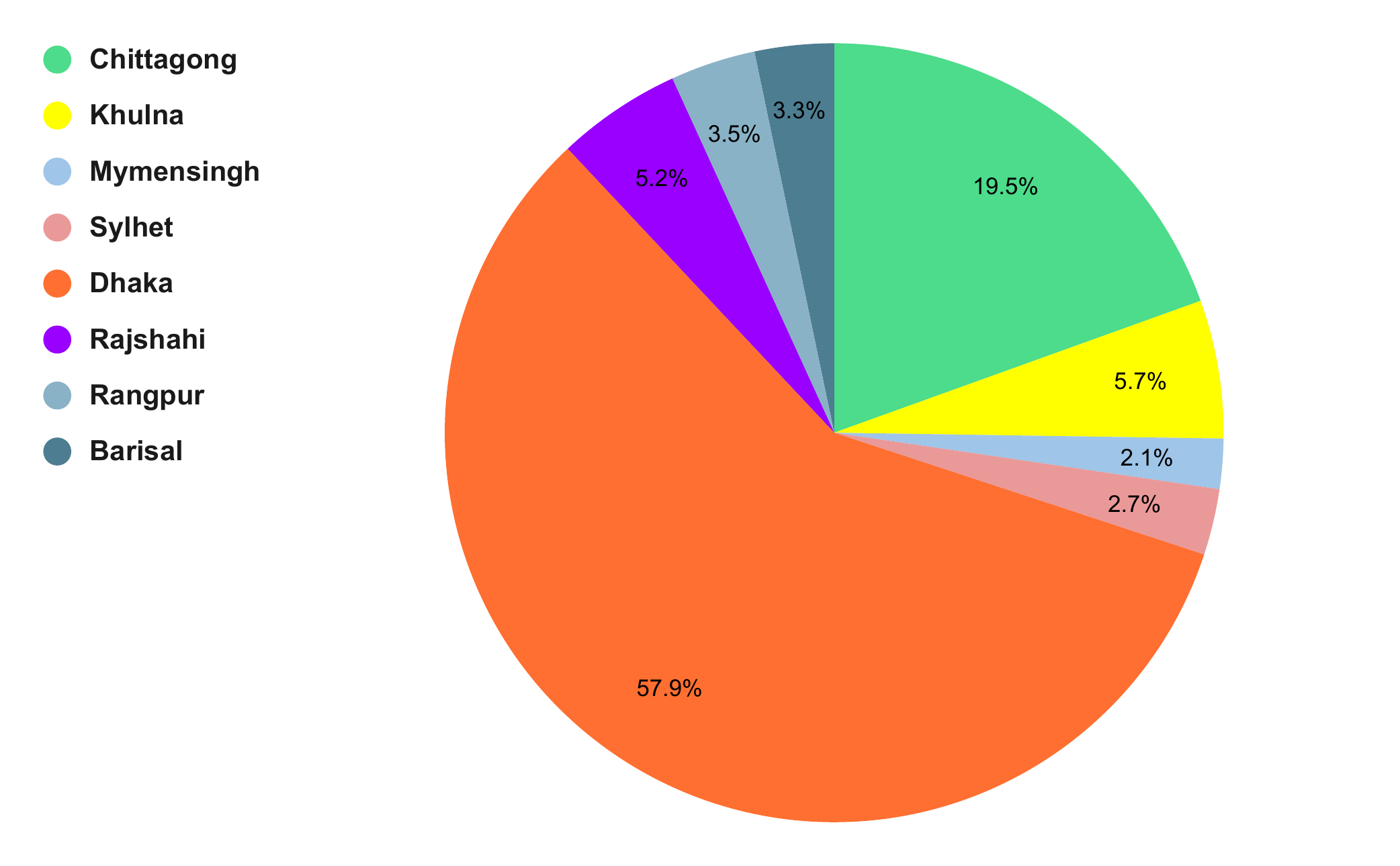}
    \caption{Division-wise Distribution of News Articles}
    \label{spatial2}
\end{figure}

\begin{figure}
    \centering
    \fbox{\includegraphics[width=0.95\textwidth]{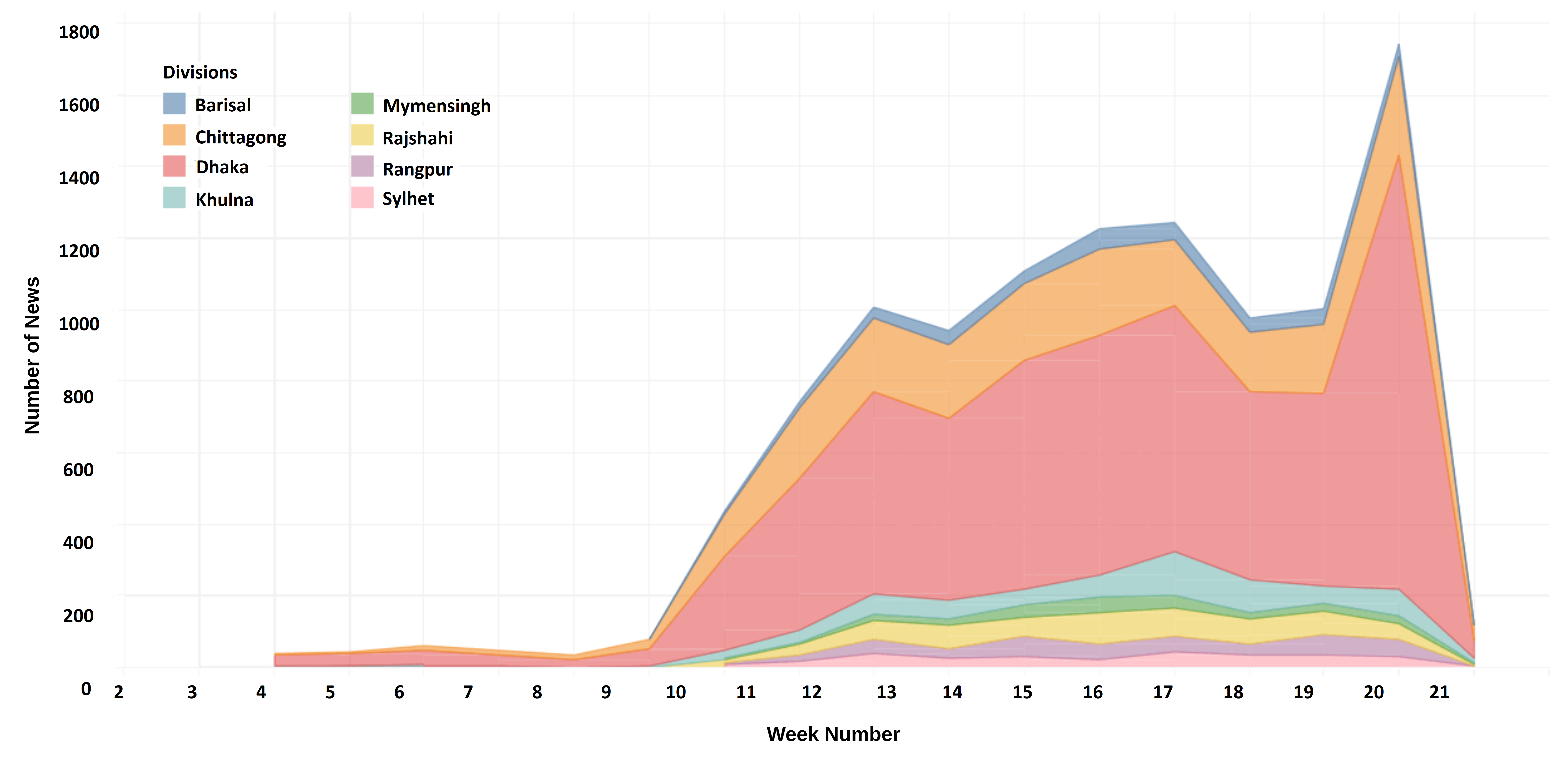}}
    \caption{Geo-spatial and Temporal Distributions of News Articles Published over Time. Horizontal axis shows consecutive weeks in the duration and vertical axis shows the volume (count of news articles).}
    \label{spatial4}
\end{figure}

The district-wise break down of published news articles for significant volume is shown in Figure~\ref{spatial3} and division-wise break down in Figure \ref{spatial2}. The figures show that most news published was related to the Dhaka district and Dhaka division. More than 57\% of the published news was related to the Dhaka division. After Dhaka, most news has been published on Chittagong. More than 19\% of the news was related to the Chittagong division. The geospatial and temporal distributions of newspaper articles are shown in Figure~\ref{spatial4}. The figure shows that the volume of published news articles related to each location significantly changed over time, lower volume before and beginning of the pandemic, and significantly increased during the pandemic.

\subsection{Topic Analysis}

\subsubsection{Topic Extraction}
\begin{figure}
\centering
\subfloat[Coherent Score\label{top1a}]{%
  \includegraphics[width=0.5\textwidth]{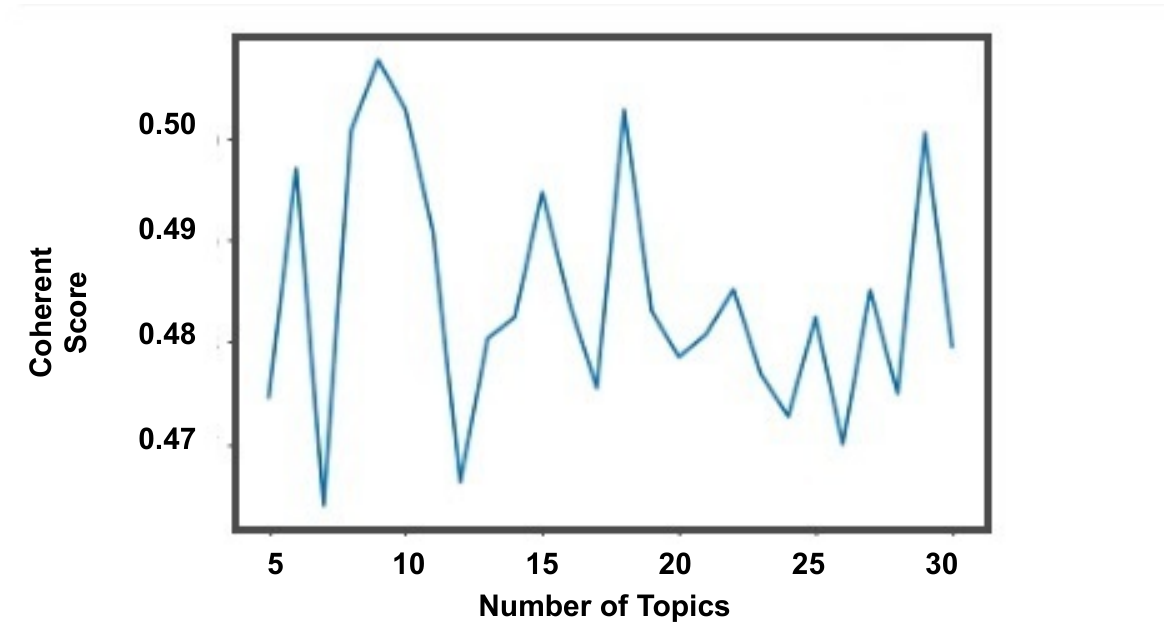}%
}\hfill 
\subfloat[Perplexity Score\label{top1b}]{%
  \includegraphics[width=0.5\textwidth]{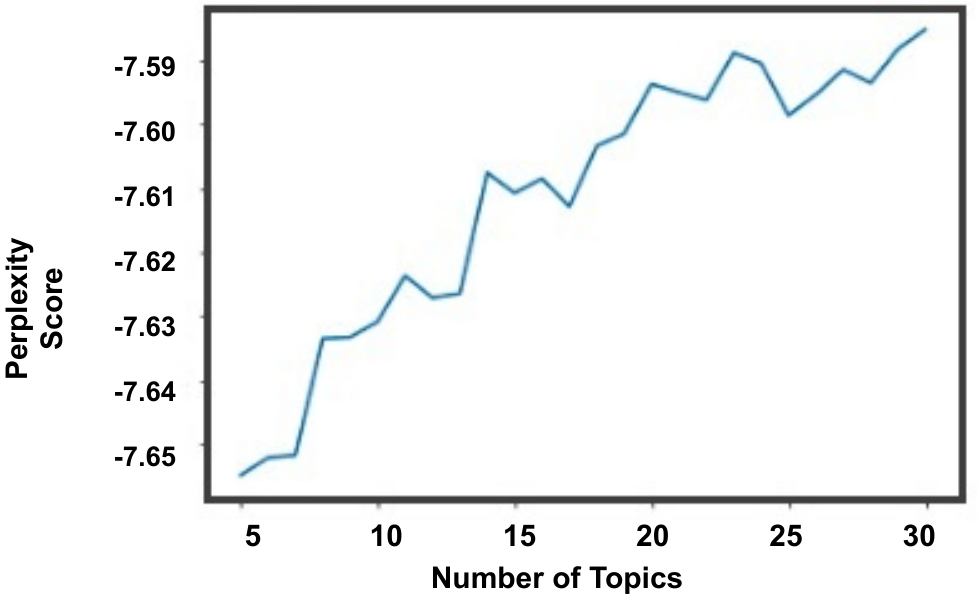}%
}\hfill

\caption{Determining optimal number of topic}
\label{optimal_number}
\end{figure}

For topic analysis through the LDA topic model, it is indispensable to decide the optimal number of topics. Seeking an appropriate LDA topic number and clarifications to examine the relationship between the COVID-19 emergency and news articles, we have given much thought. We used a coherence score and perplexity score to assess the choice of an appropriate number of topics. After preprocessing the data, we applied the LDA model to discover hidden topics in news articles. To determine the optimal number of topics, we diagnosed the coherent score and the perplexity score graph shown in Figure \ref{optimal_number}. Figure \ref{top1a} is showing the coherent score graph and Figure \ref{top1b} is showing the perplexity score graph.



\begin{figure}
\centering
\subfloat[Word cloud of Topic 1]{%
 \fbox{\includegraphics[width=0.3\textwidth]{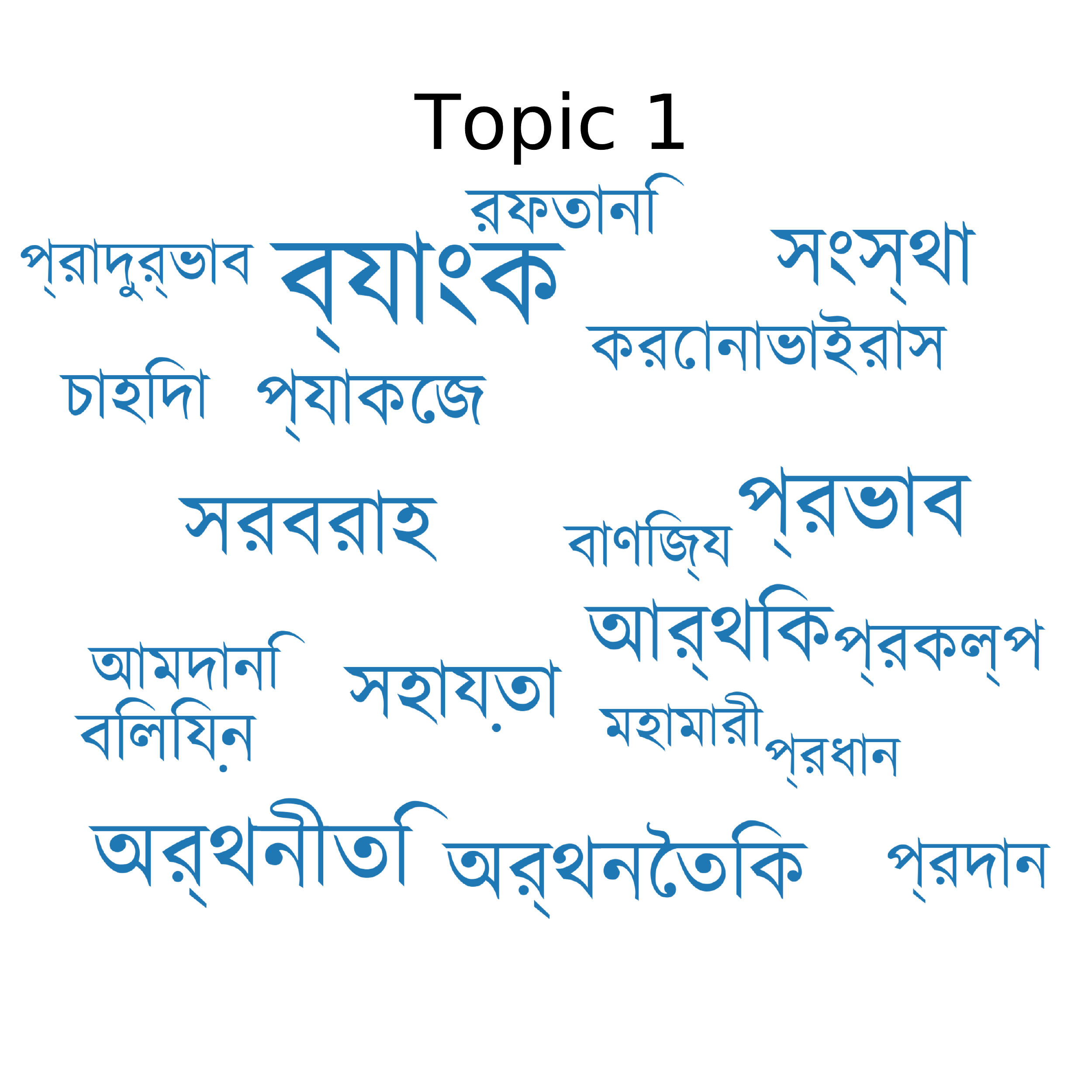}}%
}\hfill 
\subfloat[Word cloud of Topic 2]{%
 \fbox{\includegraphics[width=0.3\textwidth]{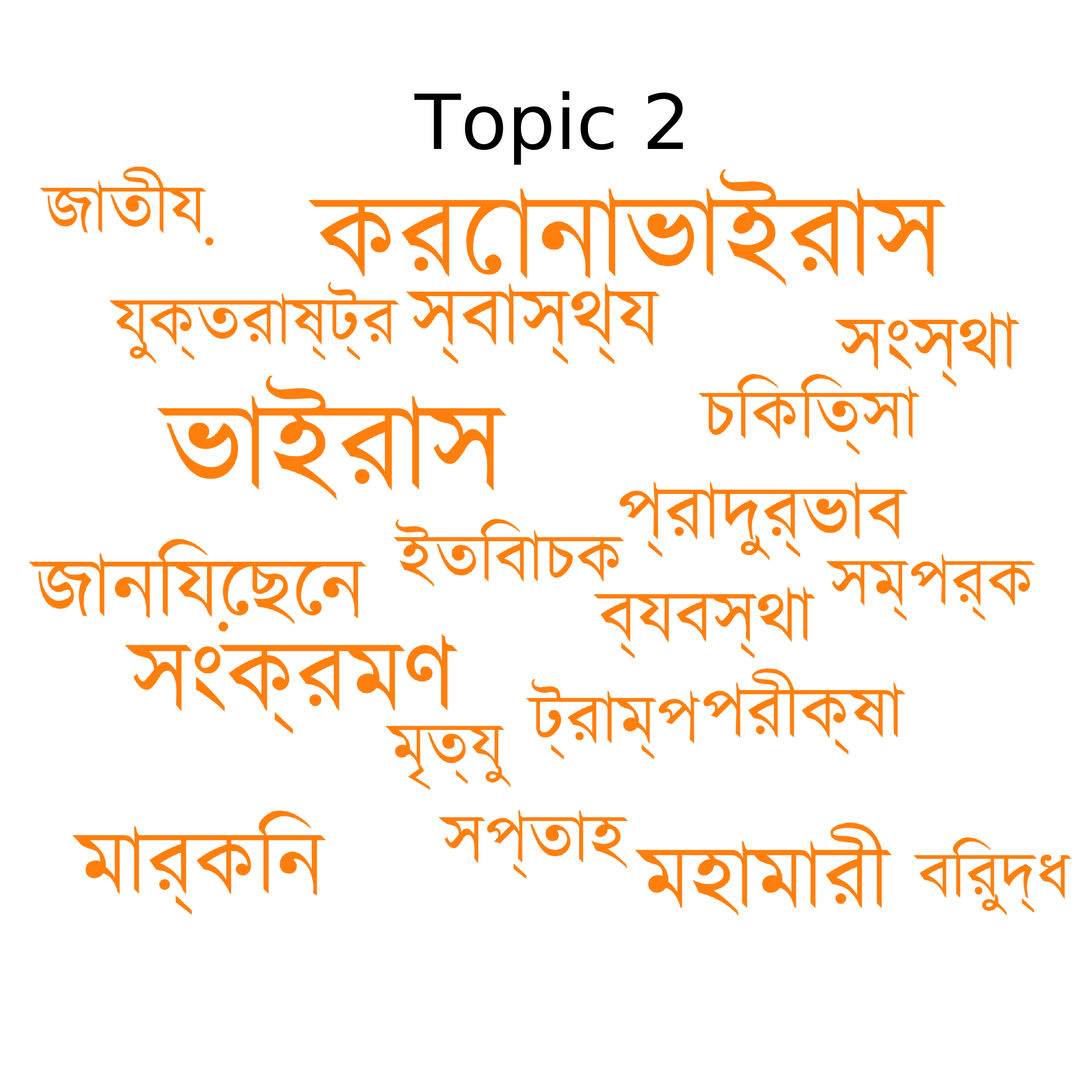}}%
}\hfill
\subfloat[Word cloud of Topic 3]{%
 \fbox{\includegraphics[width=0.3\textwidth]{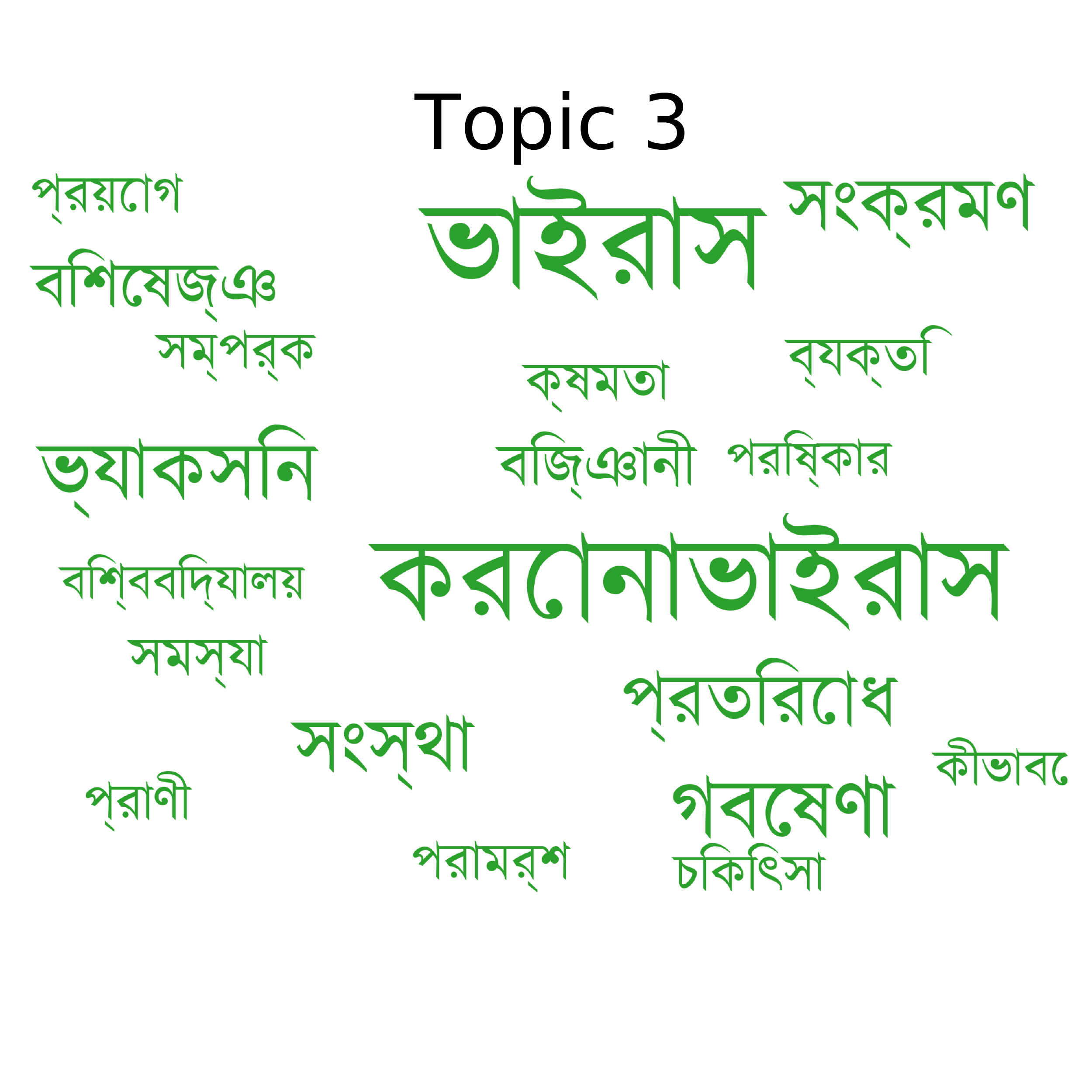}}%
}\hfill
\subfloat[Word cloud of Topic 4]{%
 \fbox{\includegraphics[width=0.3\textwidth]{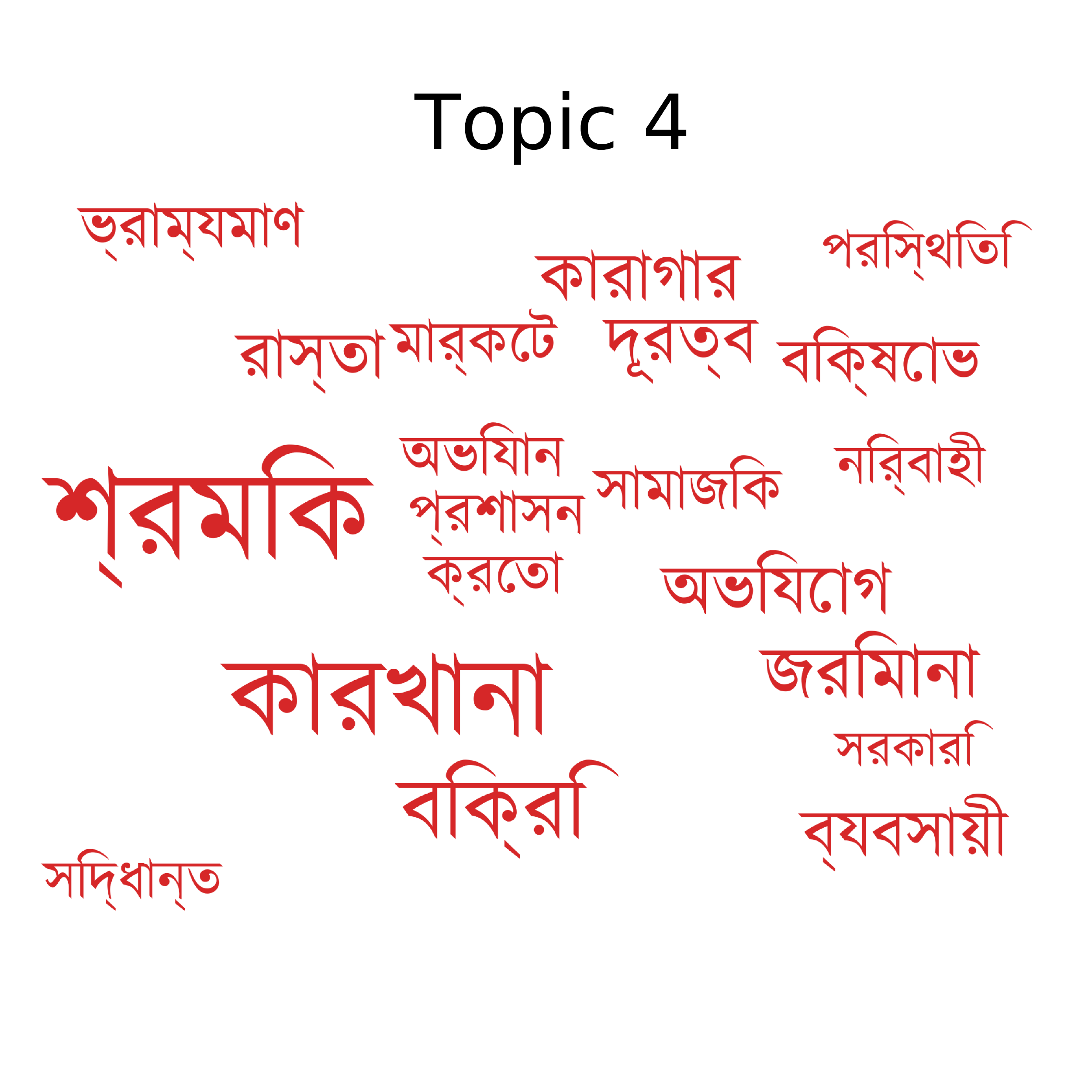}}%
}\hfill
\subfloat[Word cloud of Topic 5]{%
  \fbox{\includegraphics[width=0.3\textwidth]{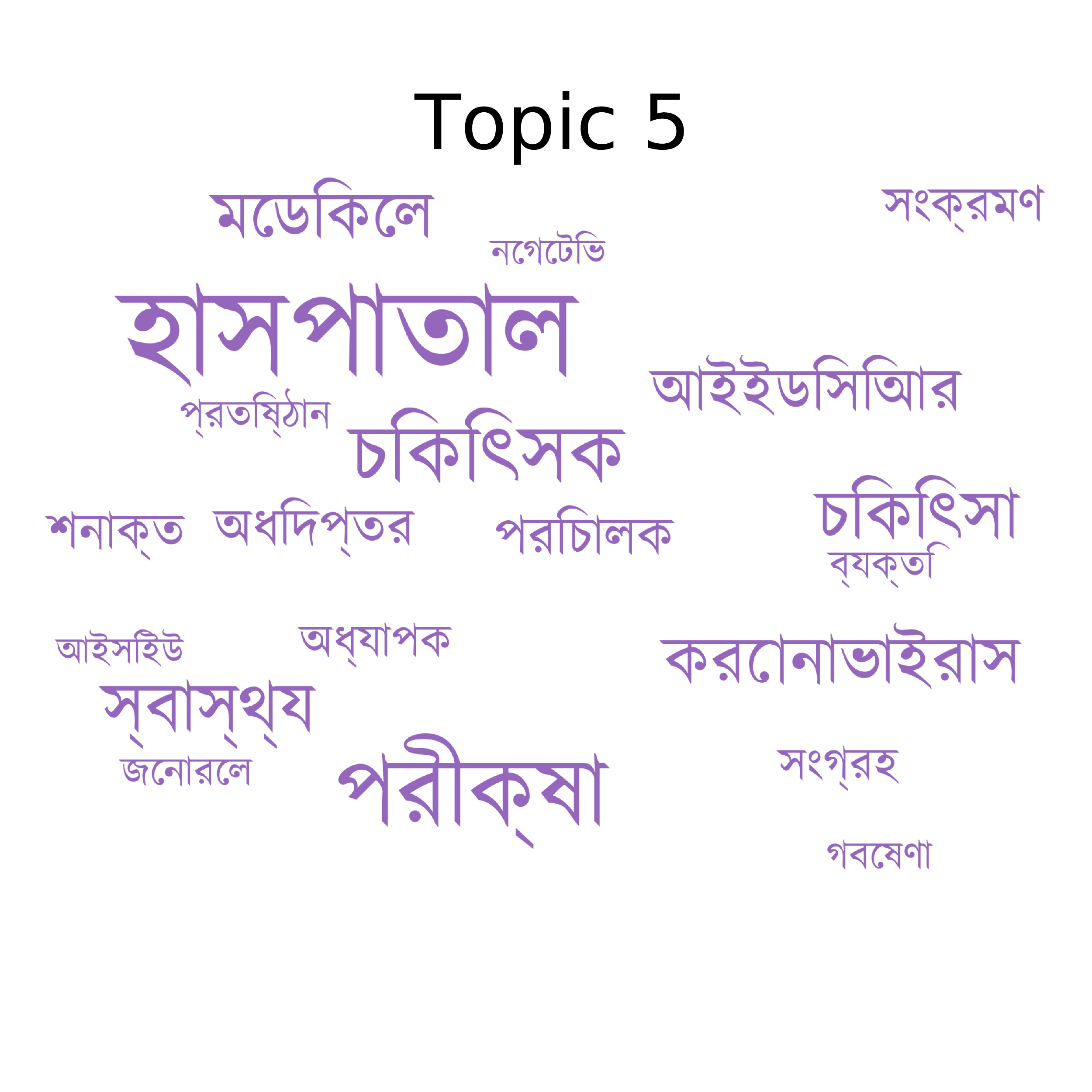}}%
}\hfill
\subfloat[Word cloud of Topic 6]{%
  \fbox{\includegraphics[width=0.3\textwidth]{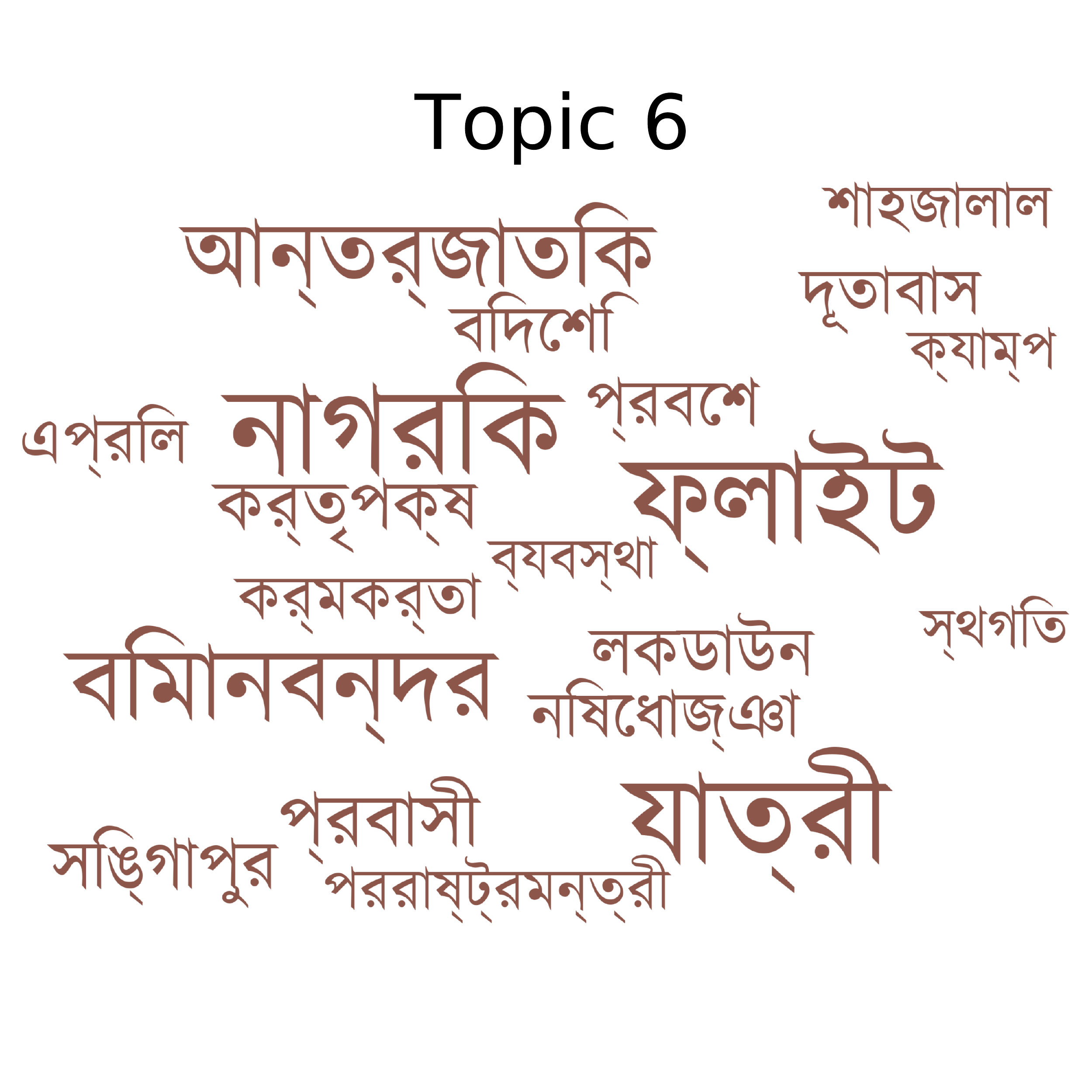}}%
}\hfill
\subfloat[Word cloud of Topic 7]{%
  \fbox{\includegraphics[width=0.3\textwidth]{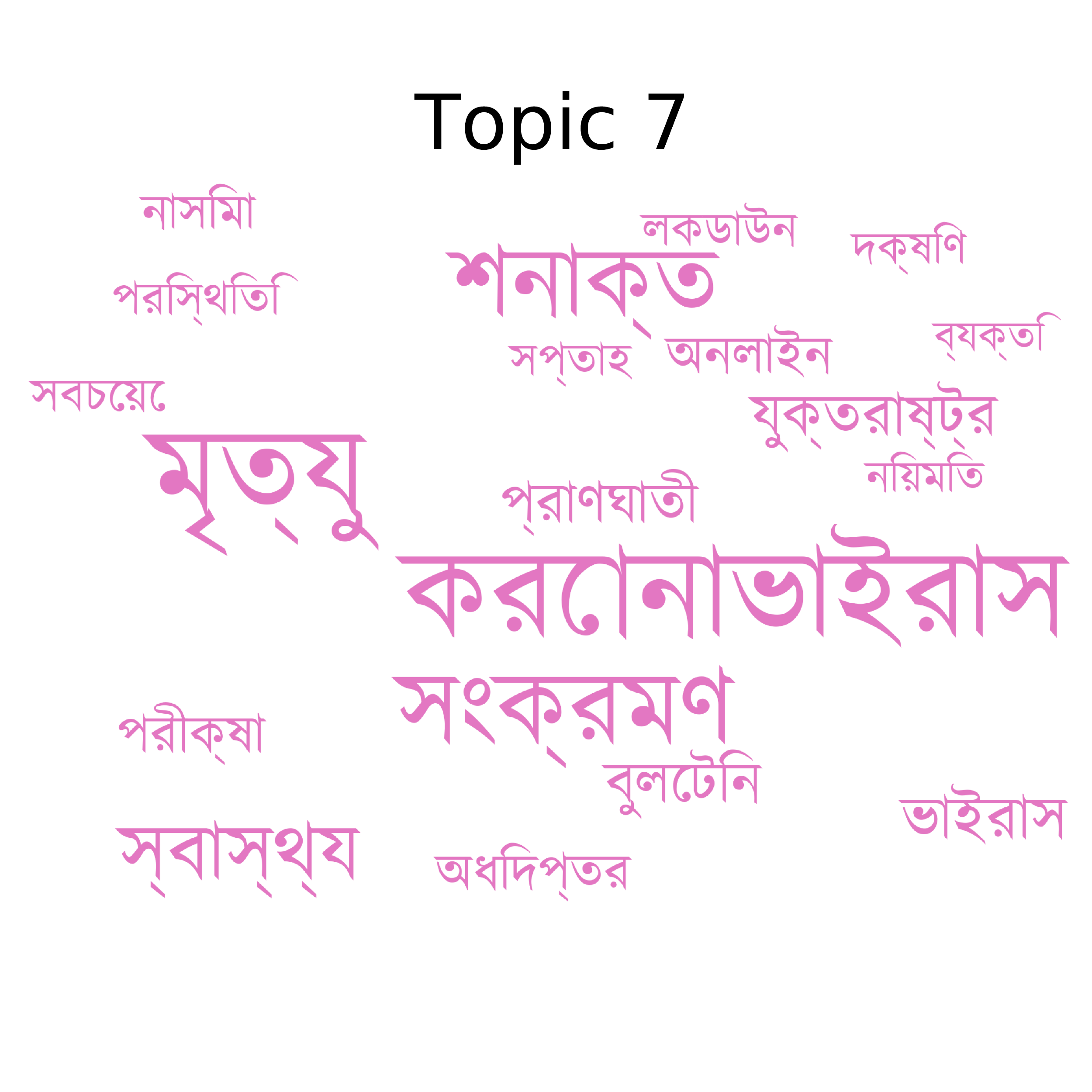}}%
}\hfill
\subfloat[Word cloud of Topic 8]{%
  \fbox{\includegraphics[width=0.3\textwidth]{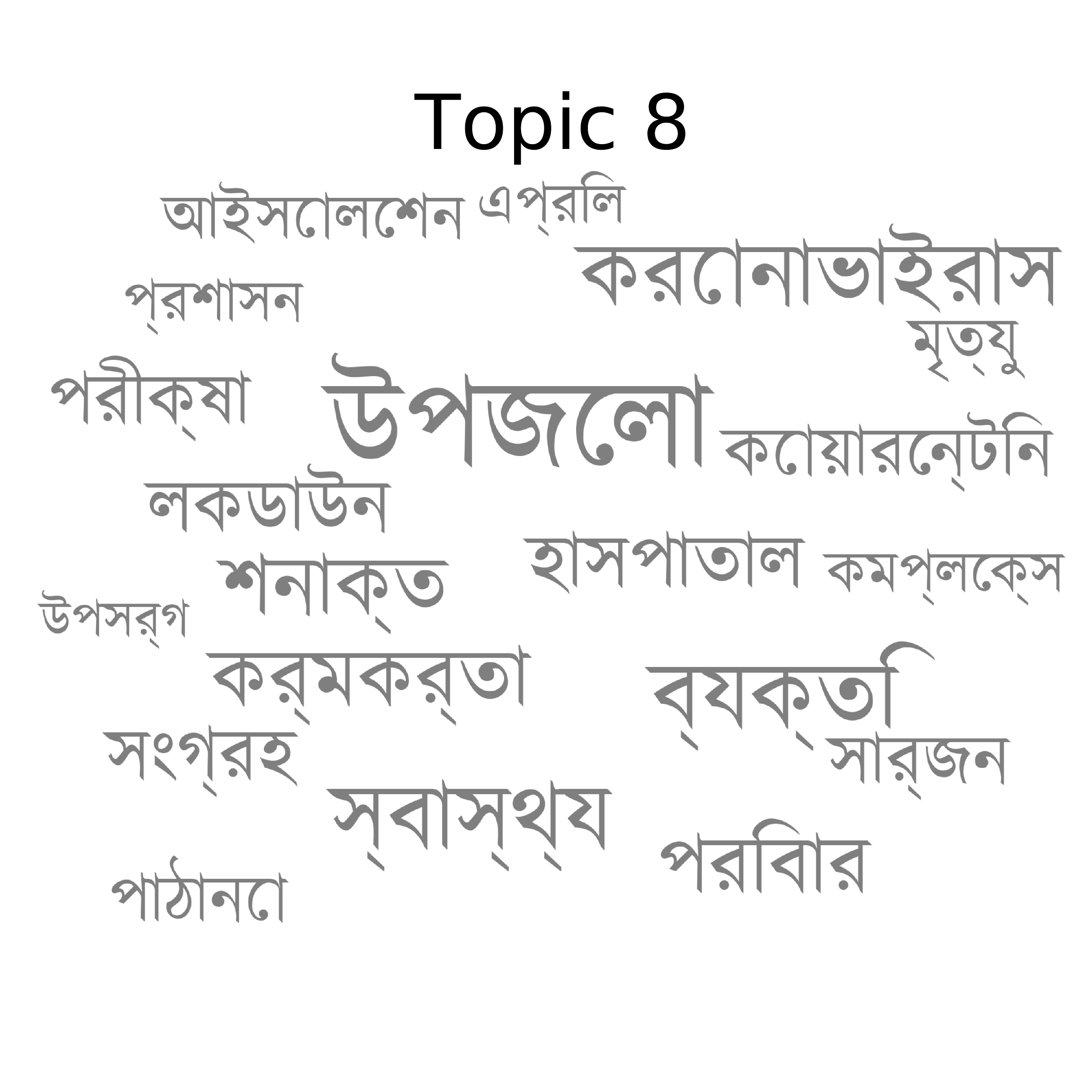}}%
}\hfill
\subfloat[Word cloud of Topic 9]{%
  \fbox{\includegraphics[width=0.3\textwidth]{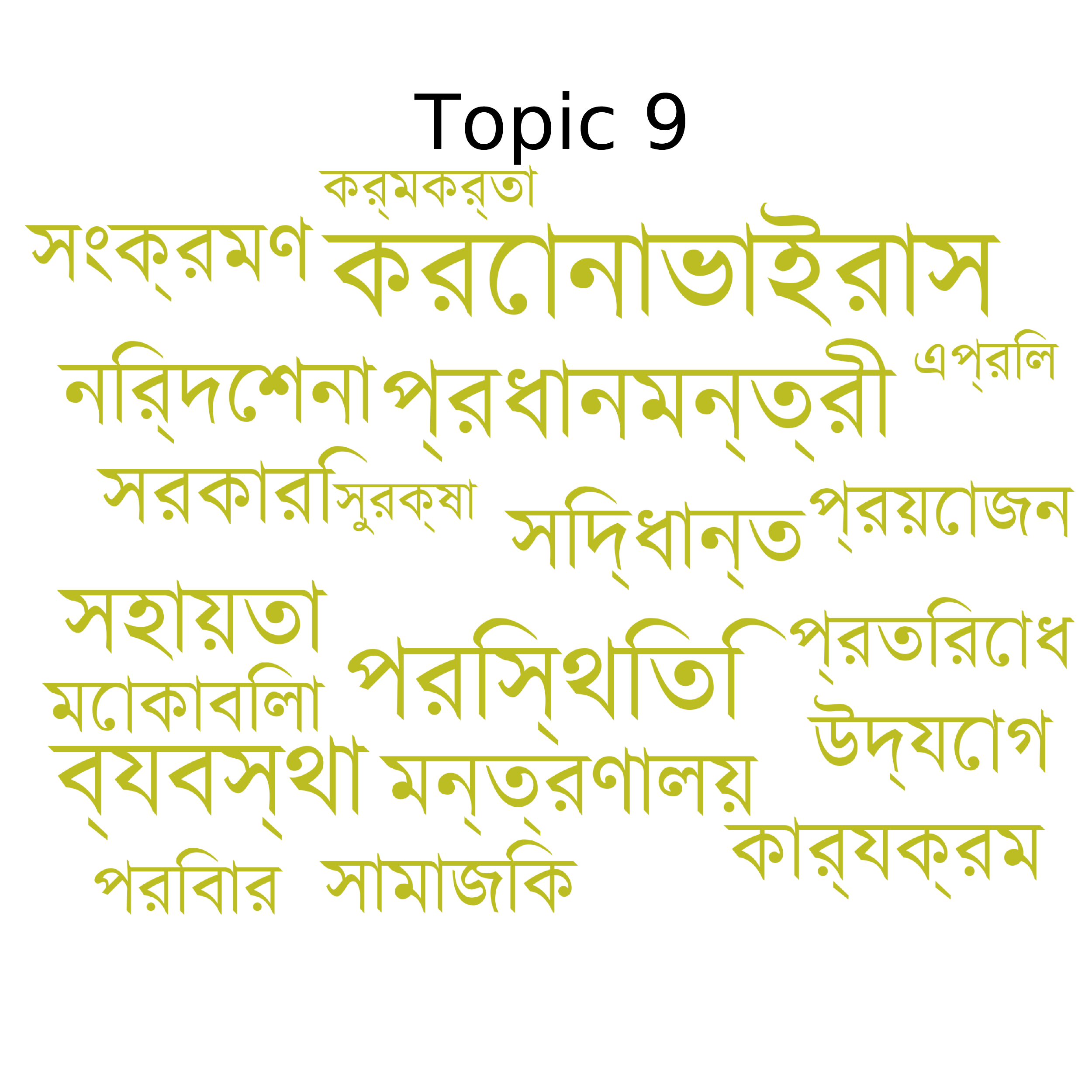}}%
}

\caption{Word Clouds for nine topics}
\label{nine_topics}
\end{figure}

\begin{figure}
\centering
\subfloat[Word counts of Topic 1]{%
 \fbox{\includegraphics[width=0.3\textwidth]{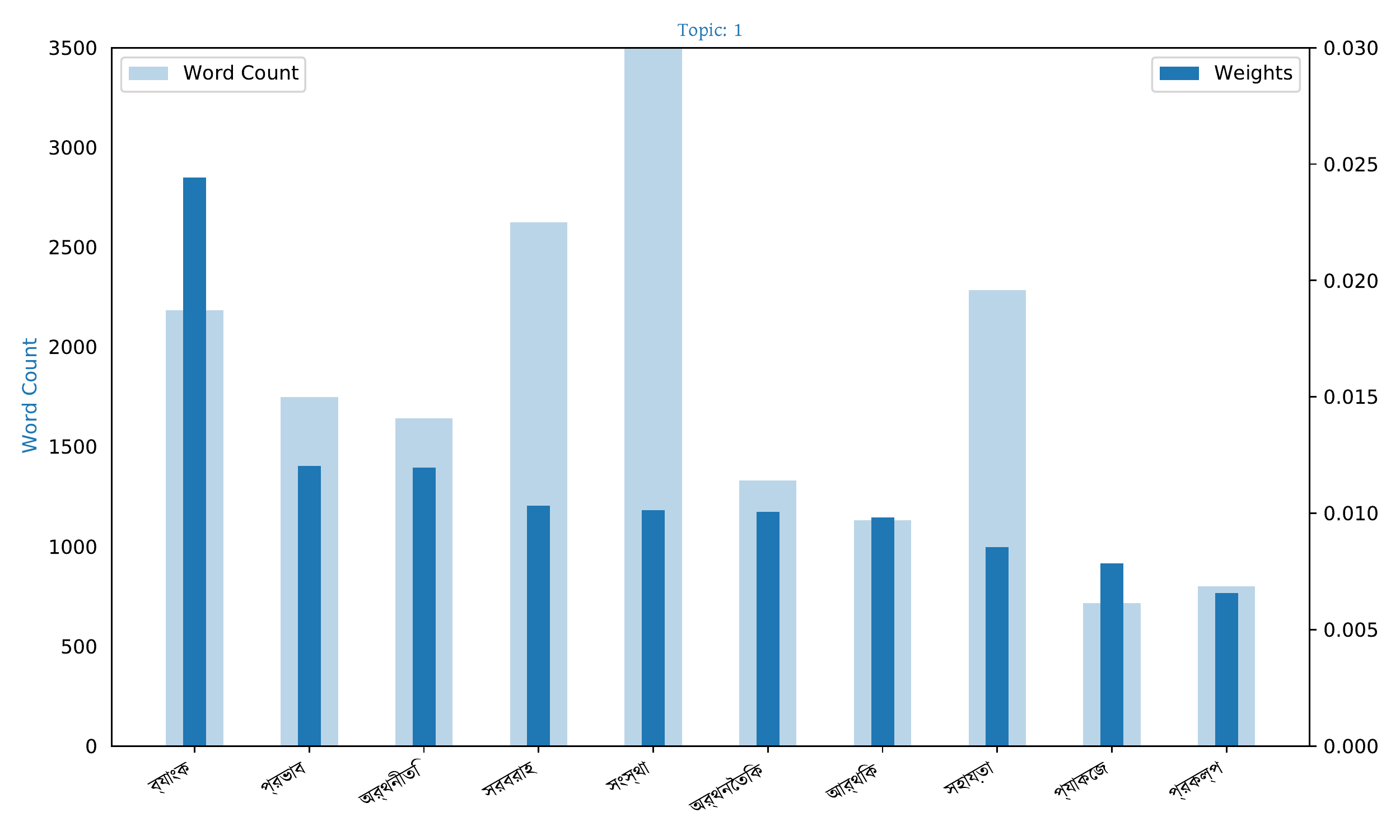}}%
}\hfill 
\subfloat[Word counts of Topic 2]{%
 \fbox{\includegraphics[width=0.3\textwidth]{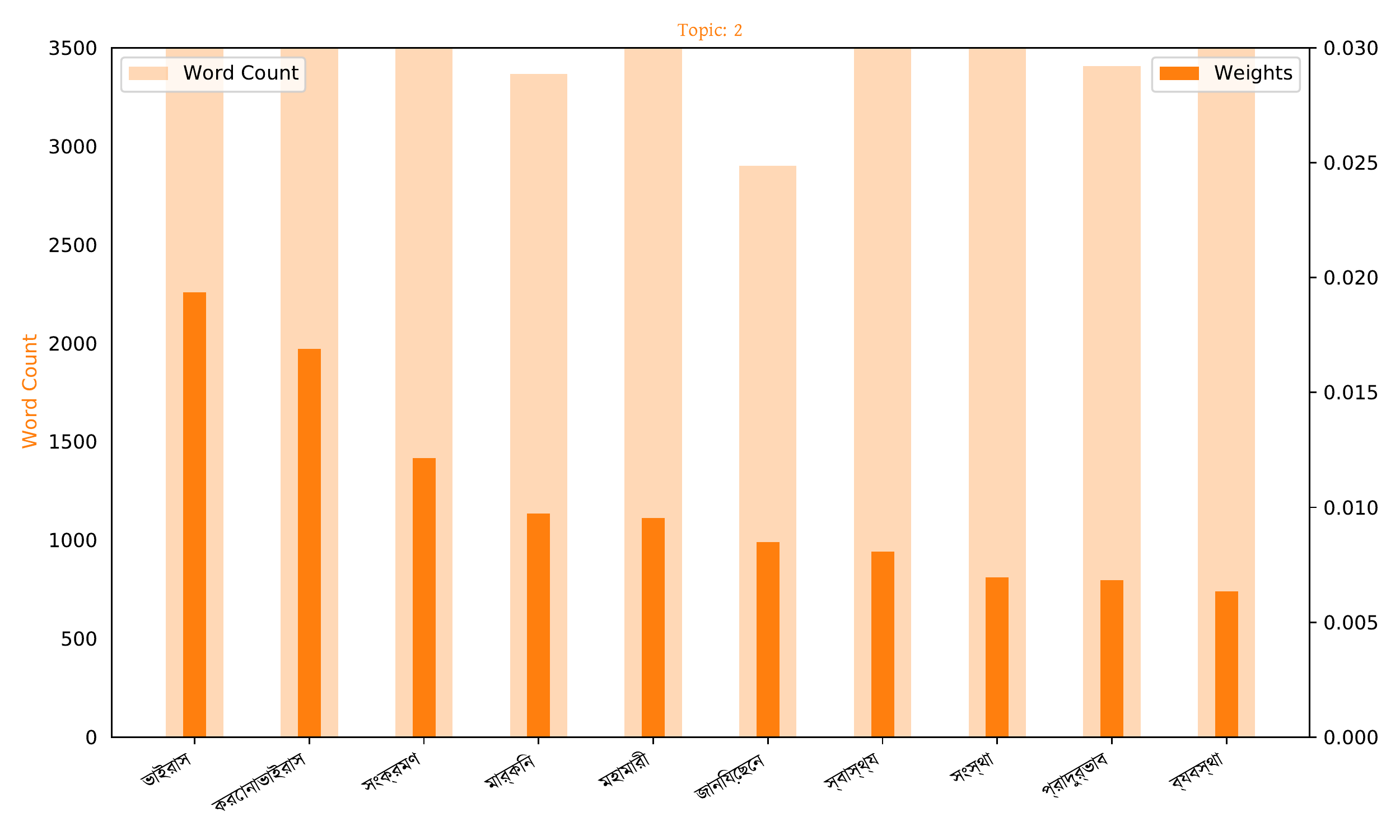}}%
}\hfill
\subfloat[Word counts of Topic 3]{%
  \fbox{\includegraphics[width=0.3\textwidth]{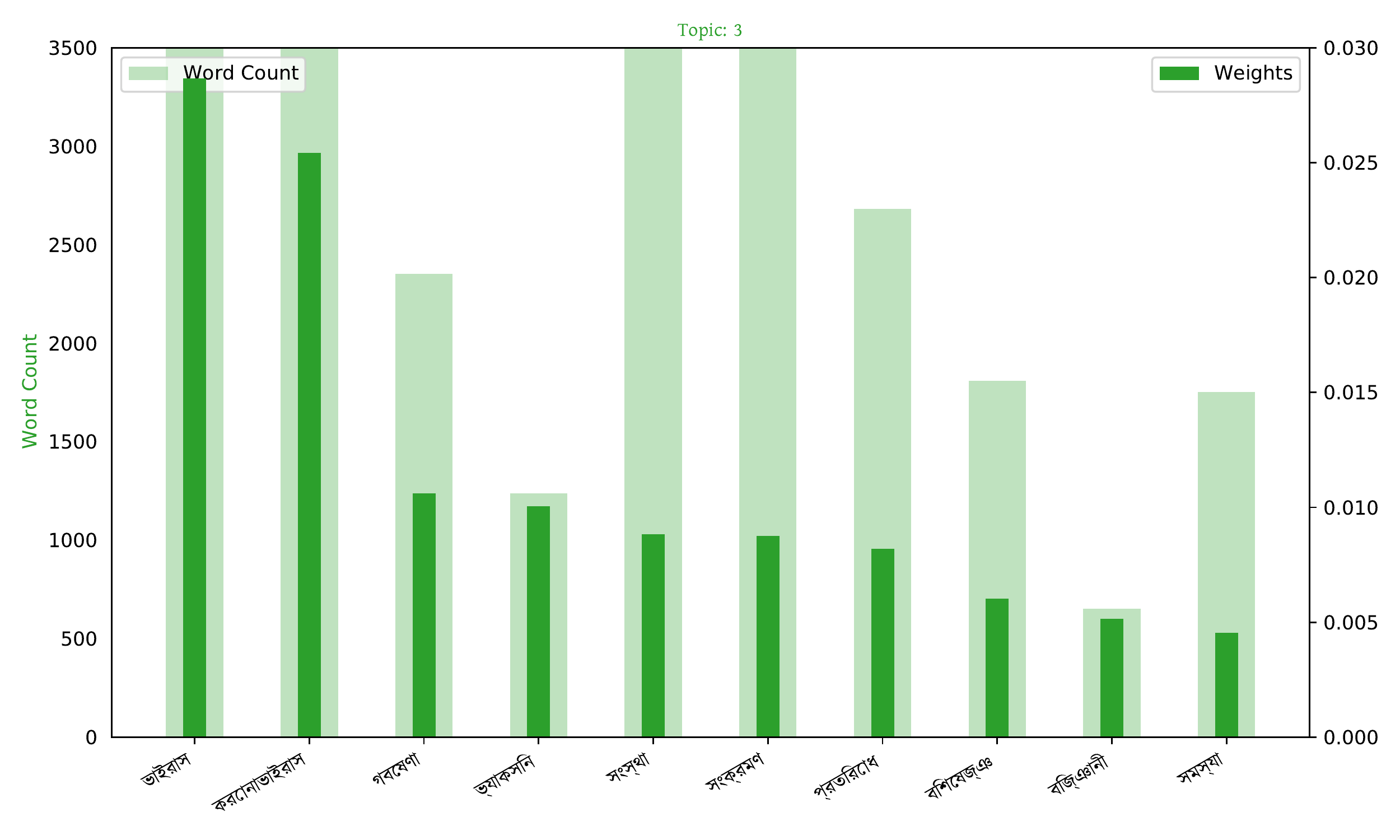}}%
}\hfill
\subfloat[Word counts of Topic 4]{%
  \fbox{\includegraphics[width=0.3\textwidth]{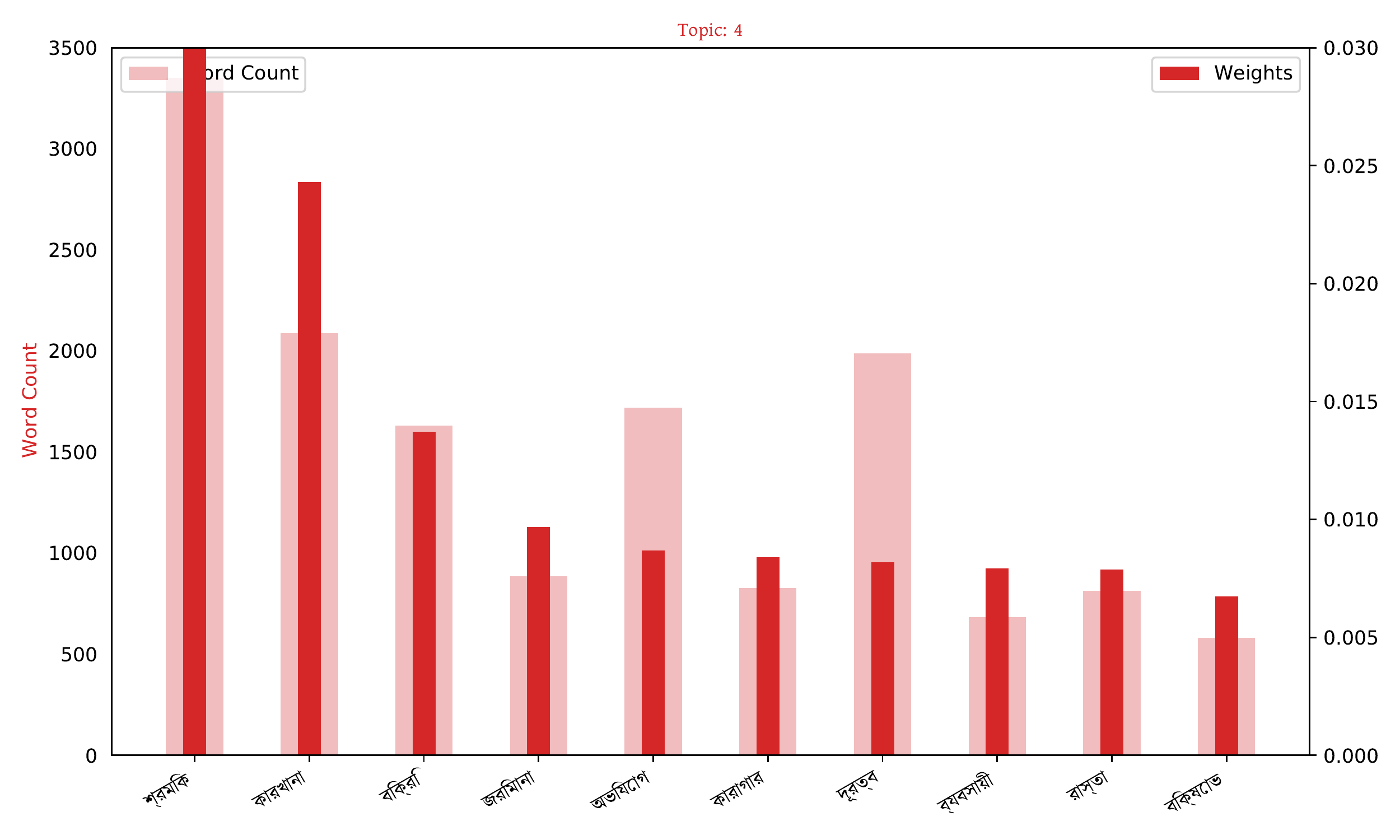}}%
}\hfill
\subfloat[Word counts of Topic 5]{%
  \fbox{\includegraphics[width=0.3\textwidth]{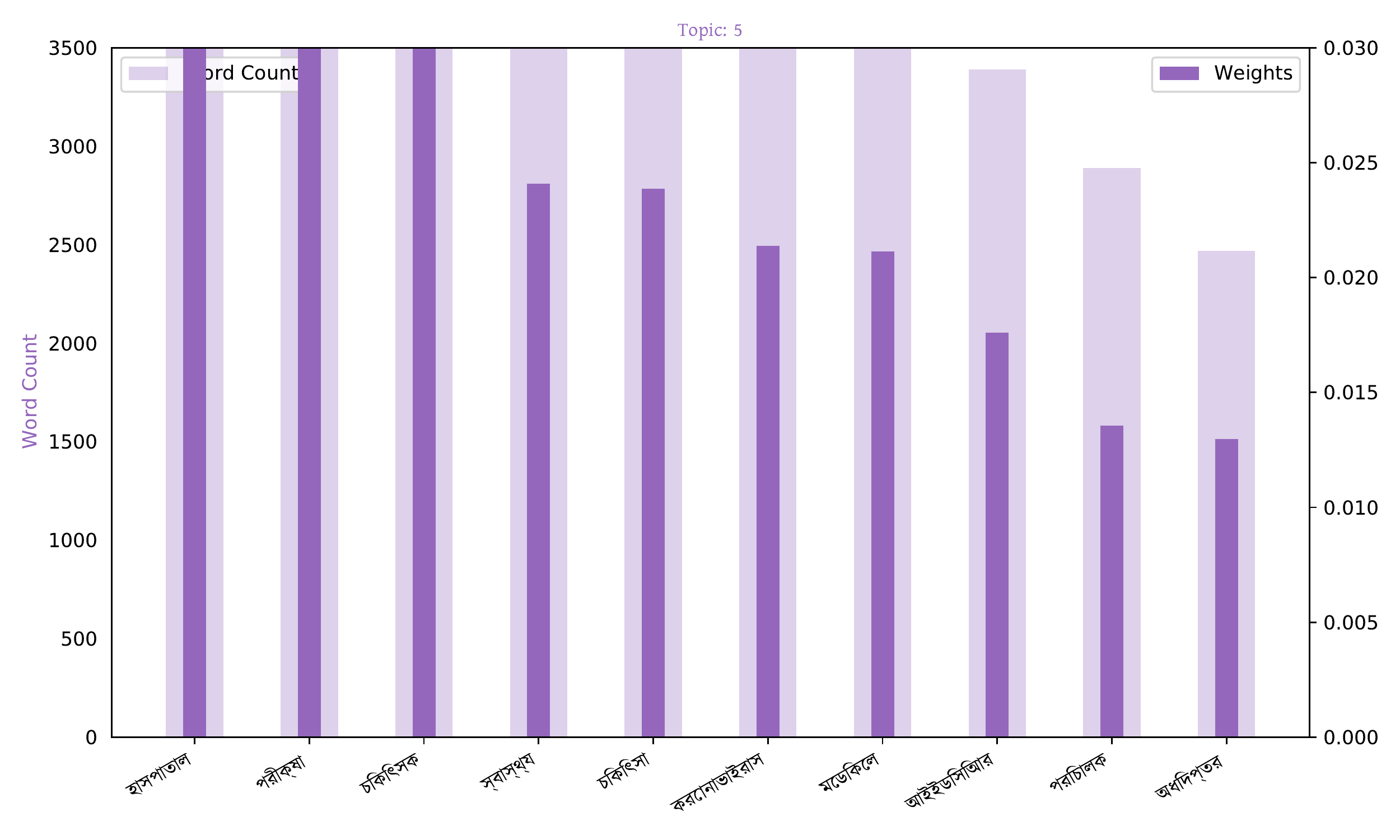}}%
}\hfill
\subfloat[Word counts of Topic 6]{%
  \fbox{\includegraphics[width=0.3\textwidth]{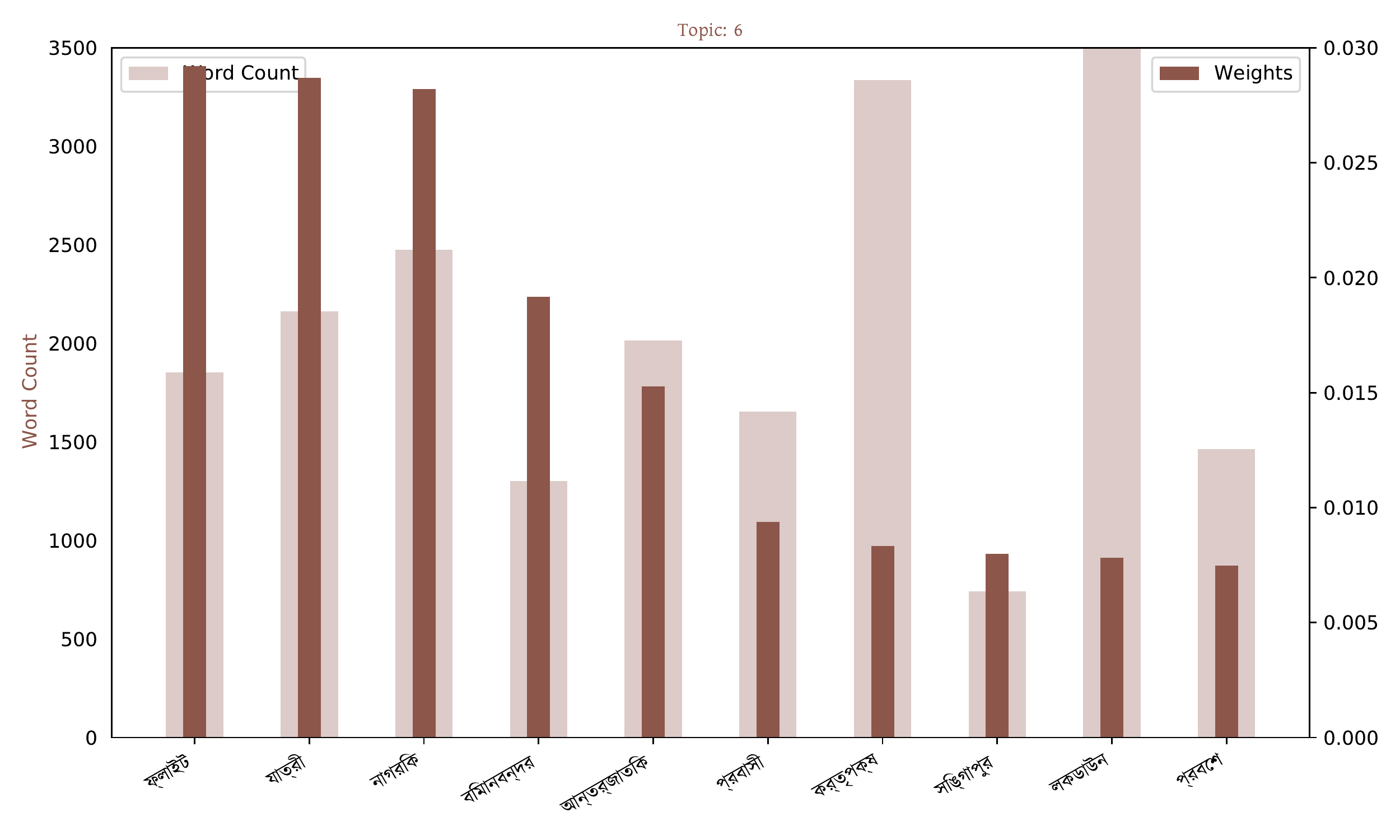}}%
}\hfill
\subfloat[Word counts of Topic 7]{%
  \fbox{\includegraphics[width=0.3\textwidth]{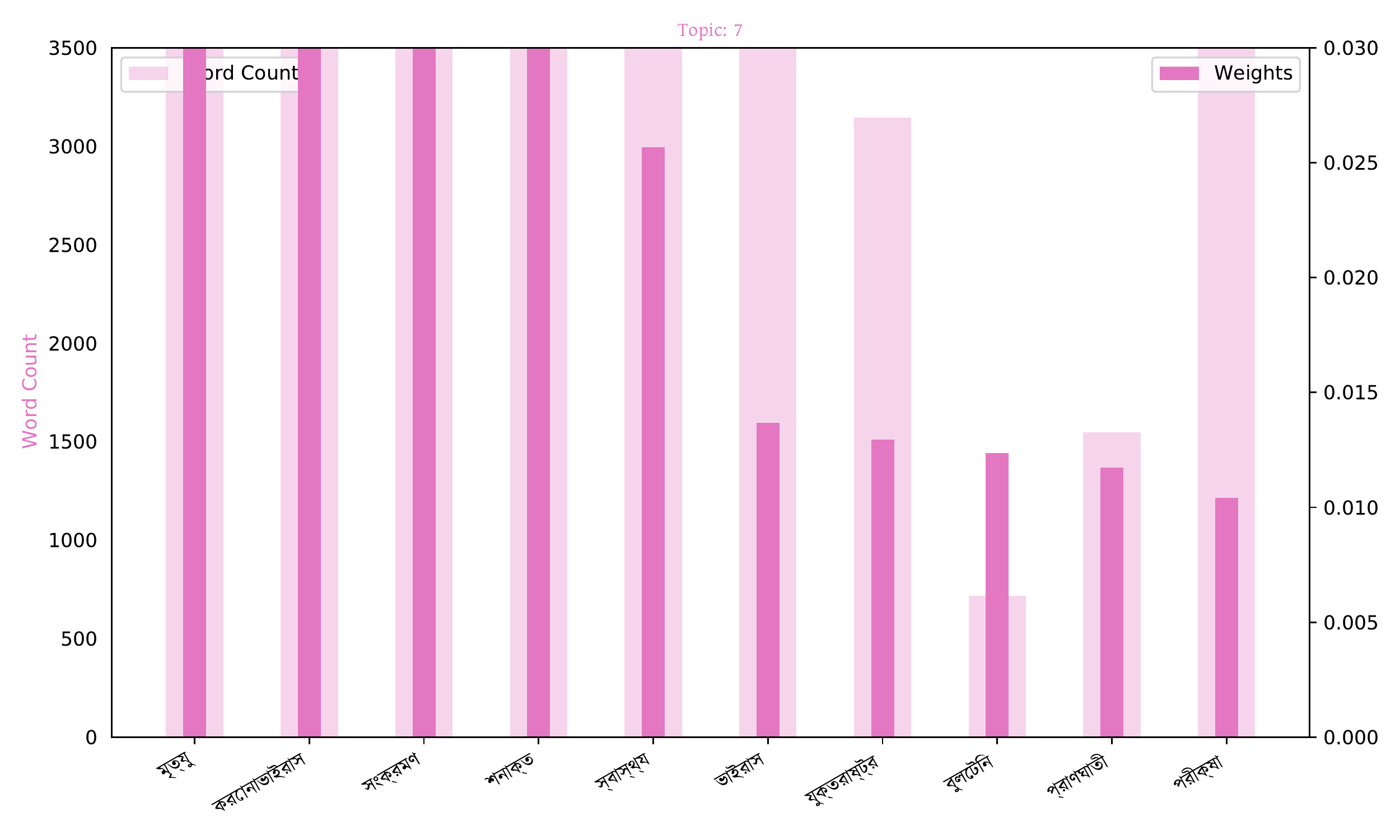}}%
}\hfill
\subfloat[Word counts of Topic 8]{%
  \fbox{\includegraphics[width=0.3\textwidth]{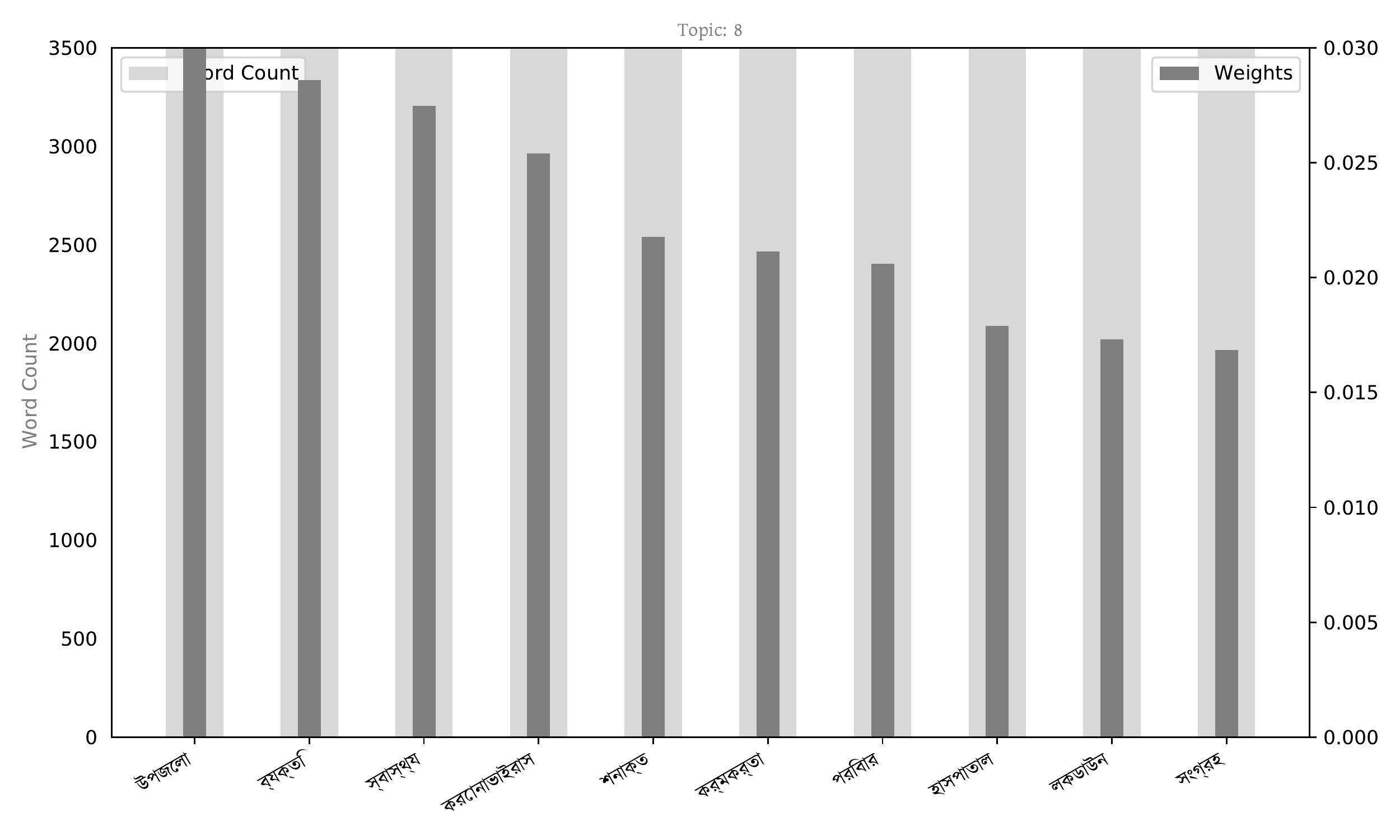}}%
}\hfill
\subfloat[Word counts of Topic 9]{%
  \fbox{\includegraphics[width=0.3\textwidth]{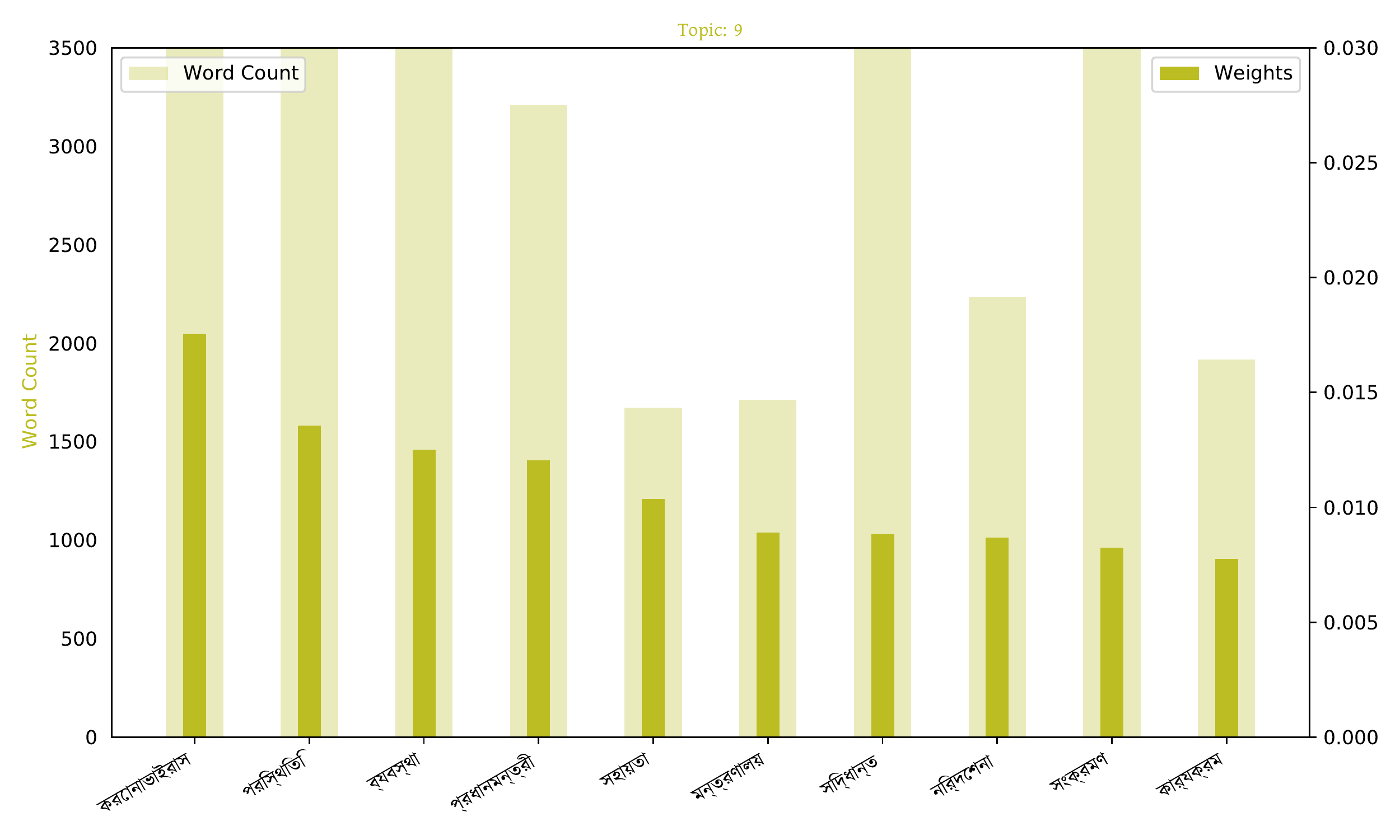}}%
}
\caption{Word Counts for nine topics}
\label{wct9}
\end{figure}

\begin{figure}
    \centering
    \fbox{\includegraphics[width=.5\textwidth]{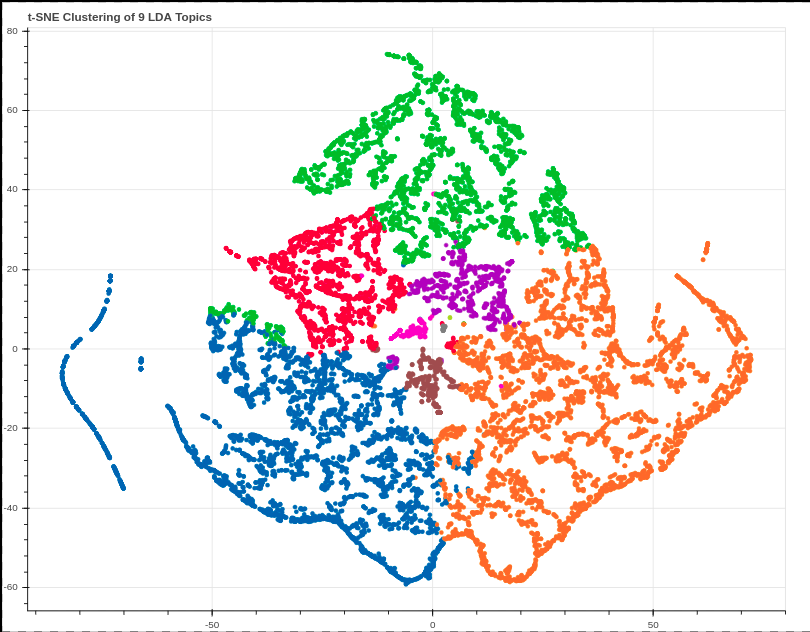}}
    \caption{t-SNE clustering chart}
    \label{top8_tsne}
\end{figure}

\begin{figure}
    \centering
    \includegraphics[width=1\textwidth]{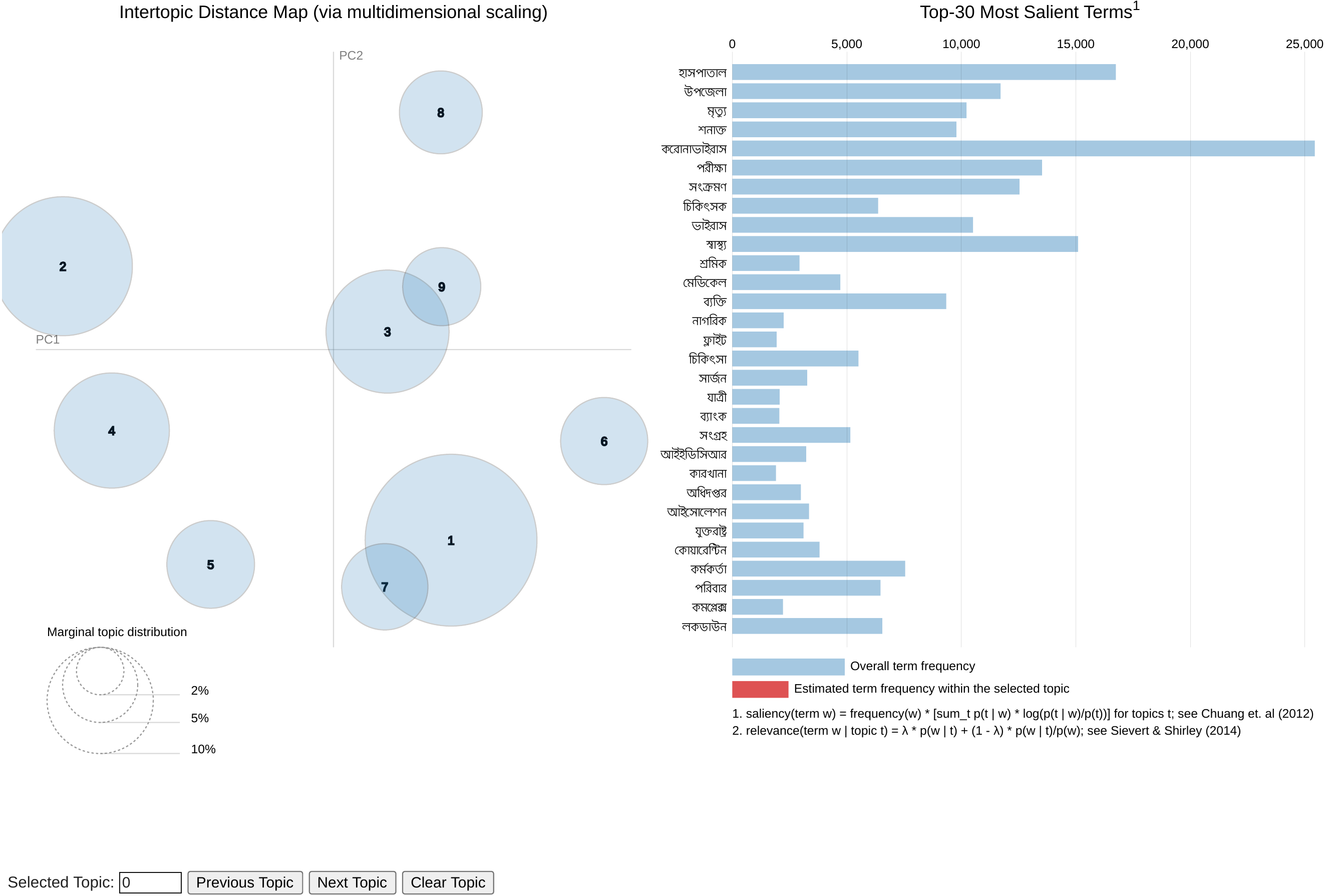}
    \caption{Interactive Topic Visualization}
    \label{top9_itv}
\end{figure}

From the coherence score graph, we got the highest coherence score (0.5077) when we set the number of topics to 9, shown in Figure~\ref{top1a}. Moreover, from the perplexity score graph, we got the highest perplexity score (-7.59) when we set the number of topics to 24, shown in Figure~\ref{top1b}. We chose the coherent score between the coherent score and perplexity score as the optimal number of topics for the coherent score is 9, which is very close to the number of manually extracted classes of 8, shown in Table \ref{tab:manual_8_topics}. So we set the number of topics for LDA topic extraction to 9.
The word clouds for top words (i.e., keywords) in each of the nine topics is shown in Figure~\ref{nine_topics}. The weights and appearance counts of the keywords in each topic is shown in Figure~\ref{wct9}. The visualization of the clusters of documents in a 2D space using the t-SNE (t-distributed stochastic neighbor embedding) algorithm is shown in Figure~\ref{top8_tsne}. In Figure~\ref{top9_itv}, inter-topic distance map and 30 relevant keywords are displayed for each topic. They discovered nine topics are listed in Table~\ref{tab:topics}.
\begin{table}[htbp]
  \centering
  \caption{Nine Topics Discovered by LDA}
    \begin{tabular}{p{\textwidth}}
    \toprule
    (1) Economic Crisis and Incentives, (2) Epidemic Situation and Outbreak, (3) Vaccine and Treatment, (4) Demonstration for Wages and Relief, (5) Medical Care and Health Organization Responses, (6) Repatriation and International Situations, (7) Daily Infected Death and Recovered Cases, (8) Strategic Preparedness, and (9) Government Announcement and Responses \\
    \bottomrule
    \end{tabular}%
  \label{tab:topics}%
\end{table}%

\begin{figure}
    \centering
    \includegraphics[width=0.6\textwidth]{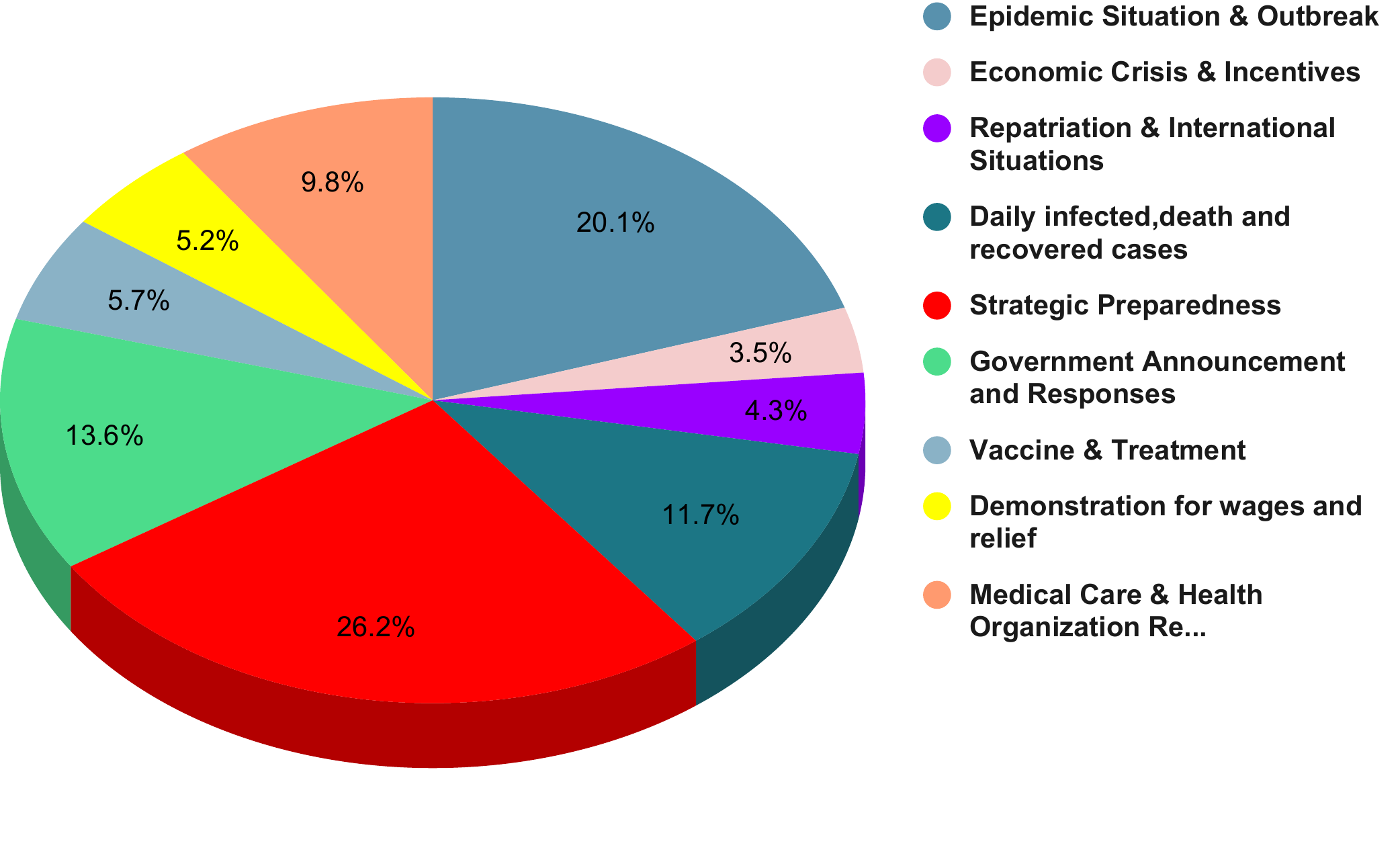}
    \caption{Proportion of Topic Frequency Distribution in the News Collection}
    \label{top11}
\end{figure}

Figure~\ref{top11} shows the topic frequency ratio in the document collection (news articles). The figure shows that Topic 8 (Strategic Preparedness) is the most frequent topic amongst all the nine topics discovered by LDA, and this topic accounted for 26.3\% of all the nine topics. The second most frequent LDA topic is Topic 2 (Epidemic Situation and Outbreak), which accounted for 20.1\%. Topic 9 (Government Announcement and Responses) and Topic 7 (Daily Infected, Death, and Recovered Cases) are 13.6\% and 11.7\%, respectively, and are the third and fourth most frequent topics. Topic 5 (Medical Care and Health Organization Responses), Topic 3 (Vaccine and Treatment), and Topic 4 (Demonstration for Wages) and Relief are at fifth, sixth, and seventh positions, and They accounted for 9.8\%, 5.7\%, and 5.2\%, respectively. Finally, Topic 6 (Repatriation and International Situations) and Topic 1 (Economic Crisis and Incentives) are the least frequent topics, and the proportion of these two topics is less than 5\%. By reviewing all these topics and analysis, we can insight into the pandemic or any important incident in a society.

\begin{figure}
\subfloat[Topic 1]{%
  \includegraphics[width=0.385\textwidth]{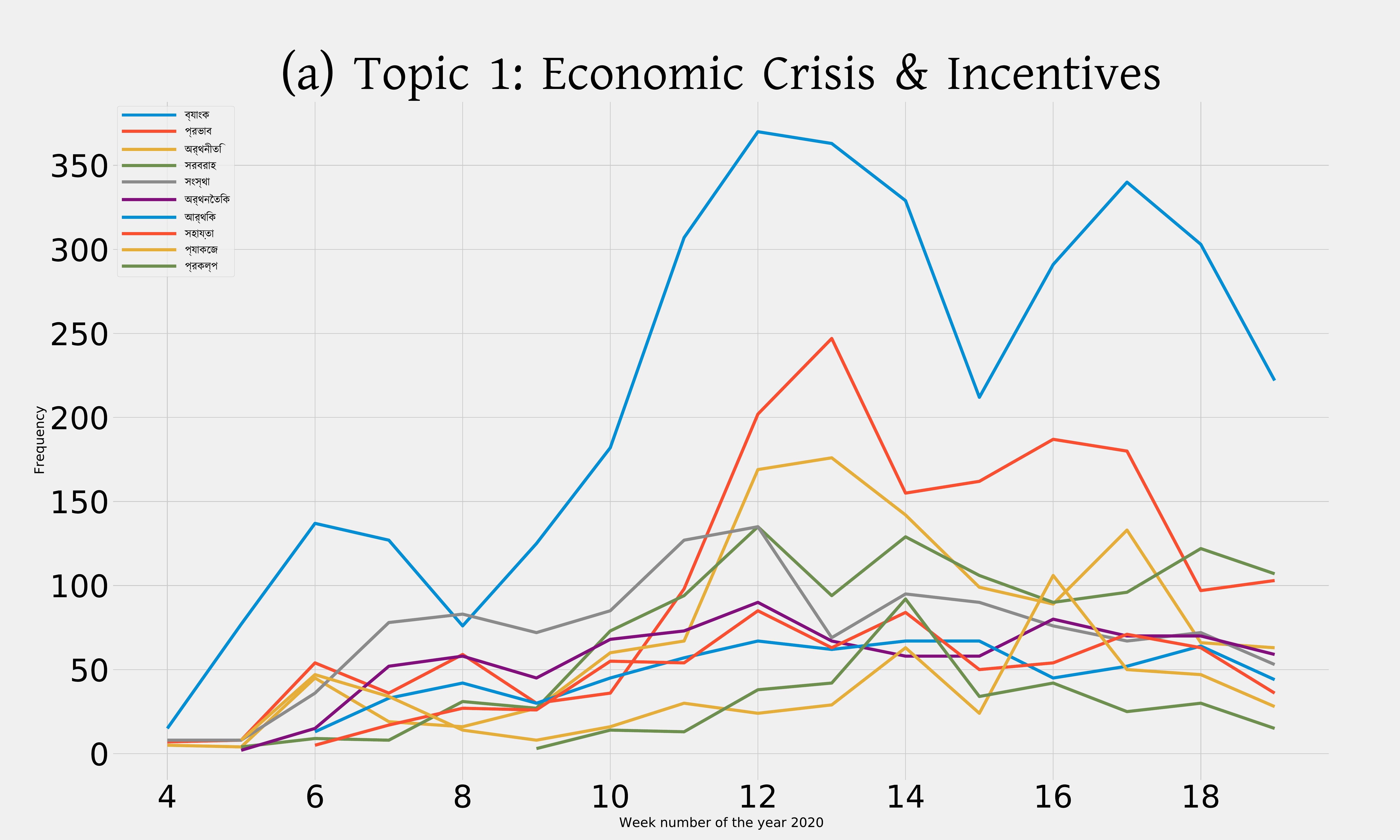}%
}\hfill 
\subfloat[Topic 2]{%
  \includegraphics[width=0.385\textwidth]{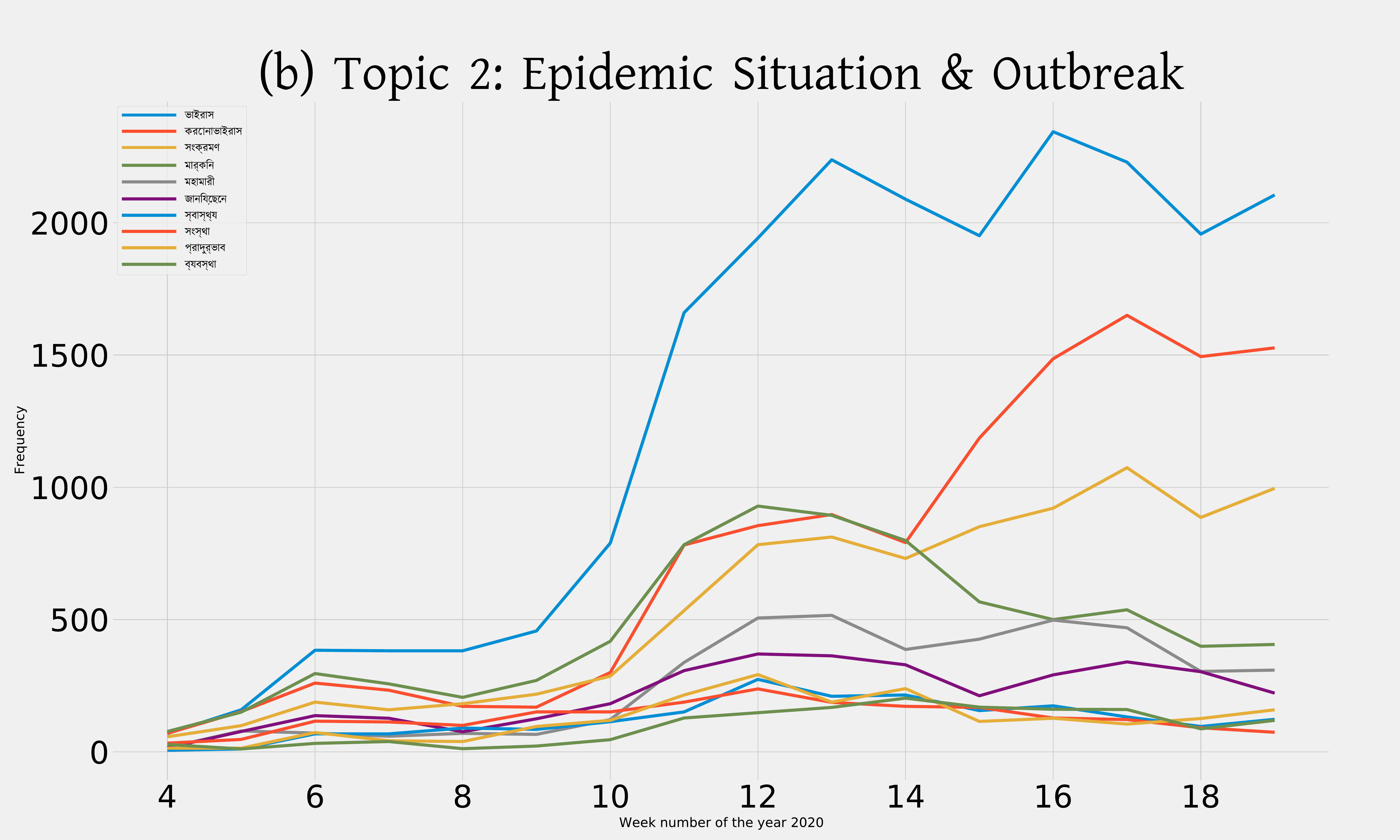}%
}\hfill
\subfloat[Topic 3]{%
  \includegraphics[width=0.385\textwidth]{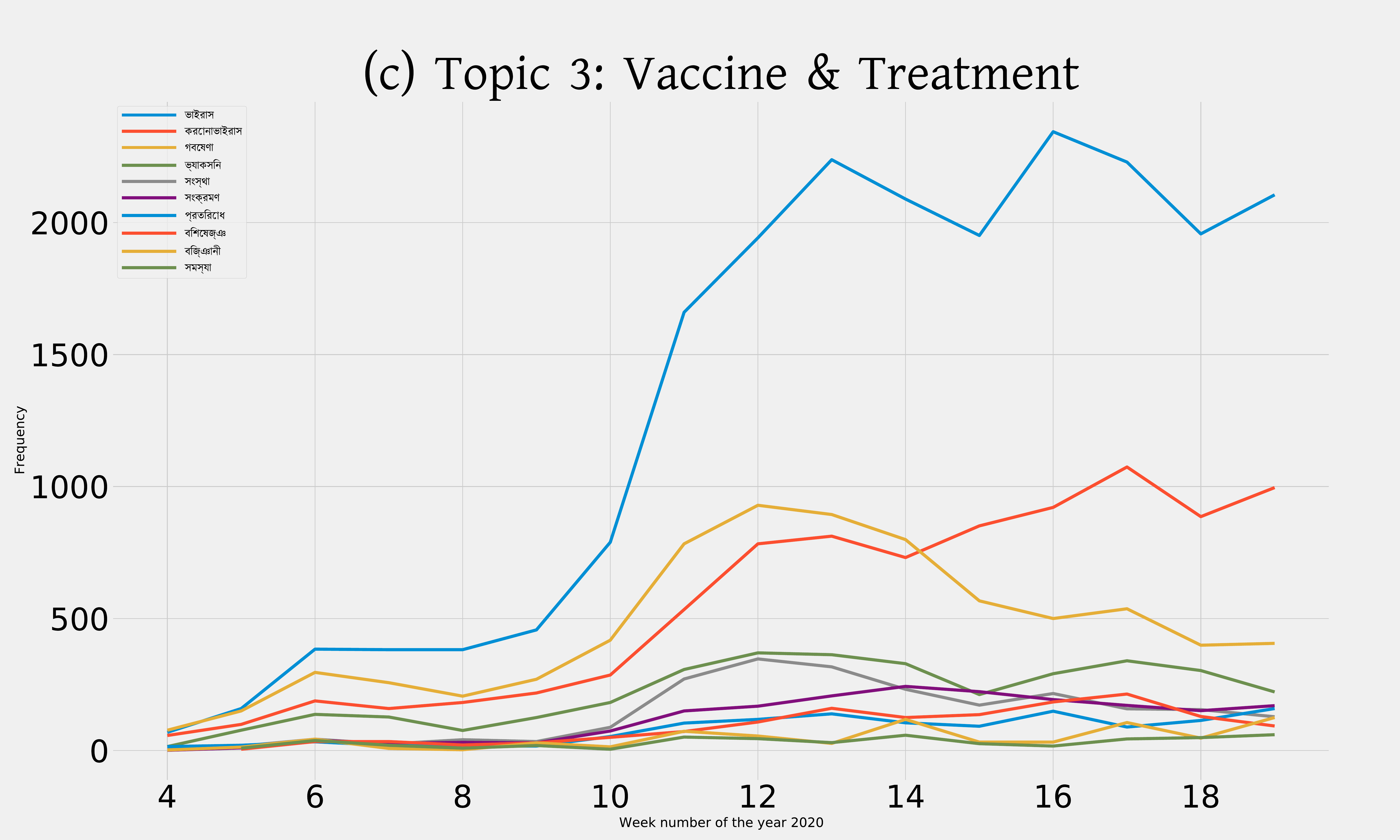}%
}\hfill
\subfloat[Topic 4]{%
  \includegraphics[width=0.385\textwidth]{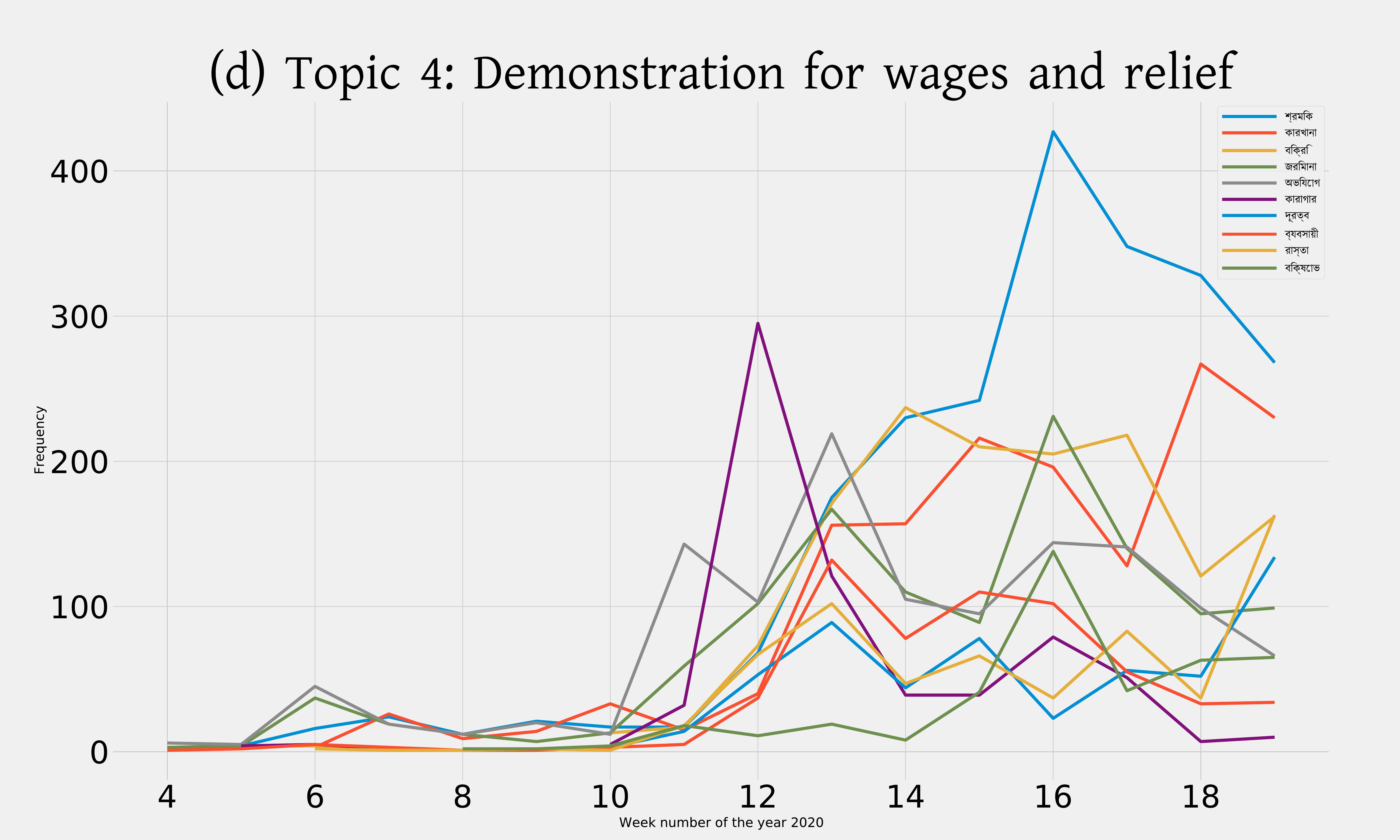}%
}\hfill
\subfloat[Topic 5]{%
  \includegraphics[width=0.385\textwidth]{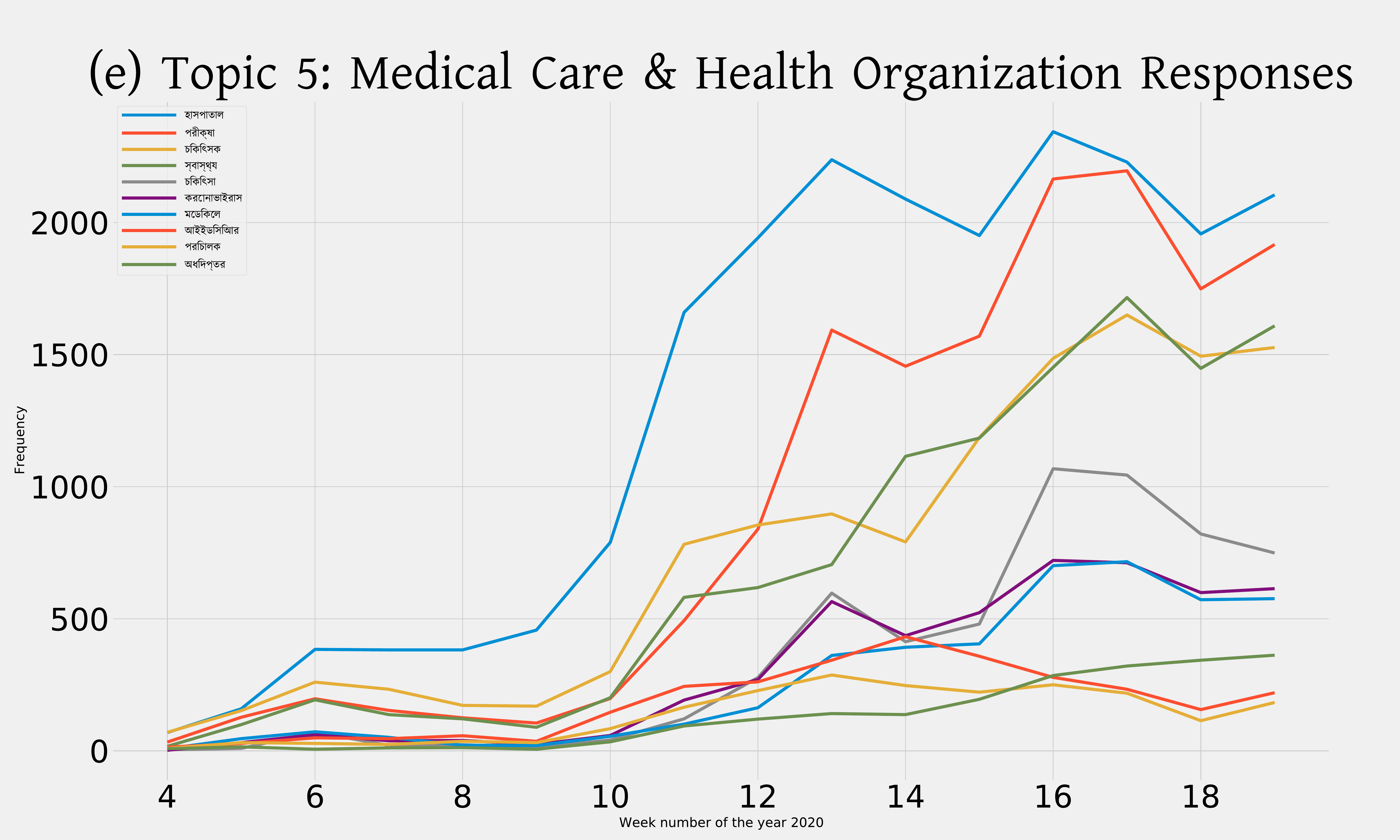}%
}\hfill
\subfloat[Topic 6]{%
  \includegraphics[width=0.385\textwidth]{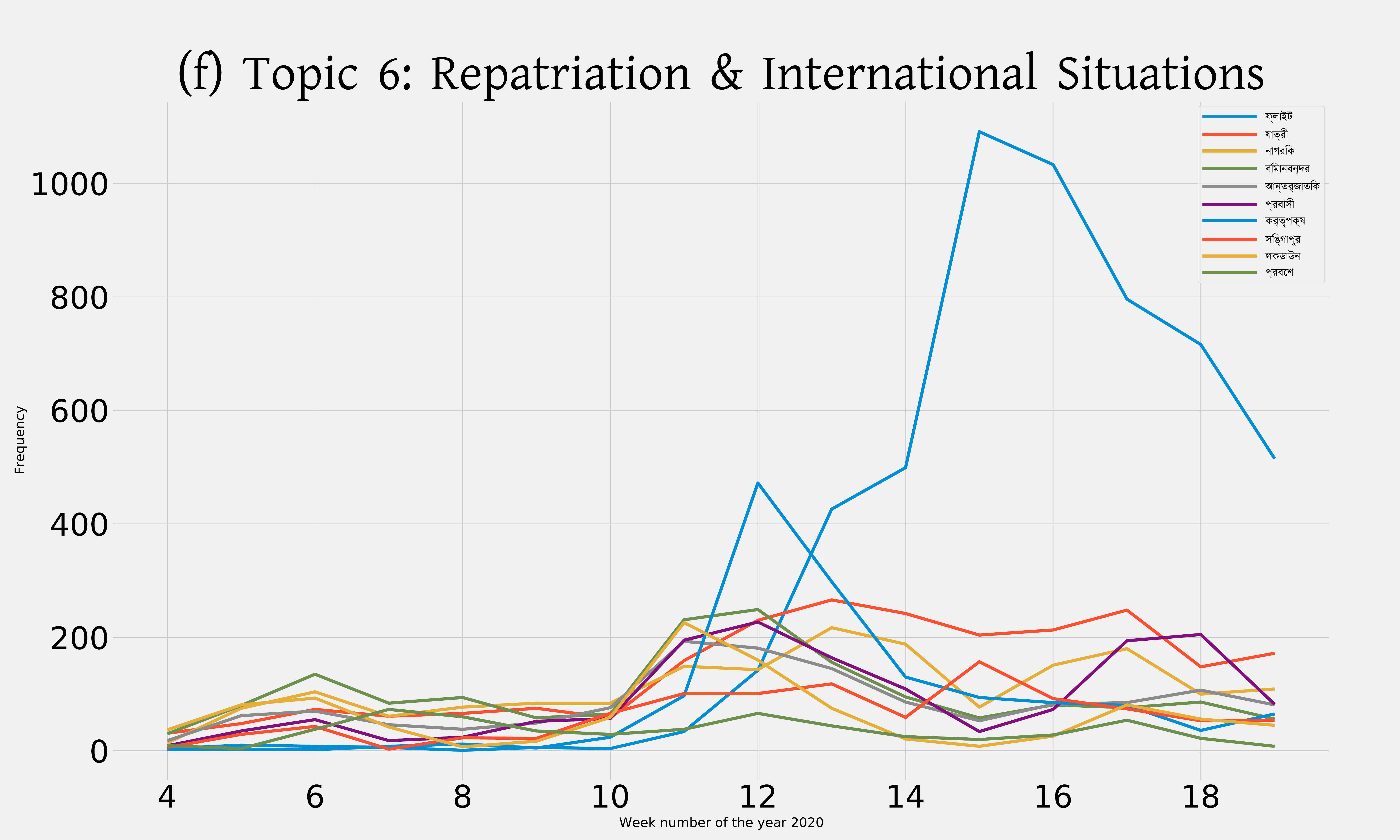}%
}\hfill
\subfloat[Topic 7]{%
  \includegraphics[width=0.385\textwidth]{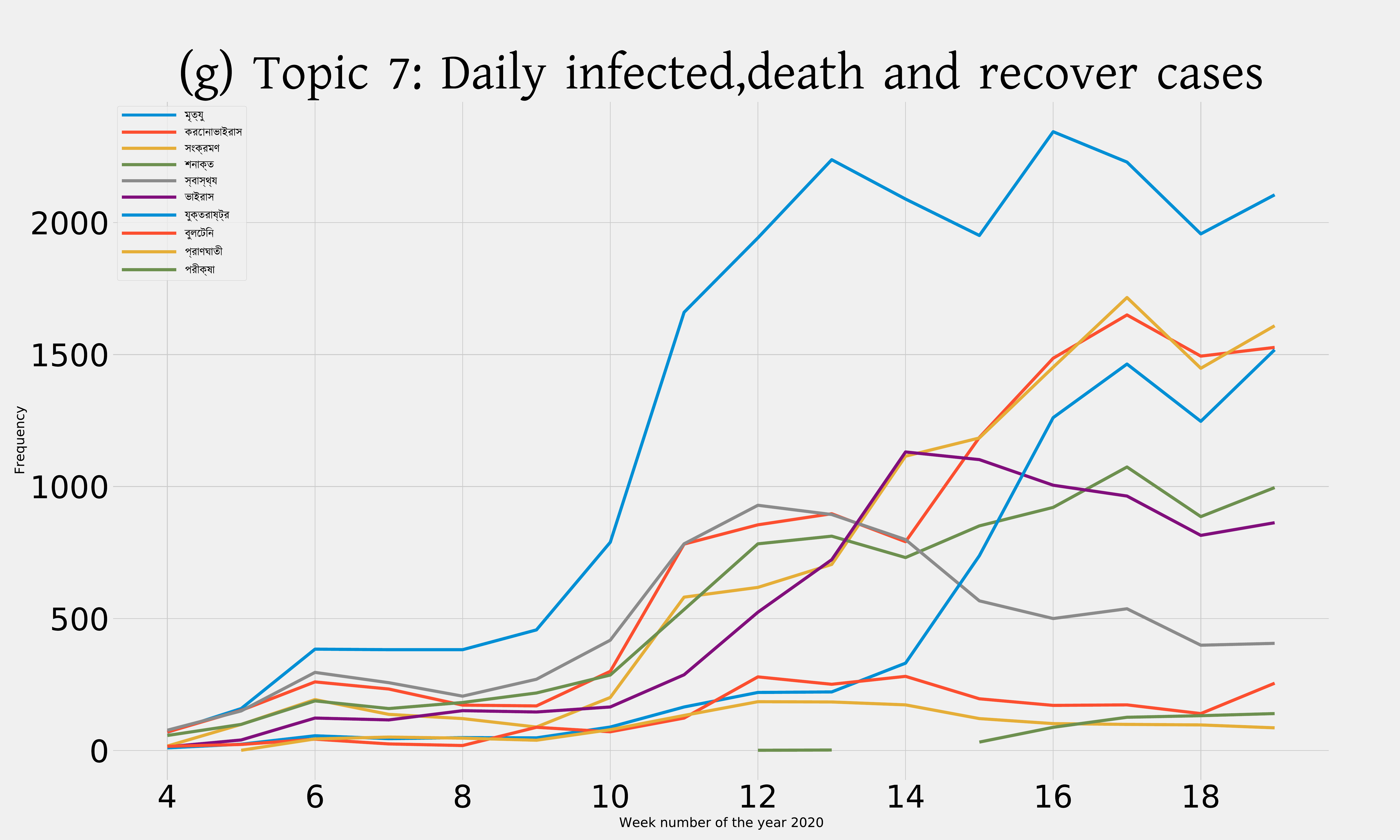}%
}\hfill
\subfloat[Topic 8]{%
  \includegraphics[width=0.385\textwidth]{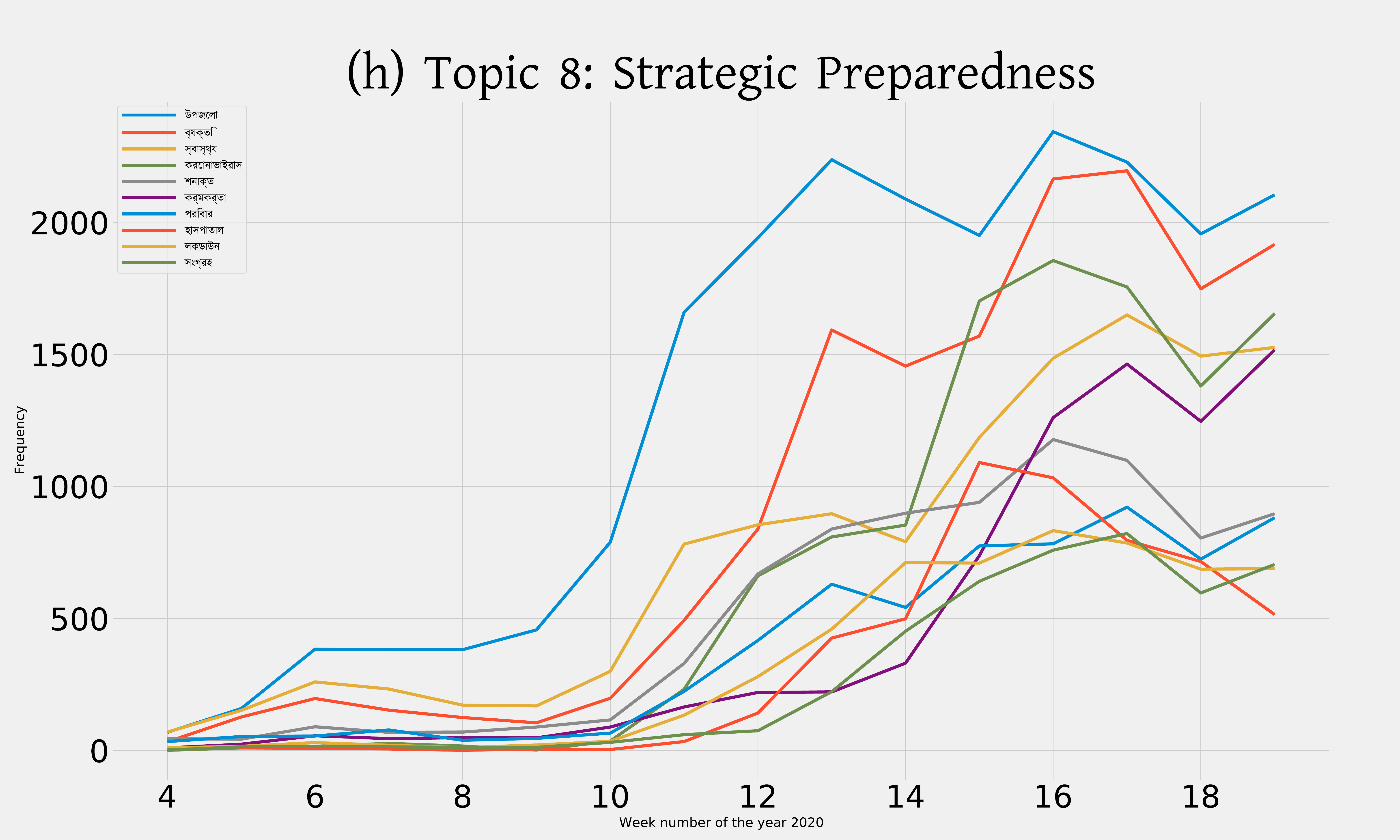}%
}\hfill
\subfloat[Topic 9]{%
  \includegraphics[width=0.385\textwidth]{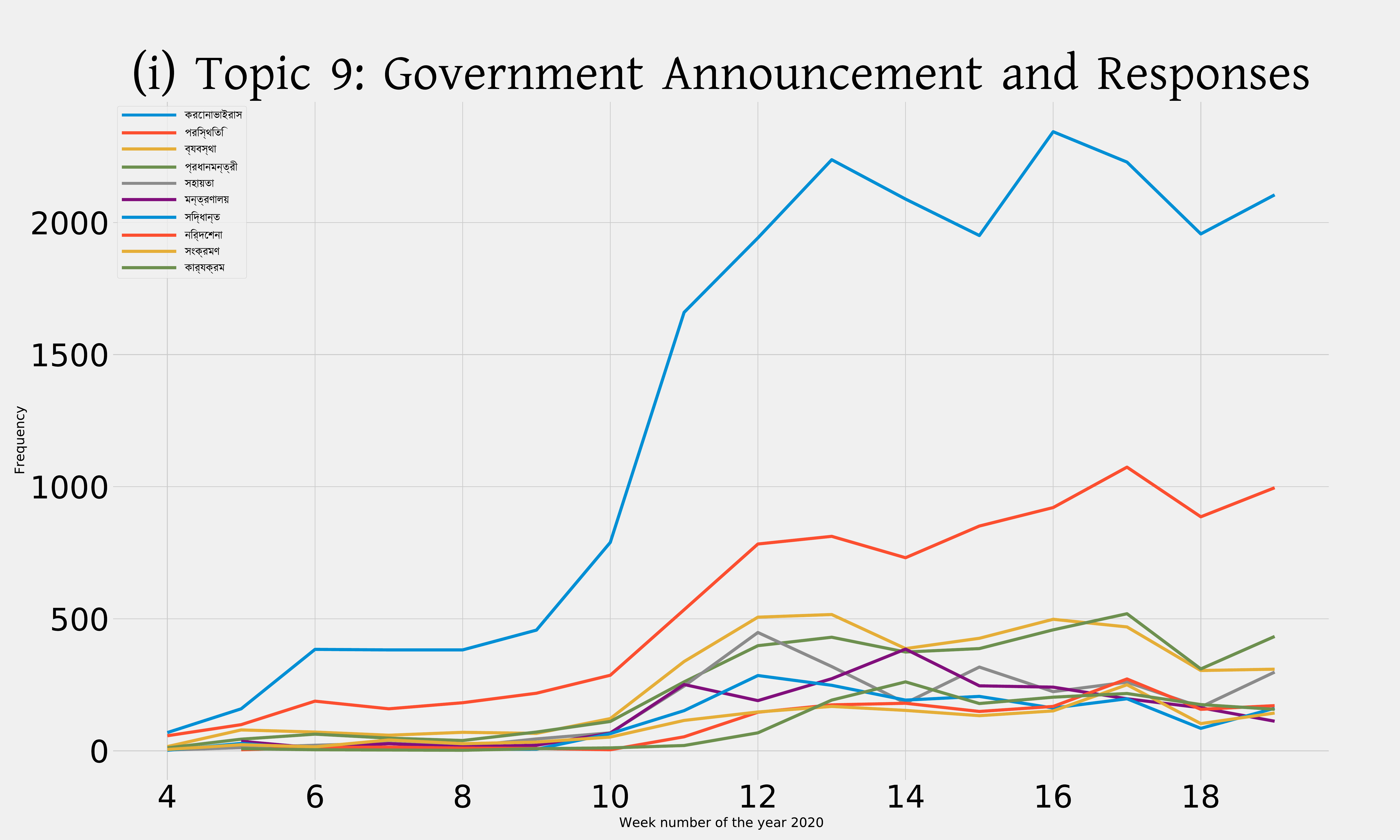}%
}

\caption{Topics Evolution of nine topics}
\label{dtm2}
\end{figure}

\subsubsection{Dynamic Topic Modeling: Temporal Trends of Topics} 
\begin{figure}
\subfloat[Temporal Trends of Topic 1\label{fig:subim1}]{%
  \fbox{\includegraphics[width=0.34\textwidth]{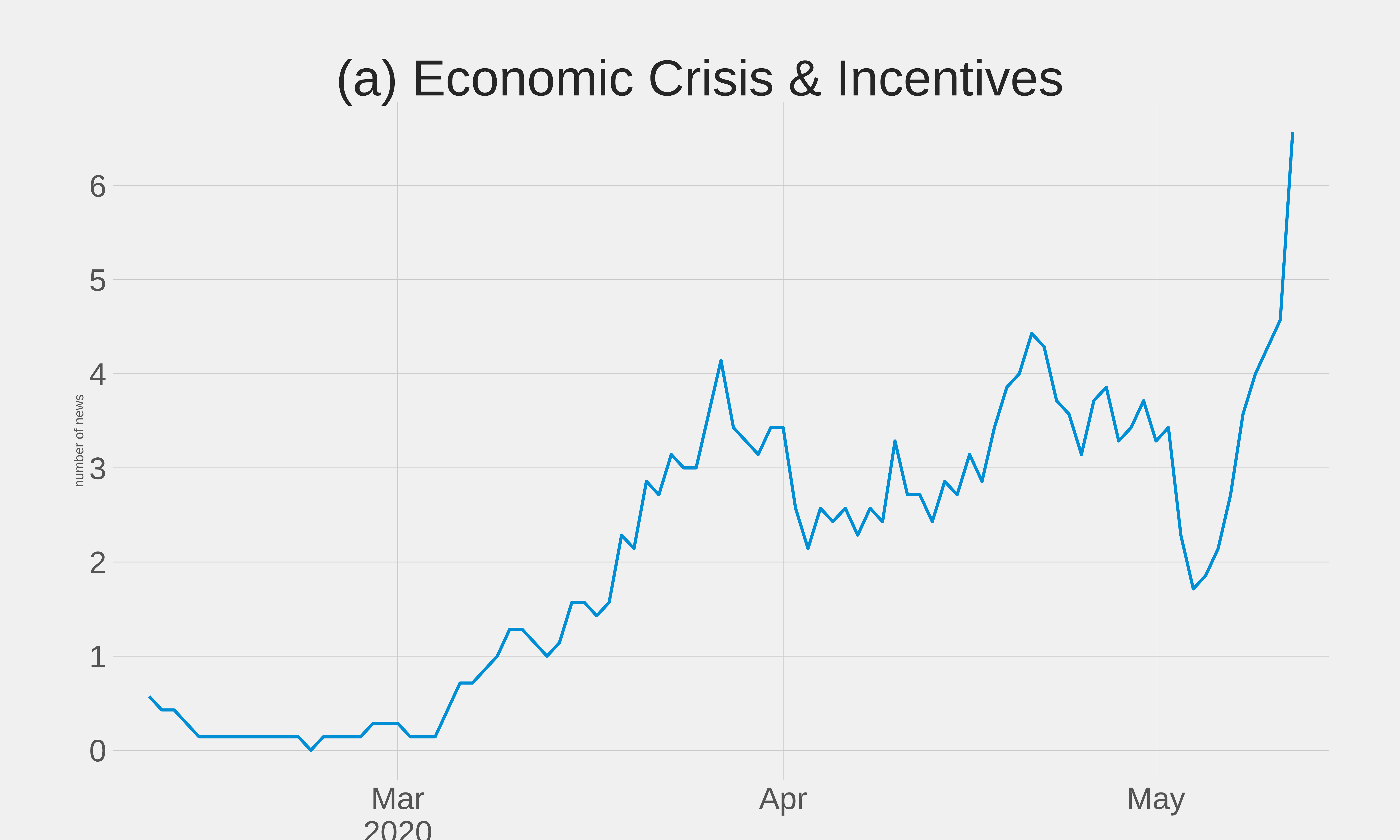}}%
}\hfill 
\subfloat[Temporal Trends of Topic 2\label{fig:subim2}]{%
  \fbox{\includegraphics[width=0.34\textwidth]{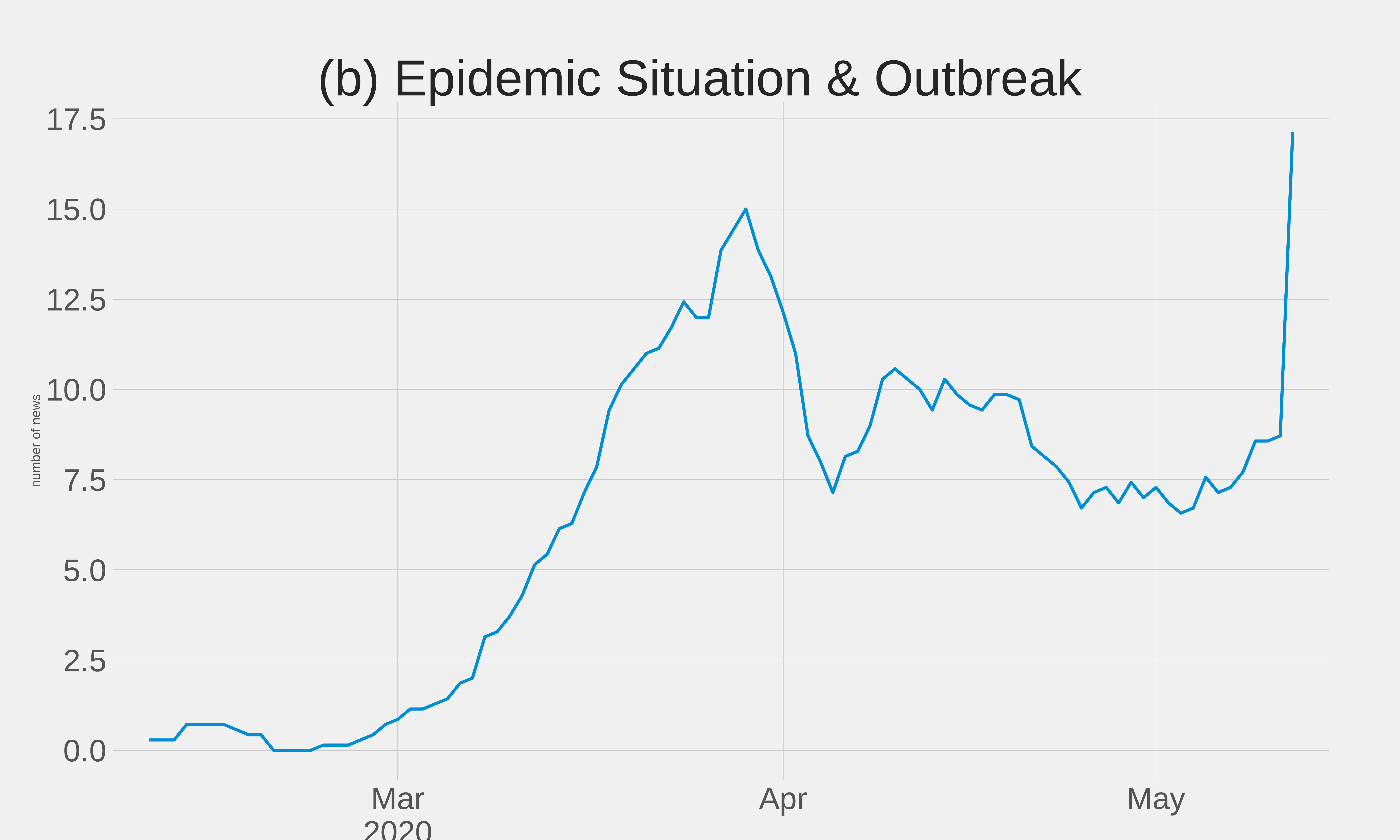}}%
}\hfill
\subfloat[Temporal Trends of Topic 3\label{fig:subim3}]{%
  \fbox{\includegraphics[width=0.34\textwidth]{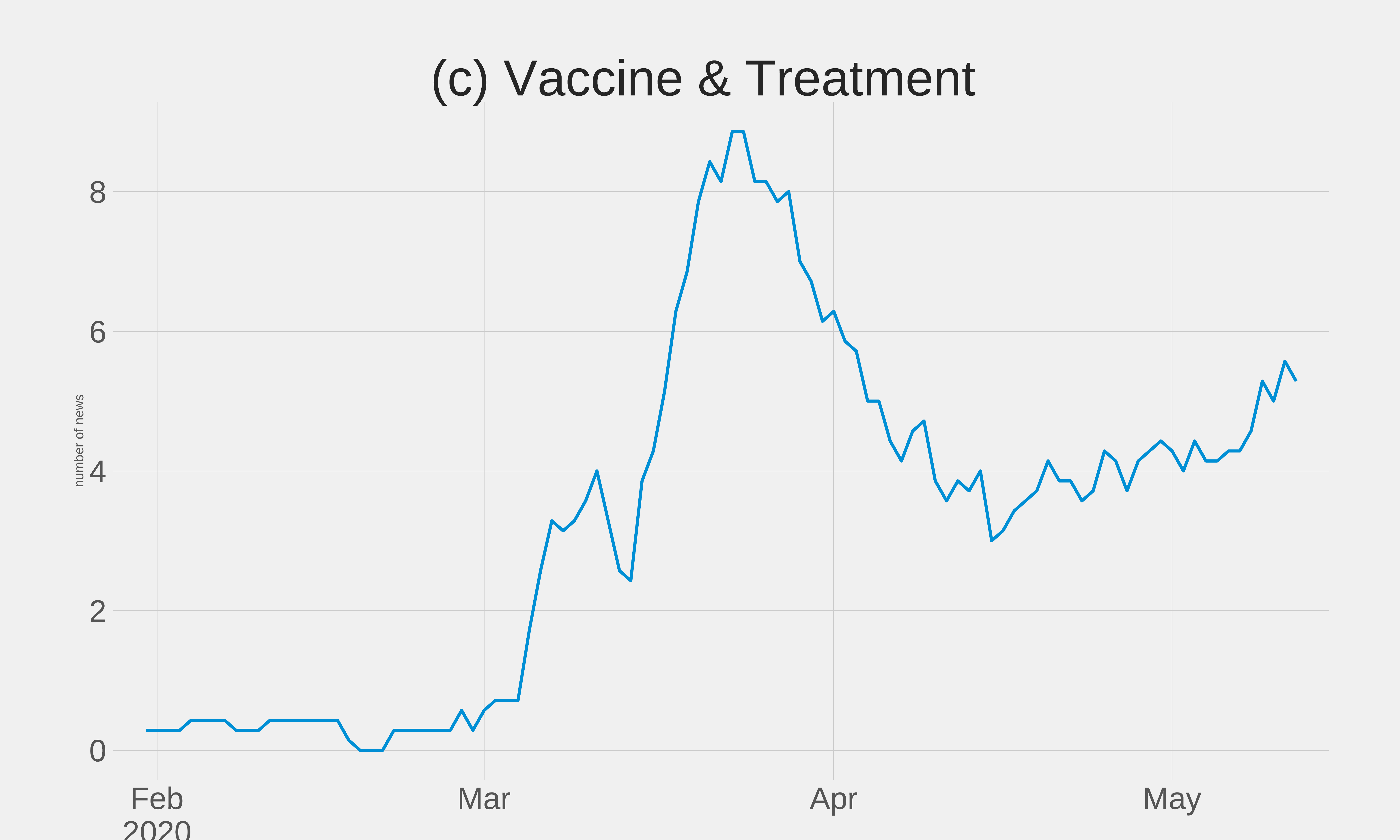}}%
}\hfill
\subfloat[Temporal Trends of Topic 4\label{fig:subim4}]{%
  \fbox{\includegraphics[width=0.34\textwidth]{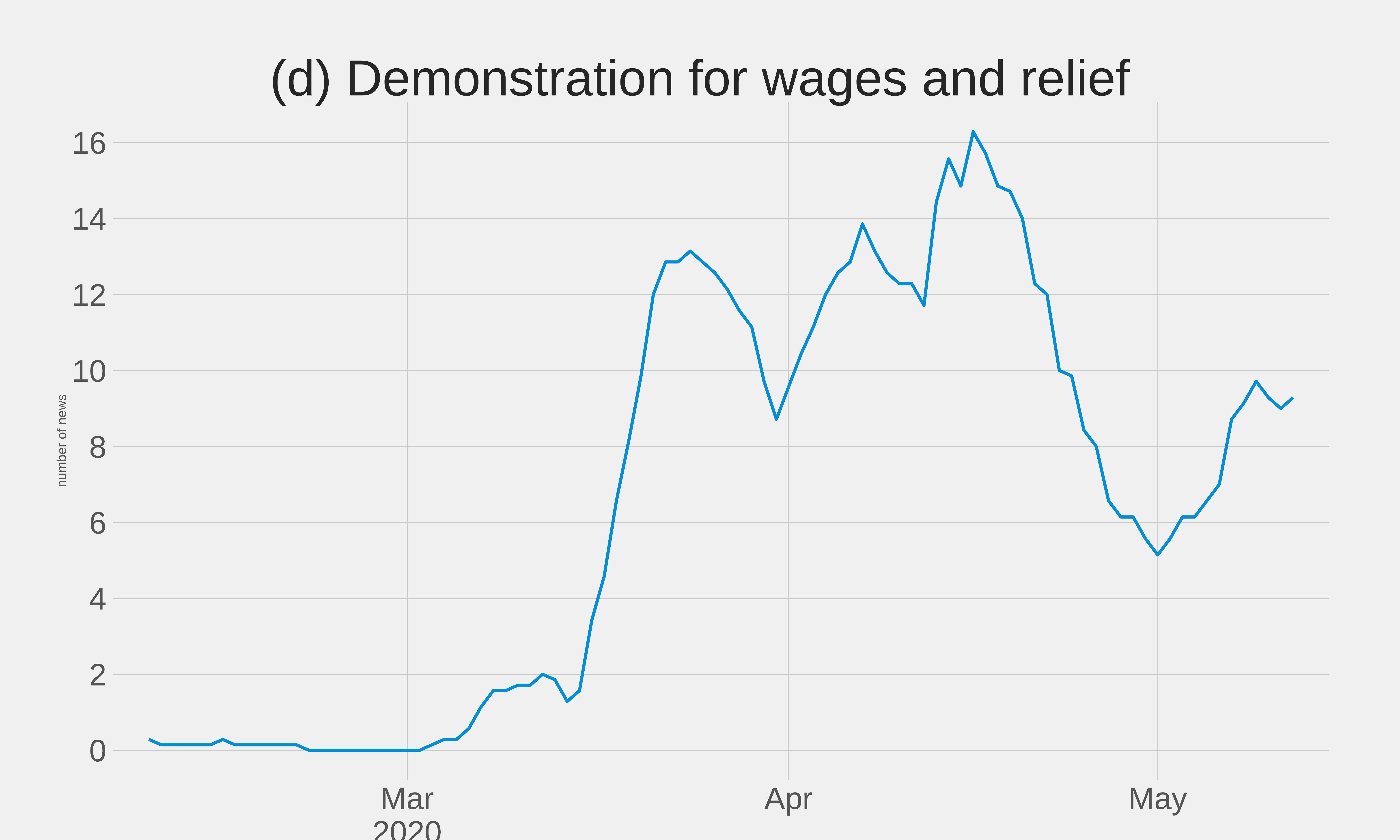}}%
}\hfill
\subfloat[Temporal Trends of Topic 5\label{fig:subim5}]{%
  \fbox{\includegraphics[width=0.34\textwidth]{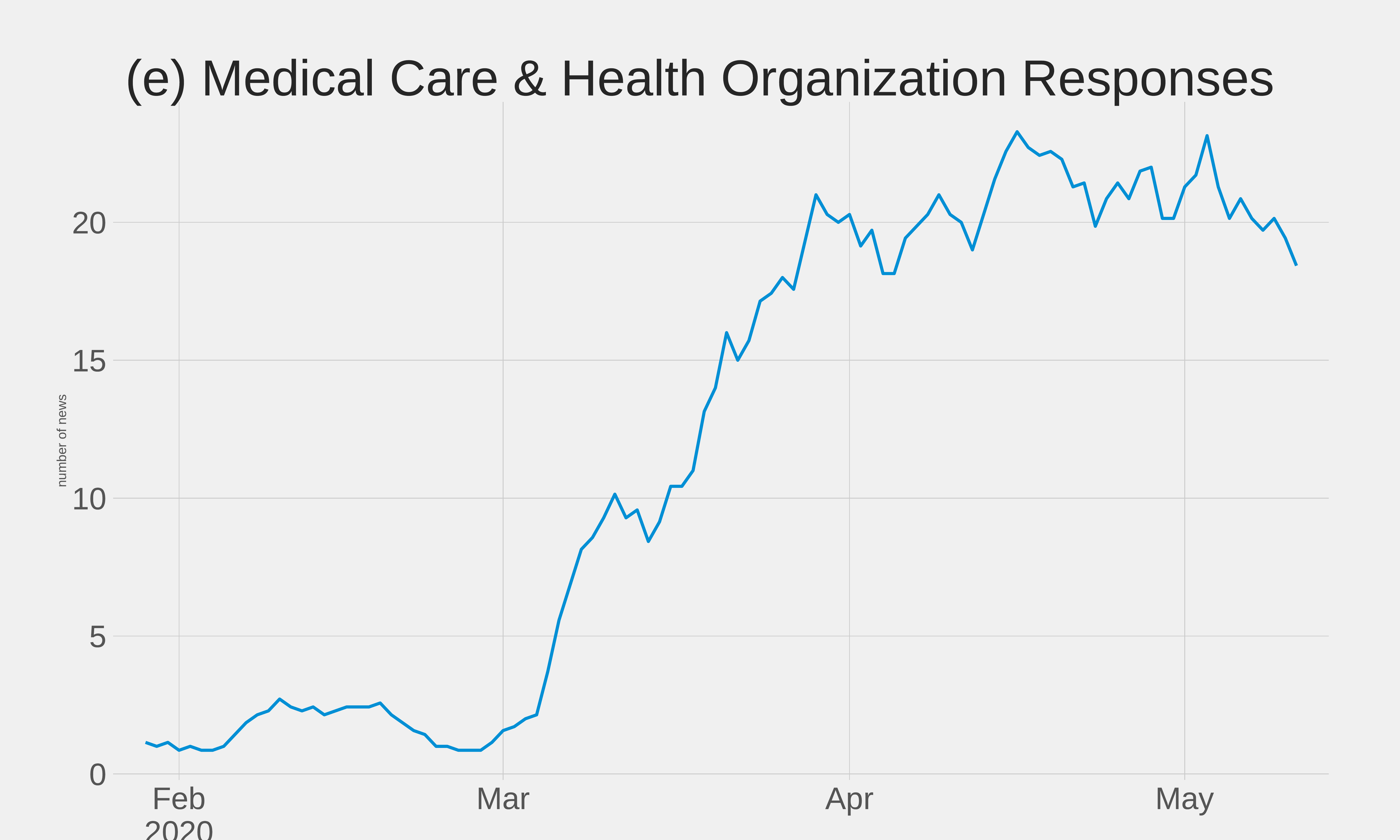}}%
}\hfill
\subfloat[Temporal Trends of Topic 6\label{fig:subim6}]{%
  \fbox{\includegraphics[width=0.34\textwidth]{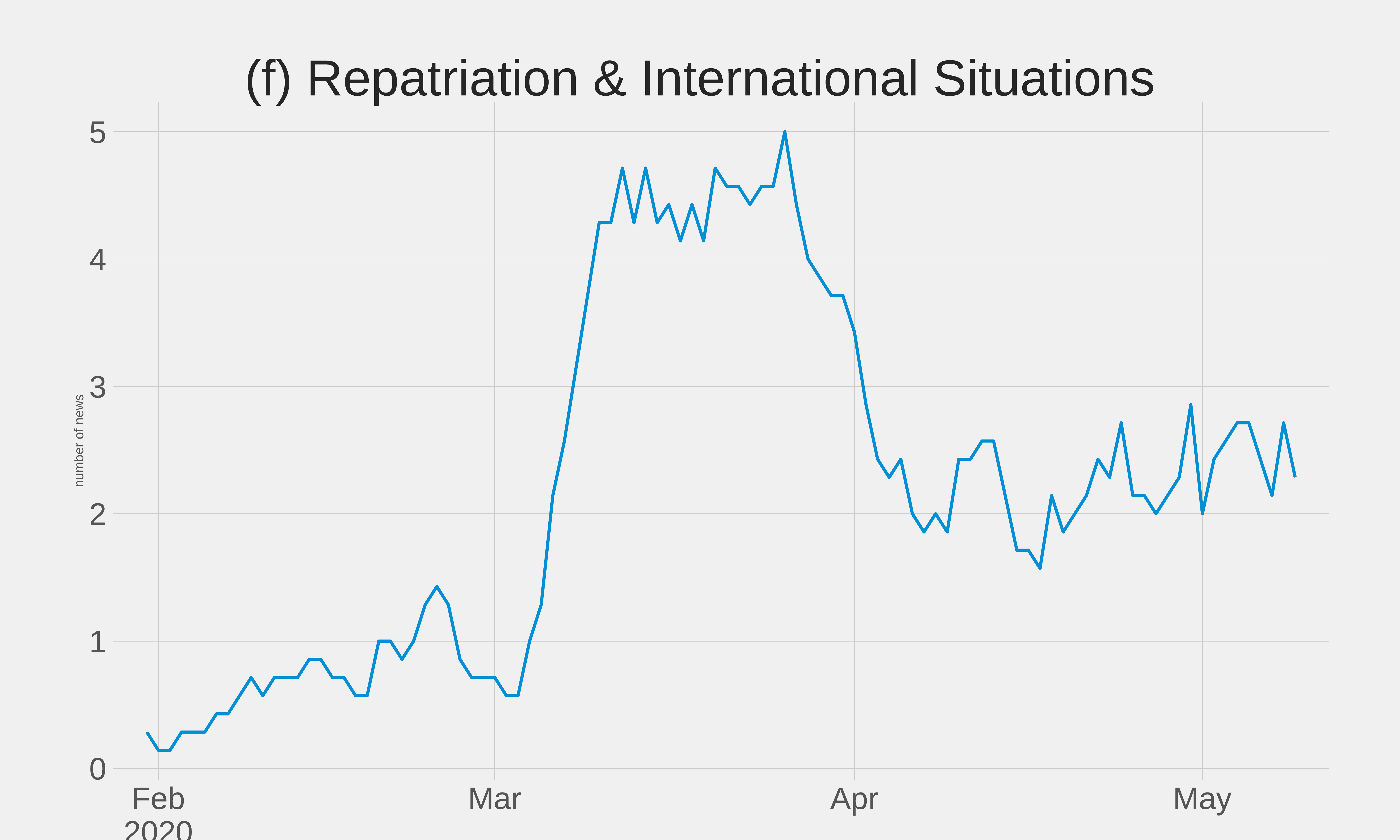}}%
}\hfill
\subfloat[Temporal Trends of Topic 7\label{fig:subim7}]{%
  \fbox{\includegraphics[width=0.34\textwidth]{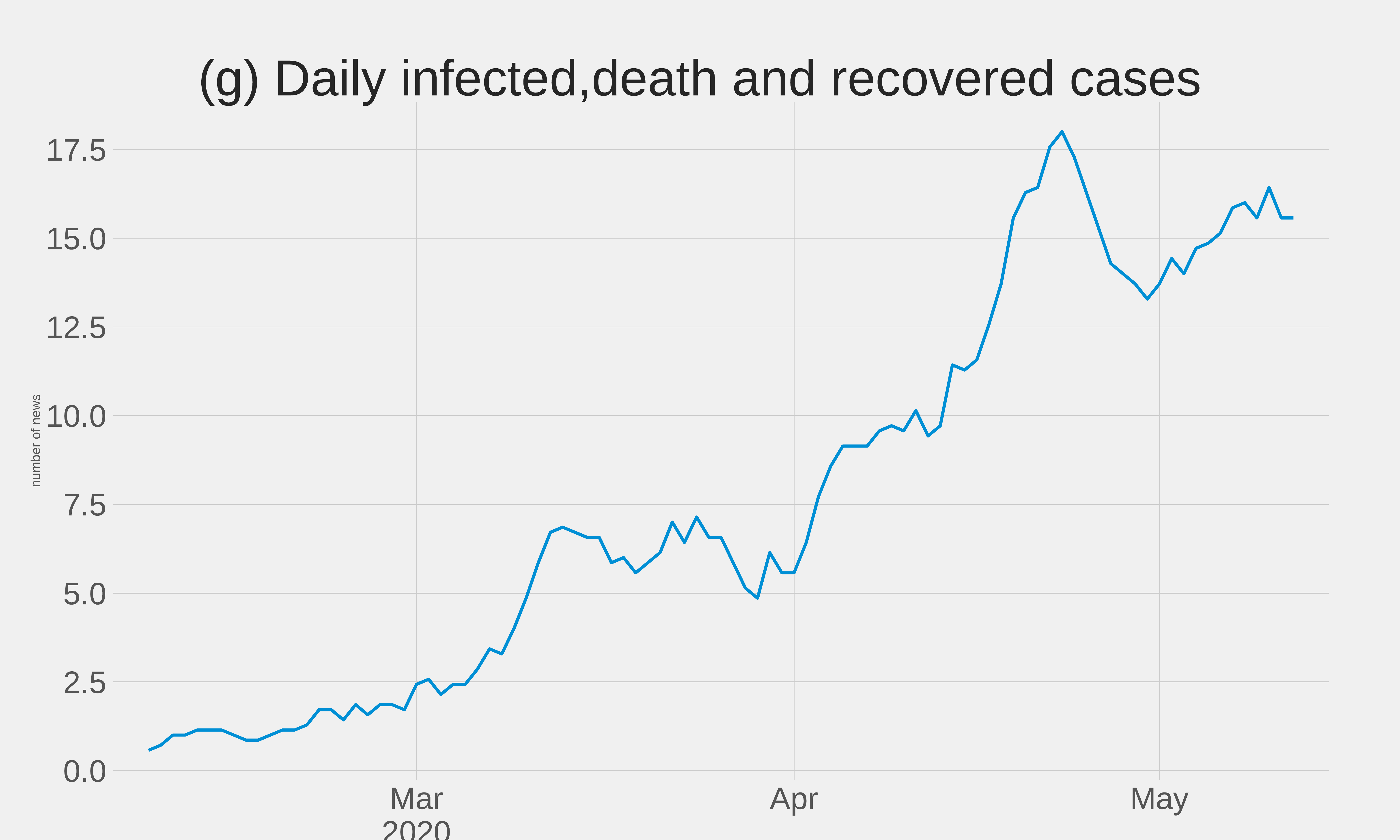}}%
}\hfill
\subfloat[Temporal Trends of Topic 8\label{fig:subim8}]{%
  \fbox{\includegraphics[width=0.34\textwidth]{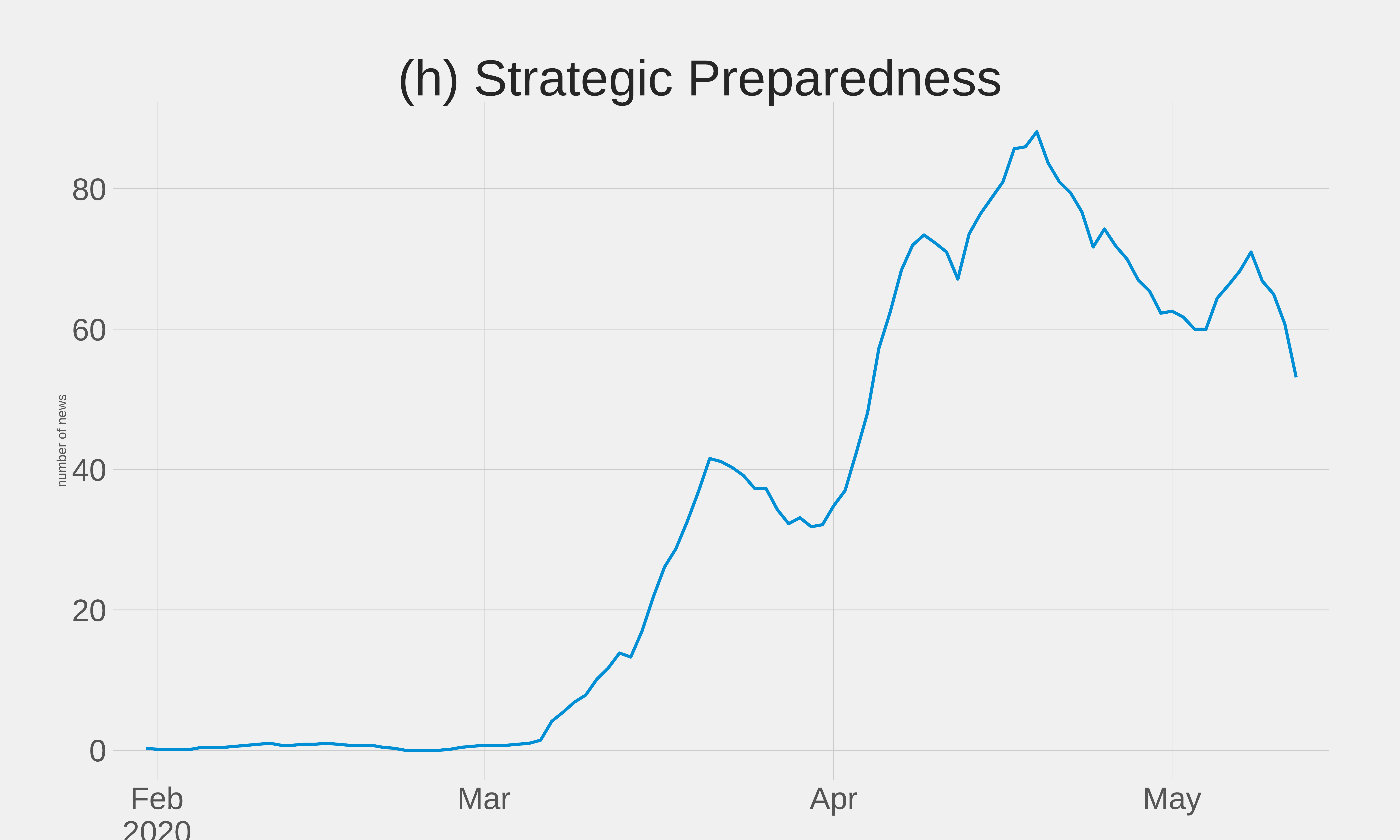}}%
}\hfill
\subfloat[Temporal Trends of Topic 9\label{fig:subim9}]{%
  \fbox{\includegraphics[width=0.34\textwidth]{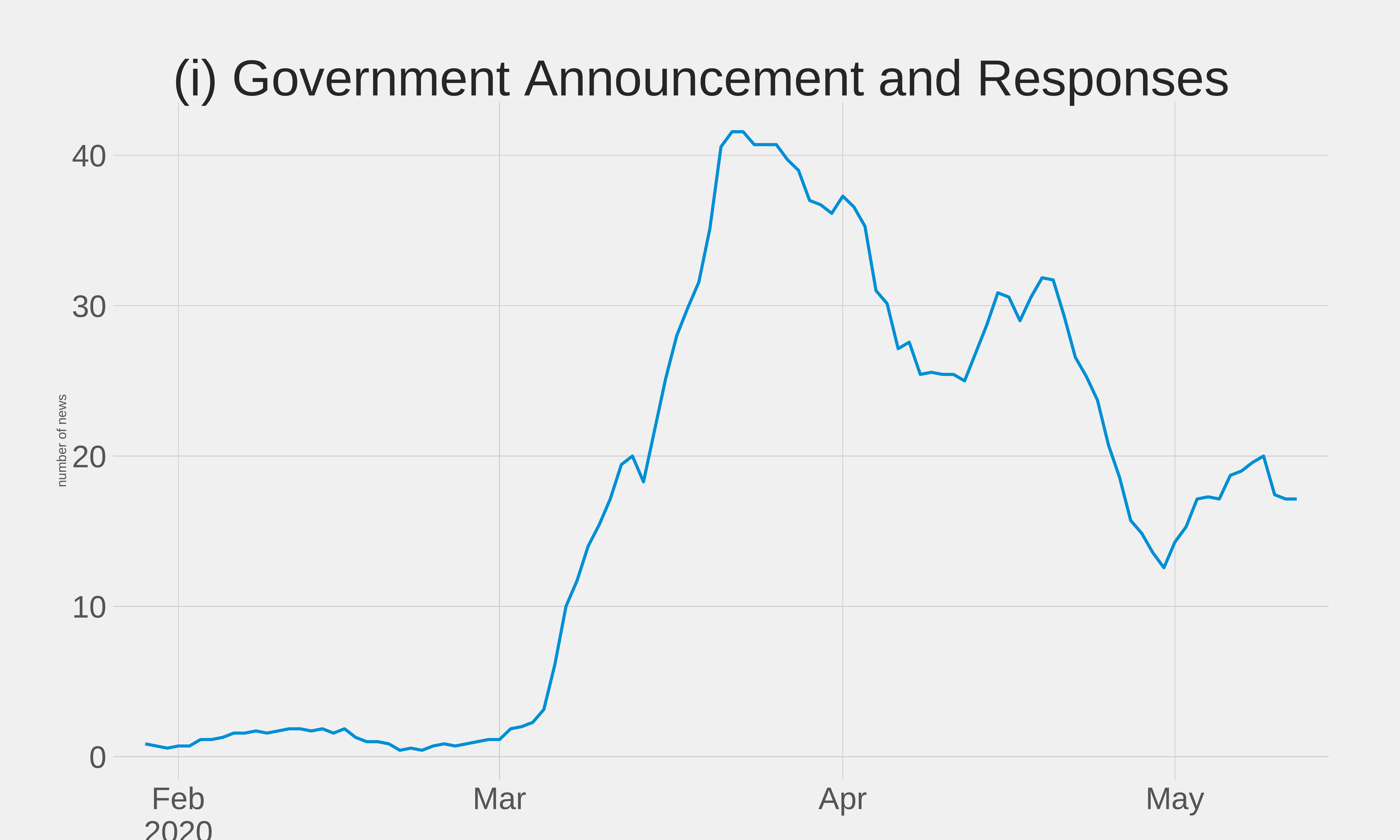}}%
}

\caption{Overall Temporal Trends of Topics}
\label{trend}
\end{figure}

To show the evolution of topics overtime during the pandemic, we used Dynamic Topic Modeling \cite{blei2006dynamic}.  Figure~\ref{dtm2} shows the evolution of nine topics over weeks during the pandemic. The figure shows how the popularity of each topic and the top words (i.e., keywords) in the topic changed over time. The overall temporal trend of these topics is shown in Figure \ref{trend}.

Topic 1: “Economic Crisis and Incentives” climbed from early March and stopped rising at the end of March. Then the curve continued to decline until the beginning of April and rose slowly in the middle of April. Then again, the curve continued to decline until the beginning of May. Then finally it reached a peak in the middle of May. 

Topic 2: “Epidemic Situation and Outbreak” climbed from the beginning of March and stopped rising at the end of April. The curve then steadily continued to decline until the beginning of April. Then slowly rose the curve to the middle of April.
Furthermore, again continued to decrease the curve until the beginning of May and then rose to a peak in the middle of May finally. 

Topic 3: “Vaccine and Treatment” climbed from the end of February and stopped growing at the beginning of March. Then the curve steadily continued to decline until the middle of March. And then again started to rise and reached a peak at the end of March. Then it also started to decline the curve until the middle of May. After that, the curve fluctuated till the end. 

Topic 4: “Demonstration for wages and relief” climbed from the beginning of March. Then the curve stopped rising in the middle of March. After that, the curve was fluctuating between the middle of March and the middle of April, and then the curve reached a peak in the middle of April. Then the curve steadily declined until the beginning of May and rose slowly in the middle of May. 

Topic 5: “Medical Care and Health Organization Responses” climbed from the beginning of February and stopped rising in early March. Then the curve started to rise from the beginning of March. In the middle of April, the curve reached its peak point. Then The slope of the curve gradually deteriorates after the middle of April. 

Topic 6: “Repatriation and International Situations” climbed from the beginning of February and stopped rising early in March. The curve then steadily declined and then again started to rise from early March to the middle of March. After that,  the curve had a steady state for a while; it went to the top at the end of March. Then again, the curve steadily declined until the beginning of April. Then the curve fluctuated till the end. 

Topic 7: “Daily infected, death and recovered cases” climbed from the beginning of February. As the number of infected and death cases was growing every day so that the number of daily infected deaths, death cases were also rising every day. So the curve was also rising. The curve reached its peak at the end of April. Then the curve declined until the beginning of May. After that, the curve again started to rise from the end of April to the end. 

Topic 8: “Strategic Preparedness” climbed from the beginning of March and stopped rising at the end of March. Then the curve declined until the end of March. After that, the curve started to rise from the end of March. Suddenly the curve downgraded for a while and then again started to rise up and then it peaked. The curve then steadily declined until the beginning of May and rose slowly in the middle of May. 

Topic 9: “Government Announcement and Responses” climbed from the beginning of March and reached a peak in the middle of March. Then the curve declined until the middle of April. After that, the curve again started to fluctuate. Then also, the curve steadily deteriorated until the end of April. Furthermore, finally, the curve rose slowly till the end.

\subsubsection{Spatial Distribution of Topics}
This subsection details the experimental results of the spatial distribution of the topics. 

\begin{figure}
\subfloat[Topic 1\label{spatialda}]{%
  \includegraphics[width=0.34\textwidth]{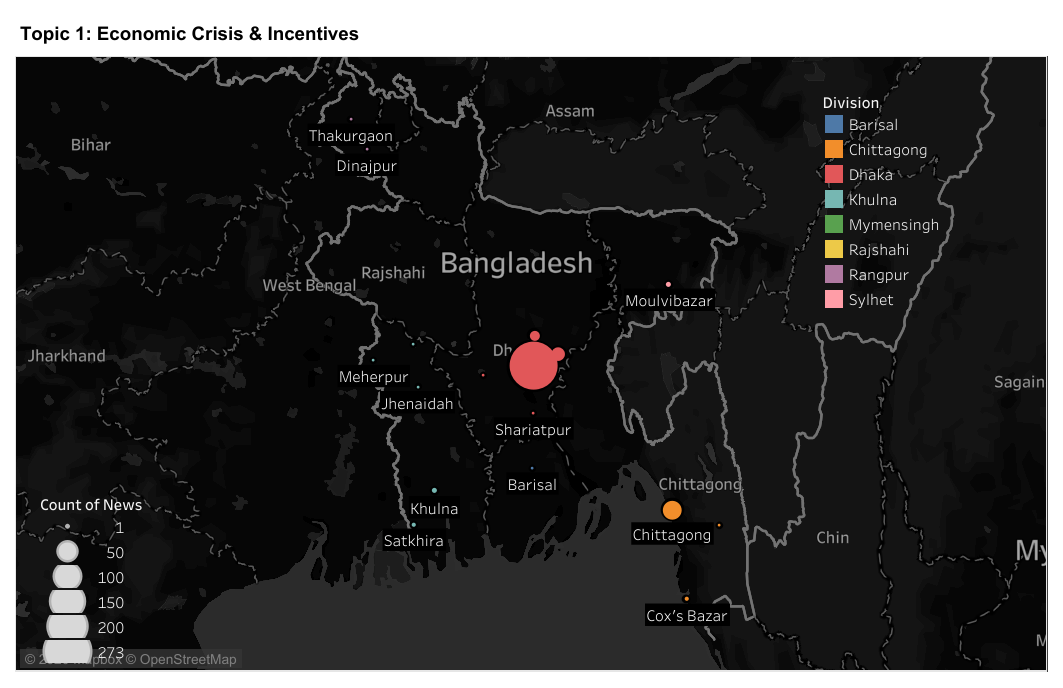}%
}\hfill 
\subfloat[Topic 2\label{spatialdb}]{%
  \includegraphics[width=0.34\textwidth]{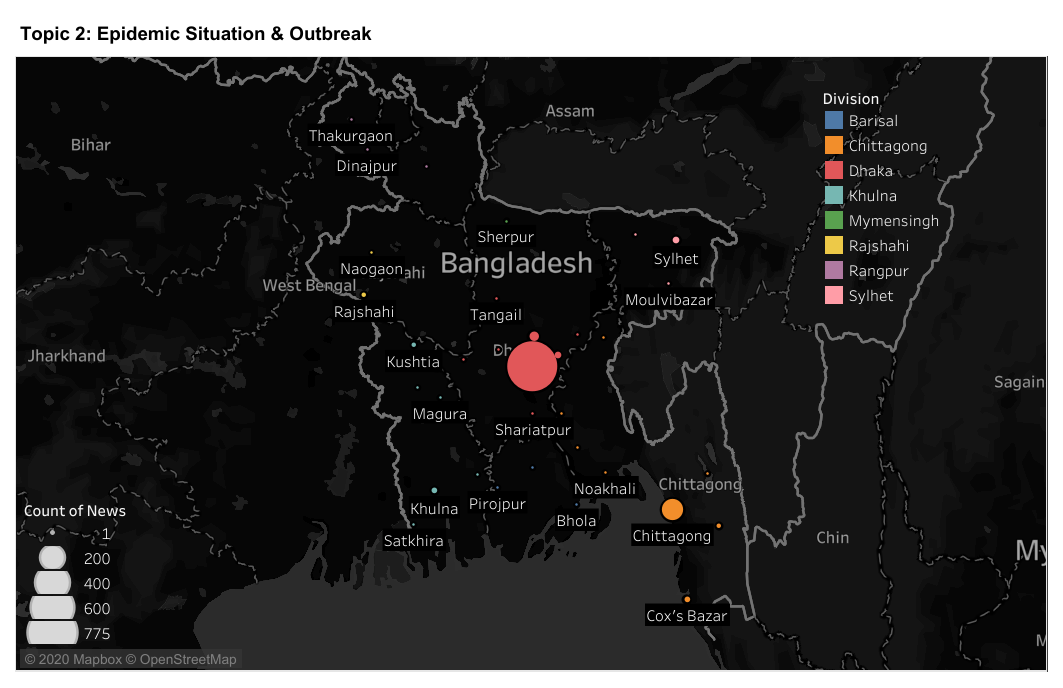}%
}\hfill
\subfloat[Topic 3\label{spatialdc}]{%
  \includegraphics[width=0.34\textwidth]{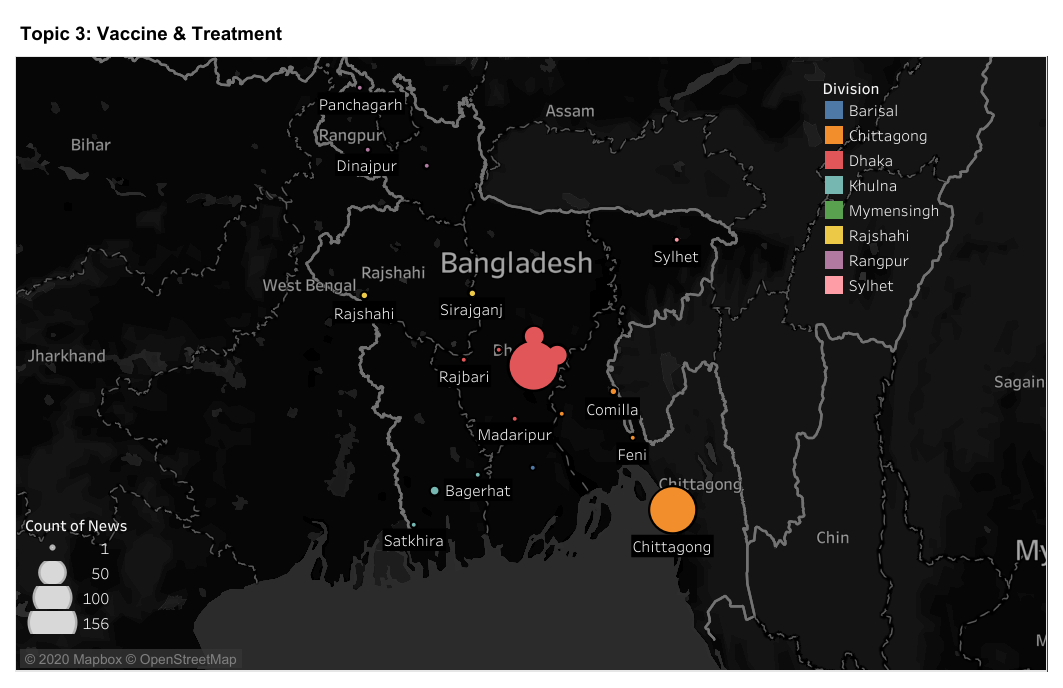}%
}\hfill
\subfloat[Topic 4\label{spatialdd}]{%
  \includegraphics[width=0.34\textwidth]{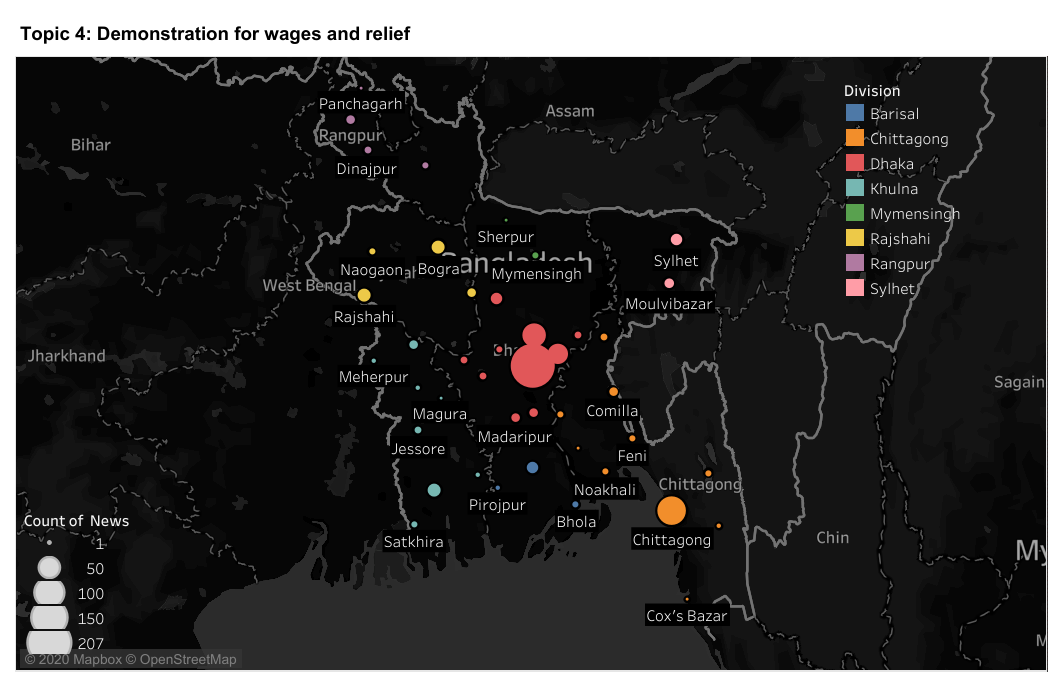}%
}\hfill
\subfloat[Topic 5\label{spatialde}]{%
  \includegraphics[width=0.34\textwidth]{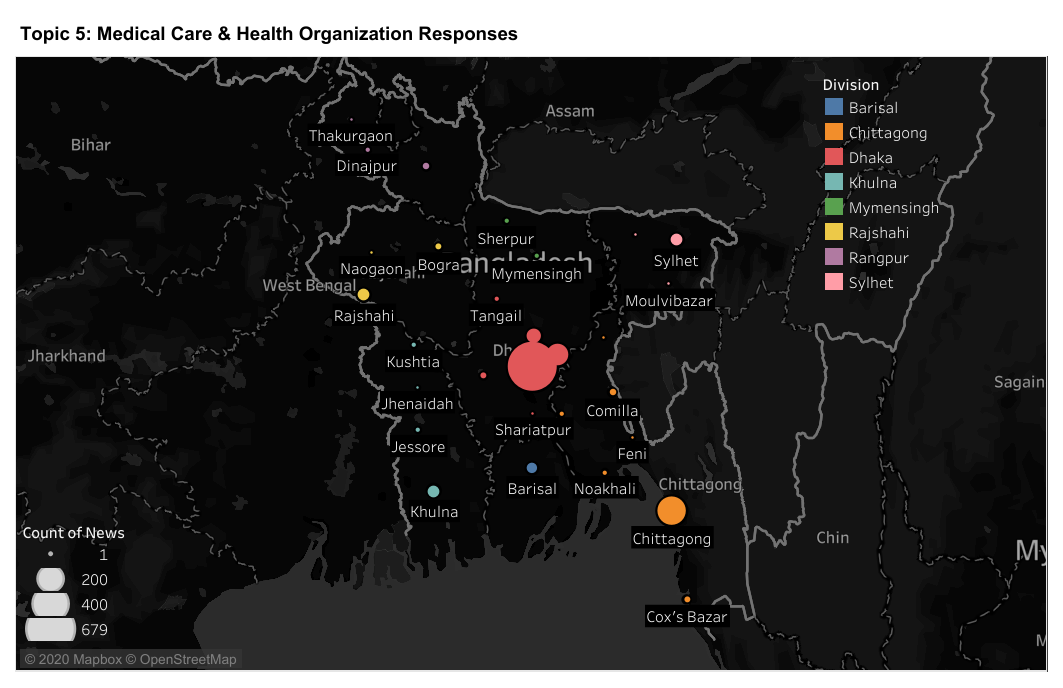}%
}\hfill
\subfloat[Topic 6\label{spatialdf}]{%
  \includegraphics[width=0.34\textwidth]{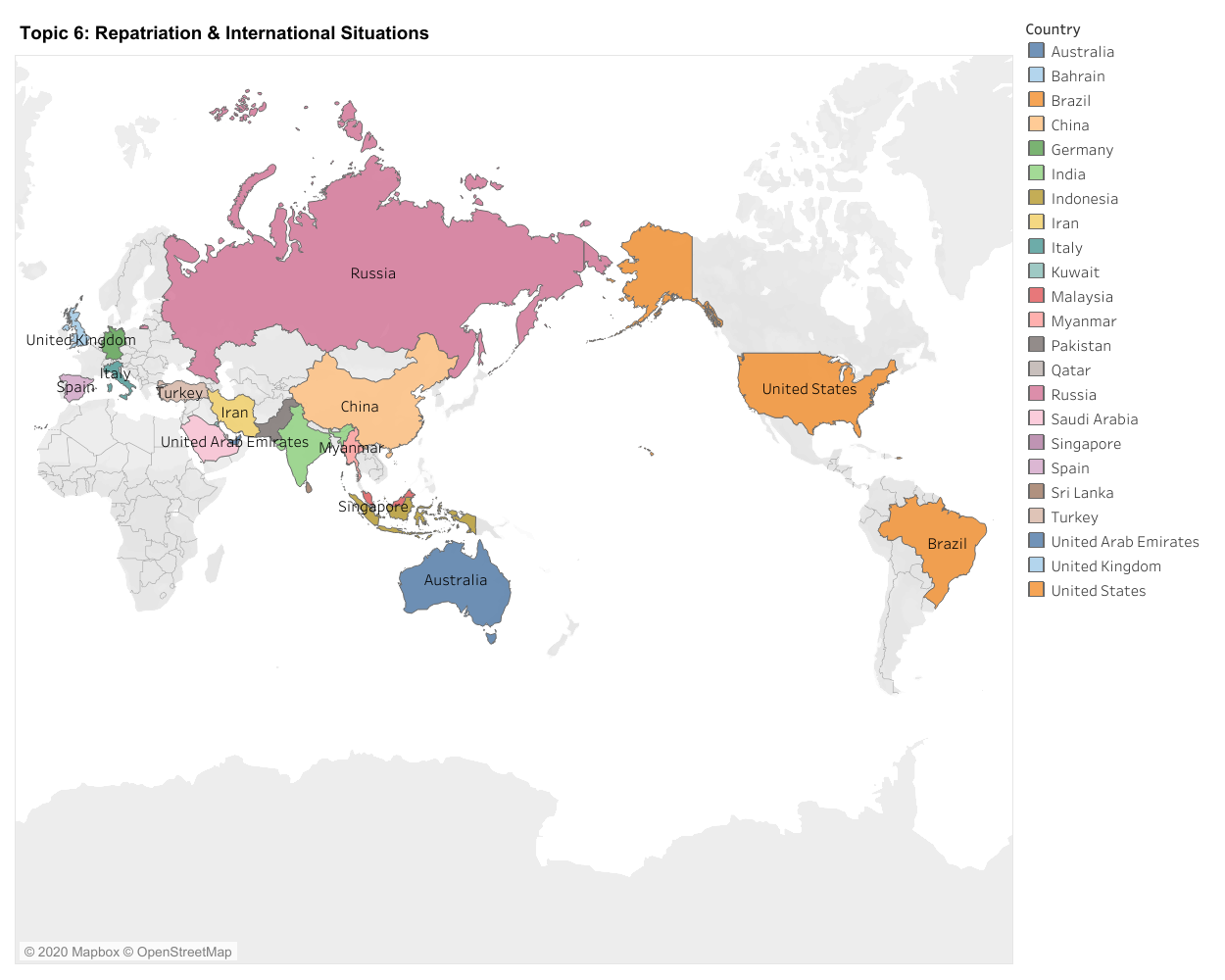}%
}\hfill
\subfloat[Topic 7\label{spatialdg}]{%
  \includegraphics[width=0.34\textwidth]{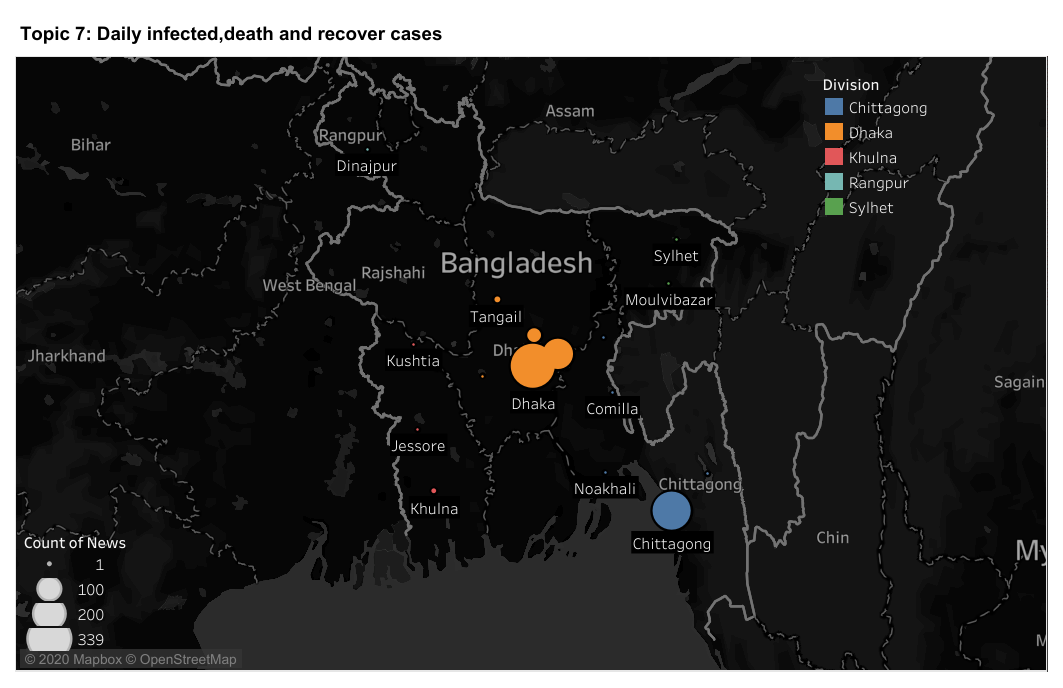}%
}\hfill
\subfloat[Topic 8\label{spatialdh}]{%
  \includegraphics[width=0.34\textwidth]{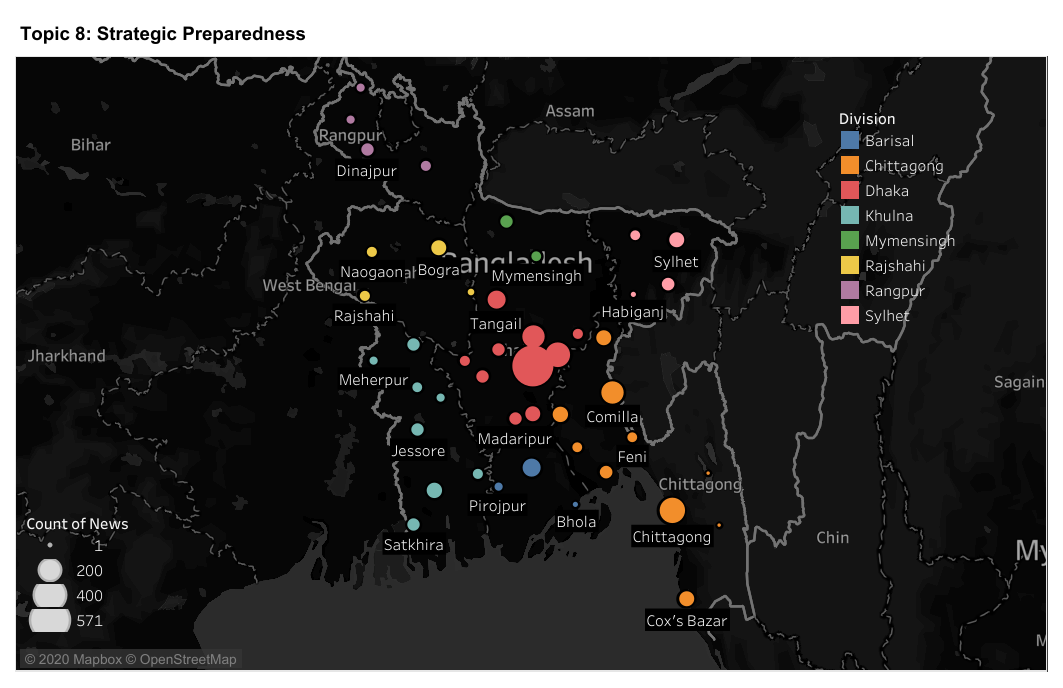}%
}\hfill
\subfloat[Topic 9\label{spatialdi}]{%
  \includegraphics[width=0.34\textwidth]{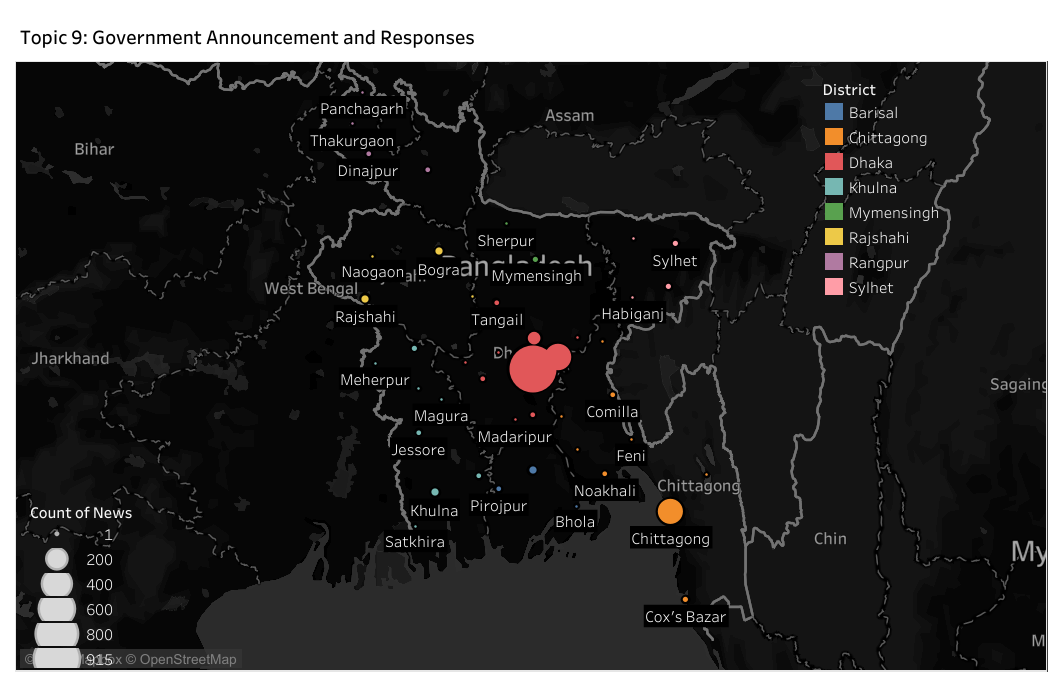}%
}

\caption{Spatial Distribution of Topics}
\label{spatiald}
\end{figure}

Topic 1: “Economic Crisis and Incentives” mainly concentrated on the Dhaka division shown in Figure~\ref{spatialda}. Most people of the Dhaka division lost their job in that period. This incident also happened in Chittagong to a small extent. The government and various agencies provided Relief and incentives to the victims. In this case, the area that has received the most relief is the Dhaka and the Chittagong division. 

Topic 2: “Epidemic Situation and Outbreak” mainly focused on the central and southern parts of Bangladesh, shown in Figure~\ref{spatialdb}. We can see that Dhaka is the most affected city in Bangladesh at that particular time. Most COVID-19 infected patients have been identified in Dhaka, more deaths have been reported in Dhaka, the situation in Dhaka was much worse than other districts and divisions at that time, and there was a much higher prevalence. Apart from Dhaka, Narayanganj, Gazipur, Chittagong has also been affected so much. 

Topic 3: “Vaccine and Treatment” mainly focused on Dhaka, Chittagong, Gazipur, Narayanganj area in Bangladesh, shown in Figure~\ref{spatialdc}. Since the prevalence of COVID-19 is higher in Dhaka, its adjoining districts like Narayanganj, Gazipur, and Chittagong so that treatments were shouted comparatively higher than in other districts. COVID-19 vaccine is also being studied in Dhaka. 

Topic 4: “Demonstration for wages and relief” is mainly concentrated all over the country shown in Figure~\ref{spatialdd}. The situation is terrible all over the country due to COVID-19. Those who are day laborers have lost their jobs; they have become destitute. For this, they had to come out of the house to survive. They had to move on the streets to provide food for their families. Since in every district of Dhaka, Chittagong, Rajshahi, Barisal, Khulna, Mymensingh, Sylhet, Rangpur, people had taken to the streets to protest for survival.  

Topic 5: “Medical Care and Health Organization Responses” also focused all over the country like Topic 4, shown in Figure~\ref{spatialde}. The state of the health system in the whole country is deplorable. Health organizations were in a very critical situation. 

Topic 6: “Repatriation and International Situations” mainly concentrated on China, USA, Italy, Russia, and other countries shown in Figure~\ref{spatialdf}. This topic talks about the situations of foreign countries and the immigrants who wanted to return to Bangladesh. Here those countries have been shown in Figure~\ref{spatialdf}.

Topic 7: “Daily infected, death and recovered cases” mainly focused on the central region, shown in Figure~\ref{spatialdg}. Dhaka division has the highest number of COVID-19 infected cases, as well as the death cases. Dhaka division includes Dhaka city, Narayanganj, Gazipur had the most cases. After Dhaka, most of the cases were found in the Chittagong division. 

Topic 8: “Strategic Preparedness” focused on all over the country shown in Figure~\ref{spatialdh}. Lockdown, isolation, home quarantine, social distancing was imposed across the country. 

Topic 9: “Government Announcement and Responses” shown in Figure~\ref{spatialdi}. It was effective almost everywhere, especially in Dhaka.

\subsection{Automatic Classification of News Articles}
We built a text classifier to automatically categorize upcoming new news articles into classes, sub-classes, and topics. We implemented an LSTM recurrent neural network model in Python utilizing Keras\footnote{https://keras.io/} deep learning library for this classification. 

We splited our data into 80\% for training, 10\% for validation, and 10\% for testing. In the data preparation, first, we cleaned the text by removing unnecessary characters and stopwords. After cleaning the text, we tokenized the data using Keras Tokenizer. After that, we built a word index from it. Then we vectorized Bengali text by turning each text into a vector. We limited the dataset to the top 50,000 words and set the max number of words in each article at 1000. After that, we added padding and truncated to our data to make the input sequences uniform and the same length for modeling.

After cleaning the data, we selected pre-trained word embeddings \footnote{Bengali Word Embeddings: \url{https://cutt.ly/KjXmEio}}. Word embedding maps each word from the vocabulary to a vector of real numbers. We used these pre-trained word embedding in the embedding layer of our LSTM model. 

In our classification model, the first layer is the embedded layer that employments 300 length vectors to represent each word. The second layer is an LSTM layer with 100 hidden units.
The final layer is a dense layer, also known as the output layer. This final layer has a length of 8, 19, and 9 for the classes, sub-classes, and LDA-discovered topics, respectively. \emph{Softmax} is used as the activation function for multi-class classification in the final layer. We used \emph{categorical cross-entropy} as the loss function, \emph{Adam} as the optimizer, and a batch size of 32. We used only five epochs as it worked quite well.

The experimental results for Precision, Accuracy, F$_1$ score and Recall are given in Table~\ref{my_table02}. The Precision, Accuracy, F1 score, and Recall are 47.80\%, 44.39\%, 45.13\%, and 42.80\%, respectively, for the eight classes. For the 19 sub-classes, Precision, Accuracy, F1 score, and Recall are 47.33\%, 38.51\%, 37.20\%, and 30.82\%, respectively. Furthermore, we also computed the performance of 9 LDA topics. For the 9 LDA topics, Precision, Accuracy, F1 score, and Recall are found 81.37\%, 79.55\%, 79.67\%, and 78.10\%, respectively.


\begin{table}
    \begin{center}
        \begin{tabular}{lccc}
    \hline
      Criteria & Classes & Sub-classes & Topics\\
    \hline
    \hline
    Precision &	47.80\% &	47.33\% &	81.37\%\\
    \hline
    Accuracy &	44.39\% &	38.51\% &	79.55\%\\
    \hline
    F1 Score &	45.13\%	& 37.20\% &	79.67\%\\
    \hline
    Recall &	42.80\%	 & 30.82\% &	78.10\%\\
    \hline
\end{tabular}
    \end{center}
    \caption{Performance in Classes, Sub-classes, and Topics}
    \label{my_table02}
\end{table}

\subsection{Sentiment Analysis}
\begin{figure}
    \centering
    \includegraphics[width=0.35\textwidth]{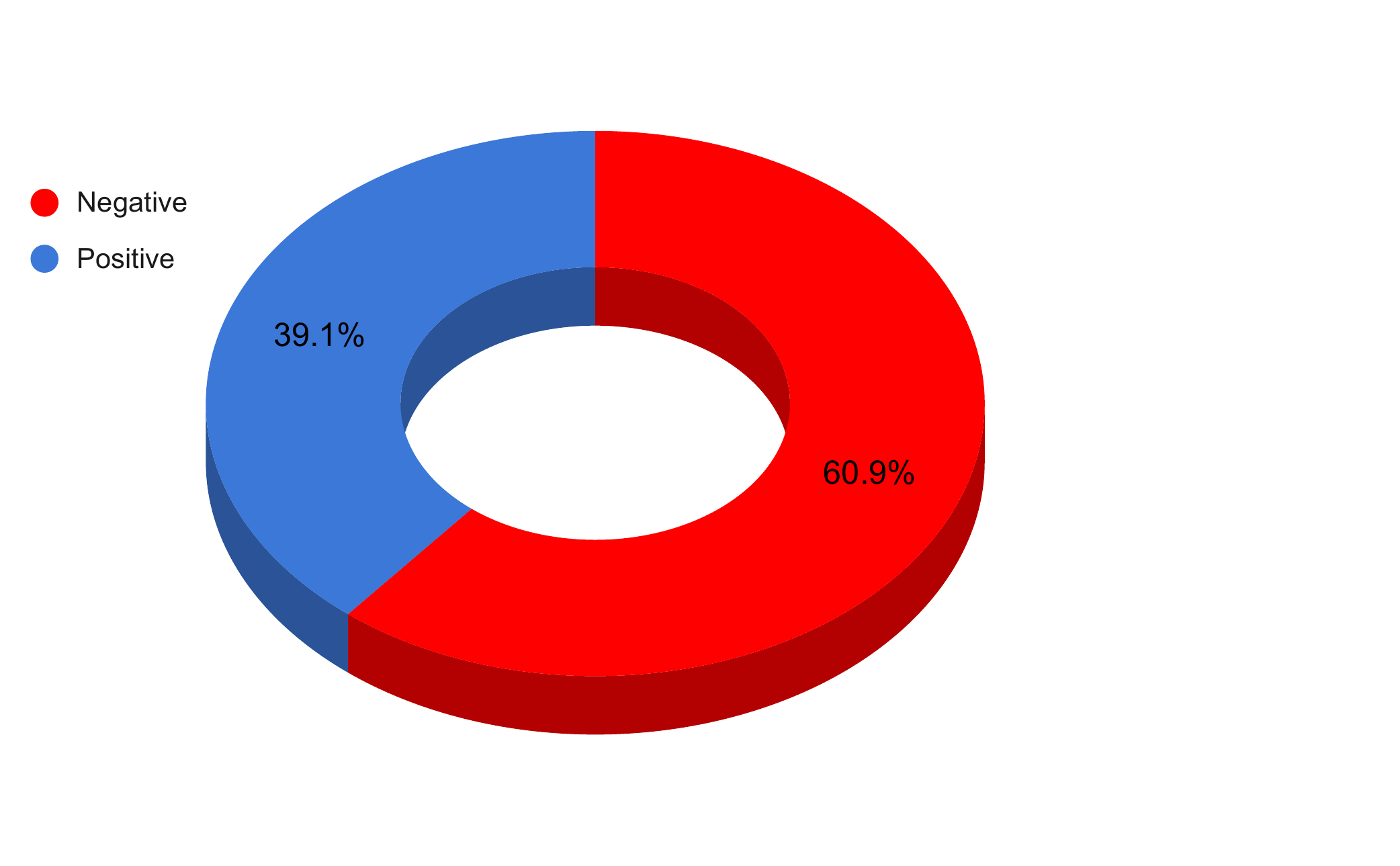}
    \caption{Proportion of Positive and Negative Sentiments}
    \label{sentiment1}
\end{figure}

We analyzed sentiment in the COVID-19 related news articles to see how positively and negatively the society was affected by the COVID-19 (or any big incident). We also analyzed the effectiveness of a hybrid CNN-BiLSTM model in identifying sentiments in Bengali texts. First, we manually labeled each news article according to positive or negative sentiment in the article. Then we trained CNN-BiLSTM to detect the sentiment of any upcoming new articles. After labeling the articles' positive/negative sensation, we visualized the results in Figure~\ref{sentiment1}. The figure shows that there were more negative sentiment news articles than positive sentiments.
\begin{figure}
    \centering
    \fbox{\includegraphics[width=0.95\textwidth]{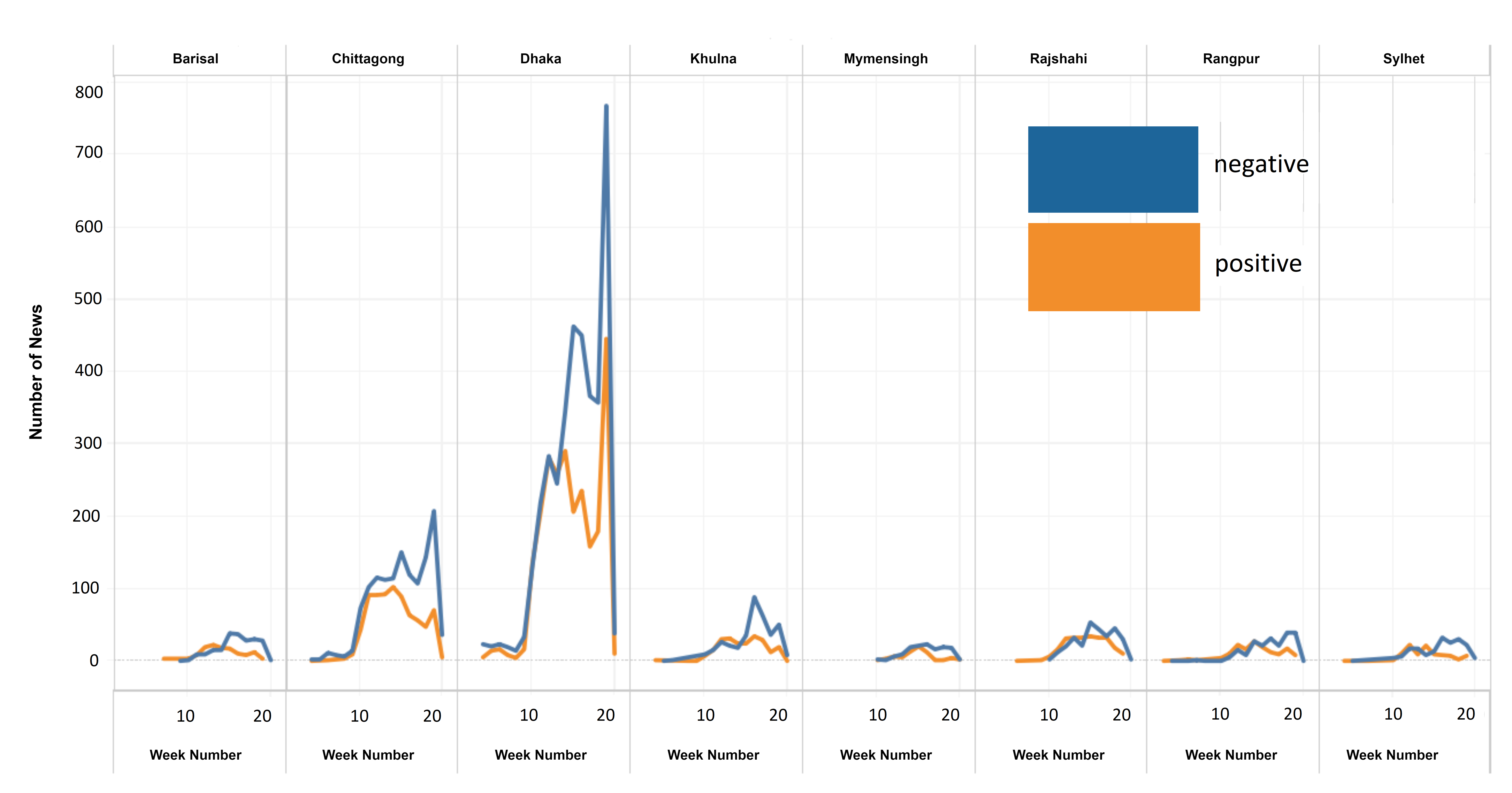}}
    \caption{Spatial and Temporal Distribution of Number of Positive and Negative Sentiment News Articles}
    \label{sentiment2}
\end{figure}

We prepared our labeled news collection as 80\% of for training, 10\% for validation, and 10\% for testing for sentiment analysis. Using Keras Tokenizer, we tokenized the data after cleaning the dataset. Then we built a word index and vectorize each text. We retrained the dataset to the 60,000 top words and set the max number of words in each article at 200 using feature selection. We added padding and truncated the data to make the input sequences uniform and the same for modeling.

After data preparation, we built our model. The first layer of the model is the embedding layer. We set the embedding dimension to 300 for embedding each word. In the second layer, we started a conv1D with 200 filters for CNN. Then in the third and fourth layers, we applied two Bi-LSTM with a dropout of 0.5. In the final layer, we used a dense network. We used \emph{Adam} as the optimizer with finely tuned hyperparameters and applied \emph{L2} regularizations to reduce overfitting. We kept the batch size of 256 as it worked quite well. We used only five epochs that gave us reasonably good results.  

After calculating the Precision, Accuracy, F1 score, and Recall, the performance and the sentiments of the classification presented in Table~\ref{my_table03}. The Precision, Accuray, F1 score, and Recall are 74.89\%, 74.94\%, 74.88\%, and 74.89\%, respectively, for sentiment analysis.

\begin{table}
    \begin{center}
        \begin{tabular}{lc}
    \hline
       Criteria         &                   Performance\\
    \hline
    \hline
    Precision	    &                         74.89\%\\
    \hline
    Accuracy	      &                        74.94\%\\
    \hline
    F1 Score	       &                       74.88\%\\
    \hline
    Recall	          &                    74.89\%\\
    \hline
\end{tabular}
    \end{center}
    \caption{Performance of Sentiment Analysis}
    \label{my_table03}
\end{table}

Figure~\ref{sentiment2} shows the spatial and temporal distribution of the number of positive and negative sentiment news articles during the pandemic. It shows how sentiments are changing over eight divisions for 20 weeks.

\section{Conclusions}
\label{sec:conclusion}
This study took an in-depth analysis of Bangladeshi daily newspaper reports from the onset of the COVID-19 pandemic. After collecting the news articles, we investigated and manually classified them into eight classes and nineteen sub-classes. We used LDA for extracting nine COVID-19 related topics from the news articles. We used the dynamic topic model to see the evaluation of topics over time. We also provided the spatial distribution of the topics. We created a text classifier that will automatically sort upcoming articles into classes, sub-classes, and topics. We also did a spatial and temporal analysis of news article volume. In the temporal analysis of volume, we decomposed the time series into four components: observed, trend, seasonal, and residual. Besides, we analyzed daily news article counts and daily infected and death cases in the temporal and spatial dimensions. Finally, we analyzed the sentiments in the news articles related to COVID-19 to understand the positive and negative impacts of events and initiatives during the COVID-19 pandemic using a CNN-BiLSTM architecture.

In a period of big social incidence, continuous analysis of newspaper articles is essential to ensure public well-being, maintain social consensus, and save lives. The automatic analysis techniques and the analysis outcomes presented in this study will help government and crisis reaction faculty improve public comprehension, evaluation, predisposition, quicken emergency reaction, and backing post-incidence administration.

\vspace{-2mm}
\bibliographystyle{plainnat}

\bibliography{References}

\begin{thebibliography}{47}
\providecommand{\natexlab}[1]{#1}
\providecommand{\url}[1]{\texttt{#1}}
\expandafter\ifx\csname urlstyle\endcsname\relax
  \providecommand{\doi}[1]{doi: #1}\else
  \providecommand{\doi}{doi: \begingroup \urlstyle{rm}\Url}\fi

\bibitem[Agarwal et~al.(2011)Agarwal, Xie, Vovsha, Rambow, and
  Passonneau]{agarwal2011sentiment}
Apoorv Agarwal, Boyi Xie, Ilia Vovsha, Owen Rambow, and Rebecca~J Passonneau.
\newblock Sentiment analysis of twitter data.
\newblock In \emph{Proceedings of the workshop on language in social media (LSM
  2011)}, pages 30--38, 2011.

\bibitem[Al~Helal and Mouhoub(2018)]{al2018topic}
Mustakim Al~Helal and Malek Mouhoub.
\newblock Topic modelling in bangla language: An lda approach to optimize
  topics and news classification.
\newblock \emph{Computer and Information Science}, 11\penalty0 (4), 2018.

\bibitem[Alm et~al.(2005)Alm, Roth, and Sproat]{alm2005emotions}
Cecilia~Ovesdotter Alm, Dan Roth, and Richard Sproat.
\newblock Emotions from text: machine learning for text-based emotion
  prediction.
\newblock In \emph{Proceedings of human language technology conference and
  conference on empirical methods in natural language processing}, pages
  579--586, 2005.

\bibitem[AlSumait et~al.(2008)AlSumait, Barbar{\'a}, and
  Domeniconi]{alsumait2008line}
Loulwah AlSumait, Daniel Barbar{\'a}, and Carlotta Domeniconi.
\newblock On-line lda: Adaptive topic models for mining text streams with
  applications to topic detection and tracking.
\newblock In \emph{2008 eighth IEEE international conference on data mining},
  pages 3--12. IEEE, 2008.

\bibitem[Asimuzzaman et~al.(2017)Asimuzzaman, Nath, Hossain, Hossain, and
  Rahman]{asimuzzaman2017sentiment}
Md~Asimuzzaman, Pinku~Deb Nath, Farah Hossain, Asif Hossain, and Rashedur~M
  Rahman.
\newblock Sentiment analysis of bangla microblogs using adaptive neuro fuzzy
  system.
\newblock In \emph{2017 13th International Conference on Natural Computation,
  Fuzzy Systems and Knowledge Discovery (ICNC-FSKD)}, pages 1631--1638. IEEE,
  2017.

\bibitem[Balasubramaniam et~al.(2020)Balasubramaniam, Nayak, and
  Bashar]{balasubramaniam2020understanding}
Thirunavukarasu Balasubramaniam, Richi Nayak, and Md~Abul Bashar.
\newblock Understanding the spatio-temporal topic dynamics of covid-19 using
  nonnegative tensor factorization: A case study.
\newblock \emph{arXiv preprint arXiv:2009.09253}, 2020.

\bibitem[Bashar and Nayak(2020)]{bashar2020qutnocturnal}
Md~Abul Bashar and Richi Nayak.
\newblock Qutnocturnal@ hasoc'19: Cnn for hate speech and offensive content
  identification in hindi language.
\newblock \emph{arXiv preprint arXiv:2008.12448}, 2020.

\bibitem[Bashar et~al.(2018)Bashar, Nayak, Suzor, and
  Weir]{bashar2018misogynistic}
Md~Abul Bashar, Richi Nayak, Nicolas Suzor, and Bridget Weir.
\newblock Misogynistic tweet detection: Modelling cnn with small datasets.
\newblock In \emph{Australasian Conference on Data Mining}, pages 3--16.
  Springer, 2018.

\bibitem[Bashar et~al.(2020{\natexlab{a}})Bashar, Nayak, and
  Balasubramaniam]{bashar2020topic}
Md~Abul Bashar, Richi Nayak, and Thirunavukarasu Balasubramaniam.
\newblock Topic, sentiment and impact analysis: Covid19 information seeking on
  social media.
\newblock \emph{arXiv preprint arXiv:2008.12435}, 2020{\natexlab{a}}.

\bibitem[Bashar et~al.(2020{\natexlab{b}})Bashar, Nayak, and
  Suzor]{bashar2020regularising}
Md~Abul Bashar, Richi Nayak, and Nicolas Suzor.
\newblock Regularising lstm classifier by transfer learning for detecting
  misogynistic tweets with small training set.
\newblock \emph{Knowledge and Information Systems}, 62\penalty0 (10):\penalty0
  4029--4054, 2020{\natexlab{b}}.

\bibitem[Bijalwan et~al.(2014)Bijalwan, Kumar, Kumari, and
  Pascual]{bijalwan2014knn}
Vishwanath Bijalwan, Vinay Kumar, Pinki Kumari, and Jordan Pascual.
\newblock Knn based machine learning approach for text and document mining.
\newblock \emph{International Journal of Database Theory and Application},
  7\penalty0 (1):\penalty0 61--70, 2014.

\bibitem[Blei and Lafferty(2006)]{blei2006dynamic}
David~M Blei and John~D Lafferty.
\newblock Dynamic topic models.
\newblock In \emph{Proceedings of the 23rd international conference on Machine
  learning}, pages 113--120, 2006.

\bibitem[Blei et~al.(2003)Blei, Ng, and Jordan]{blei2003latent}
David~M Blei, Andrew~Y Ng, and Michael~I Jordan.
\newblock Latent dirichlet allocation.
\newblock \emph{Journal of machine Learning research}, 3\penalty0
  (Jan):\penalty0 993--1022, 2003.

\bibitem[Chowdhury and Chowdhury(2014)]{chowdhury2014performing}
Shaika Chowdhury and Wasifa Chowdhury.
\newblock Performing sentiment analysis in bangla microblog posts.
\newblock In \emph{2014 International Conference on Informatics, Electronics \&
  Vision (ICIEV)}, pages 1--6. IEEE, 2014.

\bibitem[Chy et~al.(2014)Chy, Seddiqui, and Das]{chy2014bangla}
Abu~Nowshed Chy, Md~Hanif Seddiqui, and Sowmitra Das.
\newblock Bangla news classification using naive bayes classifier.
\newblock In \emph{16th Int'l Conf. Computer and Information Technology}, pages
  366--371. IEEE, 2014.

\bibitem[Cui et~al.(2006)Cui, Mittal, and Datar]{cui2006comparative}
Hang Cui, Vibhu Mittal, and Mayur Datar.
\newblock Comparative experiments on sentiment classification for online
  product reviews.
\newblock In \emph{AAAI}, volume~6, page~30, 2006.

\bibitem[Dagum(2010)]{dagum2010time}
Estela~Bee Dagum.
\newblock Time series modeling and decomposition.
\newblock \emph{Statistica}, 70\penalty0 (4):\penalty0 433--457, 2010.

\bibitem[Das and Bandyopadhyay(2010{\natexlab{a}})]{das2010sentiwordnet}
Amitava Das and Sivaji Bandyopadhyay.
\newblock Sentiwordnet for bangla.
\newblock \emph{Knowledge Sharing Event-4: Task}, 2:\penalty0 1--8,
  2010{\natexlab{a}}.

\bibitem[Das and Bandyopadhyay(2010{\natexlab{b}})]{das2010topic}
Amitava Das and Sivaji Bandyopadhyay.
\newblock Topic-based bengali opinion summarization.
\newblock In \emph{Coling 2010: Posters}, pages 232--240, 2010{\natexlab{b}}.

\bibitem[De~Santis et~al.(2020)De~Santis, Martino, and
  Rizzi]{de2020infoveillance}
Enrico De~Santis, Alessio Martino, and Antonello Rizzi.
\newblock An infoveillance system for detecting and tracking relevant topics
  from italian tweets during the covid-19 event.
\newblock \emph{IEEE Access}, 8:\penalty0 132527--132538, 2020.

\bibitem[Dieng et~al.(2019)Dieng, Ruiz, and Blei]{dieng2019dynamic}
Adji~B Dieng, Francisco~JR Ruiz, and David~M Blei.
\newblock The dynamic embedded topic model.
\newblock \emph{arXiv preprint arXiv:1907.05545}, 2019.

\bibitem[Eshan and Hasan(2017)]{eshan2017application}
Shahnoor~C Eshan and Mohammad~S Hasan.
\newblock An application of machine learning to detect abusive bengali text.
\newblock In \emph{2017 20th International Conference of Computer and
  Information Technology (ICCIT)}, pages 1--6. IEEE, 2017.

\bibitem[Han et~al.(2020)Han, Wang, Zhang, and Wang]{han2020using}
Xuehua Han, Juanle Wang, Min Zhang, and Xiaojie Wang.
\newblock Using social media to mine and analyze public opinion related to
  covid-19 in china.
\newblock \emph{International Journal of Environmental Research and Public
  Health}, 17\penalty0 (8):\penalty0 2788, 2020.

\bibitem[Hasan et~al.(2014)Hasan, Rahman, et~al.]{hasan2014sentiment}
KM~Azharul Hasan, Mosiur Rahman, et~al.
\newblock Sentiment detection from bangla text using contextual valency
  analysis.
\newblock In \emph{2014 17th International Conference on Computer and
  Information Technology (ICCIT)}, pages 292--295. IEEE, 2014.

\bibitem[Hasan et~al.(2019)Hasan, Hossain, Ahmed, and Rahman]{hasan2019topic}
M.~Hasan, M.~M. Hossain, A.~Ahmed, and M.~S. Rahman.
\newblock Topic modelling: A comparison of the performance of latent dirichlet
  allocation and lda2vec model on bangla newspaper.
\newblock In \emph{2019 International Conference on Bangla Speech and Language
  Processing (ICBSLP)}, pages 1--5. IEEE, 2019.

\bibitem[Hassan et~al.(2016)Hassan, Amin, Al~Azad, and
  Mohammed]{hassan2016sentiment}
Asif Hassan, Mohammad~Rashedul Amin, Abul~Kalam Al~Azad, and Nabeel Mohammed.
\newblock Sentiment analysis on bangla and romanized bangla text using deep
  recurrent models.
\newblock In \emph{2016 International Workshop on Computational Intelligence
  (IWCI)}, pages 51--56. IEEE, 2016.

\bibitem[Islam et~al.(2017)Islam, Jubayer, Ahmed, et~al.]{islam2017comparative}
Md~Islam, Fazla Elahi~Md Jubayer, Syed~Ikhtiar Ahmed, et~al.
\newblock A comparative study on different types of approaches to bengali
  document categorization.
\newblock \emph{arXiv preprint arXiv:1701.08694}, 2017.

\bibitem[Jagtap and Dhotre(2014)]{jagtap2014svm}
Balaji Jagtap and Virendrakumar Dhotre.
\newblock Svm and hmm based hybrid approach of sentiment analysis for teacher
  feedback assessment.
\newblock \emph{International Journal of Emerging Trends \& Technology in
  Computer Science (IJETTCS)}, 3\penalty0 (3):\penalty0 229--232, 2014.

\bibitem[Kabir et~al.(2015)Kabir, Siddique, Kotwal, and Huda]{kabir2015bangla}
Fasihul Kabir, Sabbir Siddique, Mohammed Rokibul~Alam Kotwal, and
  Mohammad~Nurul Huda.
\newblock Bangla text document categorization using stochastic gradient descent
  (sgd) classifier.
\newblock In \emph{2015 International Conference on Cognitive Computing and
  Information Processing (CCIP)}, pages 1--4. IEEE, 2015.

\bibitem[Liu et~al.(2010)Liu, Lv, Liu, and Shi]{liu2010study}
Zhijie Liu, Xueqiang Lv, Kun Liu, and Shuicai Shi.
\newblock Study on svm compared with the other text classification methods.
\newblock In \emph{2010 Second international workshop on education technology
  and computer science}, volume~1, pages 219--222. IEEE, 2010.

\bibitem[Mahtab et~al.(2018)Mahtab, Islam, and Rahaman]{mahtab2018sentiment}
Shamsul~Arafin Mahtab, Nazmul Islam, and Md~Mahfuzur Rahaman.
\newblock Sentiment analysis on bangladesh cricket with support vector machine.
\newblock In \emph{2018 International Conference on Bangla Speech and Language
  Processing (ICBSLP)}, pages 1--4. IEEE, 2018.

\bibitem[Mandal and Sen(2014)]{mandal2014supervised}
Ashis~Kumar Mandal and Rikta Sen.
\newblock Supervised learning methods for bangla web document categorization.
\newblock \emph{arXiv preprint arXiv:1410.2045}, 2014.

\bibitem[Marjanen et~al.(2020)Marjanen, Zosa, Hengchen, Pivovarova, and
  Tolonen]{marjanen2020topic}
Jani Marjanen, Elaine Zosa, Simon Hengchen, Lidia Pivovarova, and Mikko
  Tolonen.
\newblock Topic modelling discourse dynamics in historical newspapers.
\newblock \emph{arXiv preprint arXiv:2011.10428}, 2020.

\bibitem[Nguyen et~al.(2020)Nguyen, Kiyoaki, and Huynh]{nguyentopics}
Ba-Hung Nguyen, Shirai Kiyoaki, and Van-Nam Huynh.
\newblock Topics in financial filings and bankruptcy prediction with
  distributed representations of textual data.
\newblock In \emph{Proceedings of ECML-PKDD}, 2020.

\bibitem[Pal et~al.(2015)Pal, Saha, and Dash]{pal2015automatic}
Alok~Ranjan Pal, Diganta Saha, and Niladri~Sekhar Dash.
\newblock Automatic classification of bengali sentences based on sense
  definitions present in bengali wordnet.
\newblock \emph{arXiv preprint arXiv:1508.01349}, 2015.

\bibitem[Patil and Pawar(2012)]{patil2012automated}
Ajay~S Patil and BV~Pawar.
\newblock Automated classification of web sites using naive bayesian algorithm.
\newblock In \emph{Proceedings of the international multiconference of
  engineers and computer scientists}, volume~1, pages 519--523. Citeseer, 2012.

\bibitem[Pawar and Gawande(2012)]{pawar2012comparative}
Pratiksha~Y Pawar and SH~Gawande.
\newblock A comparative study on different types of approaches to text
  categorization.
\newblock \emph{International Journal of Machine Learning and Computing},
  2\penalty0 (4):\penalty0 423, 2012.

\bibitem[Rahman et~al.(2019)Rahman, Abujar, Chowdhury, Saifuzzaman, and
  Hossain]{rahman2019sentence}
Shahinur Rahman, Sheikh Abujar, SM~Mazharul~Hoque Chowdhury, Mohd Saifuzzaman,
  and Syed~Akhter Hossain.
\newblock Sentence-based topic modeling using lexical analysis.
\newblock In \emph{Emerging Technologies in Data Mining and Information
  Security}, pages 487--494. Springer, 2019.

\bibitem[Rakshit et~al.(2015)Rakshit, Ghosh, Bhattacharyya, and
  Haffari]{rakshit2015automated}
Geetanjali Rakshit, Anupam Ghosh, Pushpak Bhattacharyya, and Gholamreza
  Haffari.
\newblock Automated analysis of bangla poetry for classification and poet
  identification.
\newblock In \emph{Proceedings of the 12th International Conference on Natural
  Language Processing}, pages 247--253, 2015.

\bibitem[Tabassum and Khan(2019)]{tabassum2019design}
Nusrath Tabassum and Muhammad~Ibrahim Khan.
\newblock Design an empirical framework for sentiment analysis from bangla text
  using machine learning.
\newblock In \emph{2019 International Conference on Electrical, Computer and
  Communication Engineering (ECCE)}, pages 1--5. IEEE, 2019.

\bibitem[Tam et~al.(2002)Tam, Santoso, and Setiono]{tam2002comparative}
Vincent Tam, Ardi Santoso, and Rudy Setiono.
\newblock A comparative study of centroid-based, neighborhood-based and
  statistical approaches for effective document categorization.
\newblock In \emph{Object recognition supported by user interaction for service
  robots}, volume~4, pages 235--238. IEEE, 2002.

\bibitem[Tong and Zhang(2016)]{tong2016text}
Zhou Tong and Haiyi Zhang.
\newblock A text mining research based on lda topic modelling.
\newblock In \emph{International Conference on Computer Science, Engineering
  and Information Technology}, pages 201--210, 2016.

\bibitem[Tripto and Ali(2018)]{tripto2018detecting}
Nafis~Irtiza Tripto and Mohammed~Eunus Ali.
\newblock Detecting multilabel sentiment and emotions from bangla youtube
  comments.
\newblock In \emph{2018 International Conference on Bangla Speech and Language
  Processing (ICBSLP)}, pages 1--6. IEEE, 2018.

\bibitem[Tuhin et~al.(2019)Tuhin, Paul, Nawrine, Akter, and
  Das]{tuhin2019automated}
Rashedul~Amin Tuhin, Bechitra~Kumar Paul, Faria Nawrine, Mahbuba Akter, and
  Amit~Kumar Das.
\newblock An automated system of sentiment analysis from bangla text using
  supervised learning techniques.
\newblock In \emph{2019 IEEE 4th International Conference on Computer and
  Communication Systems (ICCCS)}, pages 360--364. IEEE, 2019.

\bibitem[Wang and Blei(2011)]{wang2011collaborative}
Chong Wang and David~M Blei.
\newblock Collaborative topic modeling for recommending scientific articles.
\newblock In \emph{Proceedings of the 17th ACM SIGKDD international conference
  on Knowledge discovery and data mining}, pages 448--456, 2011.

\bibitem[Wayasti et~al.(2018)Wayasti, Surjandari, et~al.]{wayasti2018mining}
Reggia~Aldiana Wayasti, Isti Surjandari, et~al.
\newblock Mining customer opinion for topic modeling purpose: Case study of
  ride-hailing service provider.
\newblock In \emph{2018 6th International Conference on Information and
  Communication Technology (ICoICT)}, pages 305--309. IEEE, 2018.

\bibitem[Zhao et~al.(2011)Zhao, Jiang, Weng, He, Lim, Yan, and
  Li]{zhao2011comparing}
Wayne~Xin Zhao, Jing Jiang, Jianshu Weng, Jing He, Ee-Peng Lim, Hongfei Yan,
  and Xiaoming Li.
\newblock Comparing twitter and traditional media using topic models.
\newblock In \emph{European conference on information retrieval}, pages
  338--349. Springer, 2011.

\end{thebibliography}
\end{document}